\documentclass[12pt,a4paper]{article}
\usepackage{jheplikemod,ifpdf,tocloft,rotating,amsmath,textcomp,mathrsfs,mathtools,xcolor,braket}
\usepackage{tikz}
\usetikzlibrary{calc,decorations.markings,shapes.geometric}
\usepackage[backgroundcolor=green!5, linecolor=blue!60]{todonotes}
\definecolor{mhvblue2}{rgb}{0.3,0.3,0.575}
\definecolor{mhvblue}{rgb}{0.6,0.6,0.7765}
\definecolor{nmhvred}{rgb}{0.6765,0.15,0.3}
\definecolor{ampgrey}{rgb}{0.9,0.9,0.9}
\definecolor{unord}{rgb}{0,0,0}
\definecolor{ord}{rgb}{0,0,0.575}
\definecolor{anchorLeg}{rgb}{0.575,0.0,0.225}
\definecolor{labelcolor}{rgb}{0,0,0}
\definecolor{extColor}{rgb}{0,0,0}

\newcounter{legSteps}
\newcounter{offset}

\def\figScale{0.9}
\def\legSpread{4}

\def\edgeLen{1*\figScale}

\pgfmathsetmacro{\pLen}{\edgeLen/(2*sin(72/2))}
\def\legLen{\edgeLen*0.45}

\def\labelDist{\legLen*1.5}
\def\lineThickness{(1pt)}
\def\dotSize{(\figScale*12pt)}
\def\ampSize{(1*\figScale*12pt)}
\def\eph{0.4}
\def\pageW{15.175/\figScale}

\tikzset{fullamp/.style={coordinate,minimum size=0.7*\ampSize,ball color=black!20,circle,text=white,inner sep=0}}
\tikzset{fullmhv/.style={coordinate,minimum size=0.8*\ampSize,ball color=mhvblue,circle,text=white,inner sep=0}}
\tikzset{fullmhvBig/.style={coordinate,minimum size=1*\ampSize,ball color=mhvblue,circle,text=white,inner sep=0}}
\tikzset{fullnmhv/.style={coordinate,minimum size=0.8*\ampSize,ball color=nmhvred,circle,text=white,inner sep=0}}
\tikzset{fullmhvBar/.style={coordinate,minimum size=0.8*\ampSize,ball color=white,circle,text=white,inner sep=0}}
\tikzset{ordAmp/.style={fill=ampgrey,circle,draw=black,line width=\lineThickness,minimum size=0.6*\ampSize,text=white,inner sep=0}}
\tikzset{mhv/.style={fill=mhvblue,circle,draw=black,line width=\lineThickness,minimum size=0.7*\ampSize,text=white,inner sep=0}}
\tikzset{nmhv/.style={fill=nmhvred,circle,draw=black,line width=\lineThickness,minimum size=0.7*\ampSize,text=white,inner sep=0}}
\tikzset{mhvBar/.style={fill=white,circle,draw=black,line width=\lineThickness,minimum size=0.7*\ampSize,text=white,inner sep=0}}
\tikzset{rEdge/.style={anchorLeg,line width=\lineThickness,line cap=round}}
\tikzset{fgraphEdge/.style={anchorLeg,line width=0.8*\lineThickness,line cap=round}}
\tikzset{fgraphExt/.style={ord,line width=\lineThickness,line cap=round}}
\tikzset{fgraphOpt/.style={ord,dotted,line width=\lineThickness,line cap=round}}
\tikzset{fdot/.style={fill=anchorLeg,circle,minimum size=0.35*\ampSize,inner sep=0}}
\tikzset{bdot/.style={fill=black,circle,minimum size=0.35*\ampSize,inner sep=0}}
\tikzset{ephdot/.style={fill=black,circle,minimum size=0.125*\ampSize,inner sep=0}}
\tikzset{ext/.style={black,line width=\lineThickness,line cap=round,rounded corners=0.1pt}}
\tikzset{under/.style={white,line width=4*\lineThickness,line cap=round}}
\tikzset{optExt/.style={black,dotted,line width=\lineThickness,line cap=round,rounded corners=10pt}}
\tikzset{optExtSc/.style={black,line width=\lineThickness,line cap=round,rounded corners=10pt}}
\tikzset{dashed/.style={black!70,dotted,line width=\lineThickness,line cap=round,rounded corners=10pt}}
\tikzset{ddot/.style={fill=black,circle,minimum size=0.275*\dotSize,inner sep=0}}
\tikzset{rdot/.style={fill=hred,circle,minimum size=0.275*\dotSize,inner sep=0}}
\tikzset{bldot/.style={fill=hblue,circle,minimum size=0.275*\dotSize,inner sep=0}}
\tikzset{int/.style={black,line width=\lineThickness,line cap=round,rounded corners=1.5pt}}

\tikzset{intInfR/.style={nmhvred,line width=\lineThickness,line cap=round,rounded corners=1.5pt}}
\tikzset{blueDot/.style={fill=mhvblue,circle,draw=black,line width=\lineThickness,minimum size=0.5*\ampSize,text=white,inner sep=0}}
\tikzset{whiteDot/.style={fill=white,circle,draw=black,line width=\lineThickness,minimum size=0.5*\ampSize,text=white,inner sep=0}}
\tikzset{blackDot/.style={fill=black,circle,minimum size=0.5*\ampSize,inner sep=0}}
\tikzset{compositeDot/.style={fill=none,draw=black,line width=\lineThickness,circle,minimum size=0.75*\ampSize,inner sep=0}}

\tikzset{directedEdge/.style={draw=none,decoration={markings,mark connection node=connode,mark=at position 0.5 with {\node[transform shape, scale=0.205*\figScale,shape=dart,aspect=0.5,fill=black,draw] (connode) {};}},postaction={decorate}}}

\tikzset{directedEdgeBend/.style={rounded corners=10pt,draw=none,decoration={markings,mark connection node=connode,mark=at position 0.5 with {\node[transform shape, scale=0.205,shape=dart,aspect=0.5,fill=black,draw] (connode) {};}},postaction={decorate}}}

\newcommand{\dimLines}{\tikzset{int/.style={black!25,line width=\lineThickness,line cap=round,rounded corners=1.5pt}}
\tikzset{ext/.style={black!25,line width=\lineThickness,line cap=round,rounded corners=0.1pt}}
\definecolor{extColor}{rgb}{0.75,0.75,0.75}}

\newcommand{\restoreDark}{
\tikzset{int/.style={black,line width=\lineThickness,line cap=round,rounded corners=1.5pt}}
\tikzset{ext/.style={black,line width=\lineThickness,line cap=round,rounded corners=0.1pt}}
\definecolor{extColor}{rgb}{0,0,0}}

\newcommand{\leg}[3]{\draw[ext] #1--($#1+(#2:\legLen)$);\node at ($#1+(#2:\labelDist)$)[]{{\footnotesize #3}};}

\newcommand{\legMassive}[3]{\draw[ext,fill=extColor] #1--($#1+(#2+\legSpread*2:\legLen)$)--($#1+(#2-\legSpread*2:\legLen)$)--#1;\node at ($#1+(#2:\labelDist)$)[]{{\footnotesize #3}};}

\def\boundingDraw{red}
\def\boundingDraw{none}

\newcommand{\contourVerts}[7]{
\ifthenelse{#1=1}{\node at (v1) [whiteDot] {};}{\ifthenelse{#1=2}{\node at (v1) [blueDot] {};}{\ifthenelse{#1=4}{\node at (v1) [blackDot] {};}{\ifthenelse{#1=3}{\node at (v1) [bdot] {};\node at (v1) [compositeDot] {};
}{\ifthenelse{#1=0}{\node at (v1) [bdot] {};}{}}}}}

\ifthenelse{#2=1}{\node at (v2) [whiteDot] {};}{\ifthenelse{#2=2}{\node at (v2) [blueDot] {};}{\ifthenelse{#2=4}{\node at (v2) [blackDot] {};}{\ifthenelse{#2=3}{\node at (v2) [bdot] {};\node at (v2) [compositeDot] {};
}{\ifthenelse{#2=0}{\node at (v2) [bdot] {};}{}}}}}

\ifthenelse{#3=1}{\node at (v3) [whiteDot] {};}{\ifthenelse{#3=2}{\node at (v3) [blueDot] {};}{\ifthenelse{#3=4}{\node at (v3) [blackDot] {};}{\ifthenelse{#3=3}{\node at (v3) [bdot] {};\node at (v3) [compositeDot] {};
}{\ifthenelse{#3=0}{\node at (v3) [bdot] {};}{}}}}}

\ifthenelse{#4=1}{\node at (v4) [whiteDot] {};}{\ifthenelse{#4=2}{\node at (v4) [blueDot] {};}{\ifthenelse{#4=4}{\node at (v4) [blackDot] {};}{\ifthenelse{#4=3}{\node at (v4) [bdot] {};\node at (v4) [compositeDot] {};
}{\ifthenelse{#4=0}{\node at (v4) [bdot] {};}{}}}}}

\ifthenelse{#5=1}{\node at (v5) [whiteDot] {};}{\ifthenelse{#5=2}{\node at (v5) [blueDot] {};}{\ifthenelse{#5=4}{\node at (v5) [blackDot] {};}{\ifthenelse{#5=3}{\node at (v5) [bdot] {};\node at (v5) [compositeDot] {};
}{\ifthenelse{#5=0}{\node at (v5) [bdot] {};}{}}}}}
\ifthenelse{#6=1}{\node at (v6) [whiteDot] {};}{\ifthenelse{#6=2}{\node at (v6) [blueDot] {};}{\ifthenelse{#6=4}{\node at (v6) [blackDot] {};}{\ifthenelse{#6=3}{\node at (v6) [bdot] {};\node at (v6) [compositeDot] {};
}{\ifthenelse{#6=0}{\node at (v6) [bdot] {};}{}}}}}
\ifthenelse{#7=1}{\node at (v7) [whiteDot] {};}{\ifthenelse{#7=2}{\node at (v7) [blueDot] {};}{\ifthenelse{#7=4}{\node at (v7) [blackDot] {};}{\ifthenelse{#7=3}{\node at (v7) [bdot] {};\node at (v7) [compositeDot] {};
}{\ifthenelse{#7=0}{\node at (v7) [bdot] {};}{}}}}}
}

\definecolor{legColour}{rgb}{0.35,0.35,0.35}
\definecolor{ndotColor}{rgb}{0.65,0.25,0.25}
\def\markStroke{0.65}

\tikzset{ndot/.style={transform shape,scale=0.65*\figScale,aspect=0.75,draw=ndotColor,line width=1.4*\markStroke*\figScale,shape=circle,fill=none}}
\tikzset{markedEdgeR/.style={draw=none,decoration={markings,mark connection node=connode,mark=at position 0.5 with {\node[ndotR] (connode) {};}},postaction={decorate}}}
\tikzset{ndotR/.style={transform shape,scale=0.65*\figScale,aspect=0.65,draw=ndotColor,line width=\markStroke*\figScale,shape=circle,fill=ndotColor}}
\tikzset{markedEdge/.style={draw=none,decoration={markings,mark connection node=connode,mark=at position 0.5 with {\node[ndot] (connode) {};}},postaction={decorate}}}

\newcommand{\tikzBox}[2][0.5]{\begin{tikzpicture}[scale=1,baseline=-3.05,rotate=0]\useasboundingbox ($\figScale*(-#1,-#1)$) rectangle ($\figScale*(#1,#1)$);#2\end{tikzpicture}}

\newcommand{\tikzBoxDeux}[1]{\begin{tikzpicture}[scale=\figScale,baseline=-3.05]\useasboundingbox ($(-\pageW/12,-1.2)$) rectangle ($(\pageW/12,1.2)$);\draw[int,line width=0.1,red,draw=\boundingDraw] ($(-\pageW/12,-1.2)$) rectangle ($(\pageW/12,1.2)$);
#1\end{tikzpicture}}

\newcommand{\kBox}[1]{\begin{tikzpicture}[scale=\figScale,baseline=-3.05]\useasboundingbox ($(-\pageW/9,-1.35)$) rectangle ($(\pageW/9,1.35)$);\draw[int,line width=0.1,red,draw=\boundingDraw] ($(-\pageW/9,-1.35)$) rectangle ($(\pageW/9,1.35)$);
\coordinate(v4)at(0,0);\coordinate(v1)at($\figScale*(-0.65-0.1,-0.65)$);\coordinate(v2)at($\figScale*(-1.3-0.0,0)$);\coordinate(v3)at($\figScale*(-0.65-0.1,0.65)$);\coordinate(v5)at($\figScale*(0.65+0.1,0.65)$);\coordinate(v6)at($\figScale*(1.3+0.0,-0)$);\coordinate(v7)at($\figScale*(0.65+0.1,-0.65)$);
#1\end{tikzpicture}}

\newcommand{\pBox}[1]{\begin{tikzpicture}[scale=1*\figScale,baseline=-3.05]\useasboundingbox ($(-\pageW/9,-1.4)$) rectangle ($(\pageW/9,1.4)$);\draw[int,line width=0.1,red,draw=\boundingDraw] ($(-\pageW/9,-1.4)$) rectangle ($(\pageW/9,1.4)$);\coordinate(v2)at($\figScale*(-1.23125,0)$);\coordinate(v1)at($(v2)+(-54:\figScale*0.919)$);\coordinate(v3)at($(v2)+(54:\figScale*0.919)$);\coordinate(v4)at($(v3)+(54-72:\figScale*0.919)$);\coordinate(v5)at($(v4)+(0:\figScale*0.919)$);\coordinate(v6)at($(v5)+(-90:\figScale*0.919)$);\coordinate(v7)at($(v4)+(-90:\figScale*0.919)$);
#1\end{tikzpicture}}
\newcommand{\pBoxB}[1]{\begin{tikzpicture}[scale=1.2*\figScale,baseline=-3.05]\useasboundingbox ($(-\pageW/9,-1.4)$) rectangle ($(\pageW/9,1.4)$);\draw[int,line width=0.1,red,draw=\boundingDraw] ($(-\pageW/9,-1.4)$) rectangle ($(\pageW/9,1.4)$);\coordinate(v2)at($\figScale*(-1.23125,0)$);\coordinate(v1)at($(v2)+(-54:\figScale*0.919)$);\coordinate(v3)at($(v2)+(54:\figScale*0.919)$);\coordinate(v4)at($(v3)+(54-72:\figScale*0.919)$);\coordinate(v5)at($(v4)+(0:\figScale*0.919)$);\coordinate(v6)at($(v5)+(-90:\figScale*0.919)$);\coordinate(v7)at($(v4)+(-90:\figScale*0.919)$);
#1\end{tikzpicture}}

\newcommand{\kTbox}[1]{\begin{tikzpicture}[scale=\figScale,baseline=-3.05]\useasboundingbox ($(-\pageW/9,-1.35)$) rectangle ($(\pageW/9,1.35)$);\draw[int,line width=0.1,red,draw=\boundingDraw] ($(-\pageW/9,-1.35)$) rectangle ($(\pageW/9,1.35)$);
\coordinate(v3)at(-0.3,0);\coordinate(v1)at($\figScale*(-0.65-0.6,-0.65)$);\coordinate(v2)at($\figScale*(-0.65-0.6,0.65)$);\coordinate(v4)at($\figScale*(0.65-0.1,0.65)$);\coordinate(v5)at($\figScale*(1.3-0.1,-0)$);\coordinate(v6)at($\figScale*(0.65-0.1,-0.65)$);
#1\end{tikzpicture}}

\newcommand{\dBox}[1]{\begin{tikzpicture}[scale=1.*\figScale,baseline=-3.05]\useasboundingbox ($(-\pageW/9,-1.4)$) rectangle ($(\pageW/9,1.4)$);\draw[int,line width=0.1,red,draw=\boundingDraw] ($(-\pageW/9,-1.4)$) rectangle ($(\pageW/9,1.4)$);\coordinate(v7)at($\figScale*(0,0)$);\coordinate(v6)at($(v7)+(-90:\figScale*0.65)$);\coordinate(v3)at($(v7)+(90:\figScale*0.65)$);\coordinate(v2)at($(v3)+(180:\figScale*1.25)$);\coordinate(v4)at($(v3)+(0:\figScale*1.25)$);\coordinate(v1)at($(v6)+(180:\figScale*1.25)$);\coordinate(v5)at($(v6)+(0:\figScale*1.25)$);
#1\end{tikzpicture}}

\newcommand{\pT}[1]{\begin{tikzpicture}[scale=1.*\figScale,baseline=-3.05]\useasboundingbox ($(-\pageW/9,-1.5)$) rectangle ($(\pageW/9,1.5)$);\draw[int,line width=0.1,red,draw=\boundingDraw] ($(-\pageW/9,-1.5)$) rectangle ($(\pageW/9,1.5)$);\coordinate(v2)at($\figScale*(-1.25,0)$);\coordinate(v1)at($(v2)+(-54:\figScale*1.119)$);\coordinate(v3)at($(v2)+(54:\figScale*1.119)$);\coordinate(v4)at($\figScale*(0.3,0.65)$);\coordinate(v5)at($\figScale*(1.2,-0)$);\coordinate(v6)at($\figScale*(0.3,-0.65)$);
#1\end{tikzpicture}}

\newcommand{\kT}[1]{\begin{tikzpicture}[scale=\figScale,baseline=-3.05]\useasboundingbox ($(-\pageW/9,-1.2)$) rectangle ($(\pageW/9,1.2)$);\draw[int,line width=0.1,red,draw=\boundingDraw] ($(-\pageW/9,-1.2)$) rectangle ($(\pageW/9,1.2)$);
\coordinate(v3)at(-0.0,0);\coordinate(v1)at($\figScale*(-0.65-0.2,-0.65)$);\coordinate(v2)at($\figScale*(-0.65-0.2,0.65)$);\coordinate(v4)at($\figScale*(0.65+0.2,0.65)$);\coordinate(v5)at($\figScale*(0.65+0.2,-0.65)$);
#1\end{tikzpicture}}

\newcommand{\bT}[1]{\begin{tikzpicture}[scale=\figScale,baseline=-3.05]\useasboundingbox ($(-\pageW/9,-1.2)$) rectangle ($(\pageW/9,1.2)$);\draw[int,line width=0.1,red,draw=\boundingDraw] ($(-\pageW/9,-1.2)$) rectangle ($(\pageW/9,1.2)$);\coordinate(v7)at($\figScale*(0.2,0)$);\coordinate(v5)at($(v7)+(-90:\figScale*0.65)$);\coordinate(v3)at($(v7)+(90:\figScale*0.65)$);\coordinate(v2)at($(v3)+(180:\figScale*1.25)$);\coordinate(v4)at($(v7)+(0:\figScale*0.9)$);\coordinate(v1)at($(v5)+(180:\figScale*1.25)$);
#1\end{tikzpicture}}

\newcommand{\dT}[1]{\begin{tikzpicture}[scale=\figScale,baseline=-3.05]\useasboundingbox ($(-\pageW/9,-1.2)$) rectangle ($(\pageW/9,1.2)$);\draw[int,line width=0.1,red,draw=\boundingDraw] ($(-\pageW/9,-1.2)$) rectangle ($(\pageW/9,1.2)$);\coordinate(v1)at($(180:\figScale*1.05)$);\coordinate(v2)at($(90:\figScale*0.65)$);\coordinate(v3)at($(0:\figScale*1.05)$);\coordinate(v4)at($(-90:\figScale*0.65)$);
#1\end{tikzpicture}}

\newcommand{\kBoxPlainEdges}{\draw[int](v4)--(v1);\draw[int](v1)--(v2);\draw[int](v2)--(v3);\draw[int](v3)--(v4);\draw[int](v4)--(v5);\draw[int](v5)--(v6);\draw[int](v6)--(v7);\draw[int](v7)--(v4);}

\newcommand{\pBoxPlainEdges}{\draw[int](v7)--(v1);\draw[int](v1)--(v2);\draw[int](v2)--(v3);\draw[int](v3)--(v4);\draw[int](v4)--(v5);\draw[int](v5)--(v6);\draw[int](v6)--(v7);\draw[int](v7)--(v4);}

\newcommand{\kTboxPlainEdges}{\draw[int](v3)--(v1);\draw[int](v1)--(v2);\draw[int](v2)--(v3);\draw[int](v3)--(v4);\draw[int](v4)--(v5);\draw[int](v5)--(v6);\draw[int](v6)--(v3);}

\newcommand{\dBoxPlainEdges}{\draw[int](v6)--(v1);\draw[int](v1)--(v2);\draw[int](v2)--(v3);\draw[int](v3)--(v4);\draw[int](v4)--(v5);\draw[int](v5)--(v6);\draw[int](v6)--(v3);}

\newcommand{\pTPlainEdges}{\draw[int](v6)--(v1);\draw[int](v1)--(v2);\draw[int](v2)--(v3);\draw[int](v3)--(v4);\draw[int](v4)--(v5);\draw[int](v5)--(v6);\draw[int](v6)--(v4);}

\newcommand{\kBoxLegs}[7]{
\setcounter{legSteps}{0}
\def\zeroAngle{-90}\def\spread{45}\setcounter{offset}{-1}\addtocounter{offset}{#1}
\ifthenelse{#1=0}{}{\ifthenelse{#1=1}{\stepcounter{legSteps}\leg{(v1)}{\zeroAngle}{\arabic{legSteps}};}{
\foreach\n in {1,...,#1}{\def\eph{\arabic{offset}}\def\angle{\zeroAngle-2*\n*\spread/\eph+#1*\spread/\eph+\spread/\eph}\stepcounter{legSteps}\leg{(v1)}{\angle}{\arabic{legSteps}}}}}

\def\zeroAngle{180}\def\spread{45}\setcounter{offset}{-1}\addtocounter{offset}{#2}
\ifthenelse{#2=0}{}{\ifthenelse{#2=1}{\stepcounter{legSteps}\leg{(v2)}{\zeroAngle}{\arabic{legSteps}};}{
\foreach\n in {1,...,#2}{\def\eph{\arabic{offset}}\def\angle{\zeroAngle-2*\n*\spread/\eph+#2*\spread/\eph+\spread/\eph}\stepcounter{legSteps}\leg{(v2)}{\angle}{\arabic{legSteps}}}}}

\def\zeroAngle{90}\def\spread{45}\setcounter{offset}{-1}\addtocounter{offset}{#3}
\ifthenelse{#3=0}{}{\ifthenelse{#3=1}{\stepcounter{legSteps}\leg{(v3)}{\zeroAngle}{\arabic{legSteps}};}{
\foreach\n in {1,...,#3}{\def\eph{\arabic{offset}}\def\angle{\zeroAngle-2*\n*\spread/\eph+#3*\spread/\eph+\spread/\eph}\stepcounter{legSteps}\leg{(v3)}{\angle}{\arabic{legSteps}}}}}

\def\zeroAngle{90}\def\spread{15}\setcounter{offset}{-1}\addtocounter{offset}{#4}
\ifthenelse{#4=0}{}{\ifthenelse{#4=1}{\stepcounter{legSteps}\leg{(v4)}{\zeroAngle}{\arabic{legSteps}};}{
\foreach\n in {1,...,#4}{\def\eph{\arabic{offset}}\def\angle{\zeroAngle-2*\n*\spread/\eph+#4*\spread/\eph+\spread/\eph}\stepcounter{legSteps}\leg{(v4)}{\angle}{\arabic{legSteps}}}}}

\def\zeroAngle{90}\def\spread{45}\setcounter{offset}{-1}\addtocounter{offset}{#5}
\ifthenelse{#5=0}{}{\ifthenelse{#5=1}{\stepcounter{legSteps}\leg{(v5)}{\zeroAngle}{\arabic{legSteps}};}{
\foreach\n in {1,...,#5}{\def\eph{\arabic{offset}}\def\angle{\zeroAngle-2*\n*\spread/\eph+#5*\spread/\eph+\spread/\eph}\stepcounter{legSteps}\leg{(v5)}{\angle}{\arabic{legSteps}}}}}

\def\zeroAngle{0}\def\spread{45}\setcounter{offset}{-1}\addtocounter{offset}{#6}
\ifthenelse{#6=0}{}{\ifthenelse{#6=1}{\stepcounter{legSteps}\leg{(v6)}{\zeroAngle}{\arabic{legSteps}};}{
\foreach\n in {1,...,#6}{\def\eph{\arabic{offset}}\def\angle{\zeroAngle-2*\n*\spread/\eph+#6*\spread/\eph+\spread/\eph}\stepcounter{legSteps}\leg{(v6)}{\angle}{\arabic{legSteps}}}}}

\def\zeroAngle{-90}\def\spread{45}\setcounter{offset}{-1}\addtocounter{offset}{#7}
\ifthenelse{#7=0}{}{\ifthenelse{#7=1}{\stepcounter{legSteps}\leg{(v7)}{\zeroAngle}{\arabic{legSteps}};}{
\foreach\n in {1,...,#7}{\def\eph{\arabic{offset}}\def\angle{\zeroAngle-2*\n*\spread/\eph+#7*\spread/\eph+\spread/\eph}\stepcounter{legSteps}\leg{(v7)}{\angle}{\arabic{legSteps}}}}}
}

\newcommand{\kTboxLegs}[6]{
\setcounter{legSteps}{0}
\def\zeroAngle{-135}\def\spread{45}\setcounter{offset}{-1}\addtocounter{offset}{#1}
\ifthenelse{#1=0}{}{\ifthenelse{#1=1}{\stepcounter{legSteps}\leg{(v1)}{\zeroAngle}{\arabic{legSteps}};}{
\foreach\n in {1,...,#1}{\def\eph{\arabic{offset}}\def\angle{\zeroAngle-2*\n*\spread/\eph+#1*\spread/\eph+\spread/\eph}\stepcounter{legSteps}\leg{(v1)}{\angle}{\arabic{legSteps}}}}}

\def\zeroAngle{135}\def\spread{45}\setcounter{offset}{-1}\addtocounter{offset}{#2}
\ifthenelse{#2=0}{}{\ifthenelse{#2=1}{\stepcounter{legSteps}\leg{(v2)}{\zeroAngle}{\arabic{legSteps}};}{
\foreach\n in {1,...,#2}{\def\eph{\arabic{offset}}\def\angle{\zeroAngle-2*\n*\spread/\eph+#2*\spread/\eph+\spread/\eph}\stepcounter{legSteps}\leg{(v2)}{\angle}{\arabic{legSteps}}}}}

\def\zeroAngle{90}\def\spread{15}\setcounter{offset}{-1}\addtocounter{offset}{#3}
\ifthenelse{#3=0}{}{\ifthenelse{#3=1}{\stepcounter{legSteps}\leg{(v3)}{\zeroAngle}{\arabic{legSteps}};}{
\foreach\n in {1,...,#3}{\def\eph{\arabic{offset}}\def\angle{\zeroAngle-2*\n*\spread/\eph+#3*\spread/\eph+\spread/\eph}\stepcounter{legSteps}\leg{(v3)}{\angle}{\arabic{legSteps}}}}}

\def\zeroAngle{90}\def\spread{45}\setcounter{offset}{-1}\addtocounter{offset}{#4}
\ifthenelse{#4=0}{}{\ifthenelse{#4=1}{\stepcounter{legSteps}\leg{(v4)}{\zeroAngle}{\arabic{legSteps}};}{
\foreach\n in {1,...,#4}{\def\eph{\arabic{offset}}\def\angle{\zeroAngle-2*\n*\spread/\eph+#4*\spread/\eph+\spread/\eph}\stepcounter{legSteps}\leg{(v4)}{\angle}{\arabic{legSteps}}}}}

\def\zeroAngle{0}\def\spread{45}\setcounter{offset}{-1}\addtocounter{offset}{#5}
\ifthenelse{#5=0}{}{\ifthenelse{#5=1}{\stepcounter{legSteps}\leg{(v5)}{\zeroAngle}{\arabic{legSteps}};}{
\foreach\n in {1,...,#5}{\def\eph{\arabic{offset}}\def\angle{\zeroAngle-2*\n*\spread/\eph+#5*\spread/\eph+\spread/\eph}\stepcounter{legSteps}\leg{(v5)}{\angle}{\arabic{legSteps}}}}}

\def\zeroAngle{-90}\def\spread{45}\setcounter{offset}{-1}\addtocounter{offset}{#6}
\ifthenelse{#6=0}{}{\ifthenelse{#6=1}{\stepcounter{legSteps}\leg{(v6)}{\zeroAngle}{\arabic{legSteps}};}{
\foreach\n in {1,...,#6}{\def\eph{\arabic{offset}}\def\angle{\zeroAngle-2*\n*\spread/\eph+#6*\spread/\eph+\spread/\eph}\stepcounter{legSteps}\leg{(v6)}{\angle}{\arabic{legSteps}}}}}
}

\newcommand{\kTLegs}[5]{
\setcounter{legSteps}{0}
\def\zeroAngle{-135}\def\spread{45}\setcounter{offset}{-1}\addtocounter{offset}{#1}
\ifthenelse{#1=0}{}{\ifthenelse{#1=1}{\stepcounter{legSteps}\leg{(v1)}{\zeroAngle}{\arabic{legSteps}};}{
\foreach\n in {1,...,#1}{\def\eph{\arabic{offset}}\def\angle{\zeroAngle-2*\n*\spread/\eph+#1*\spread/\eph+\spread/\eph}\stepcounter{legSteps}\leg{(v1)}{\angle}{\arabic{legSteps}}}}}

\def\zeroAngle{135}\def\spread{45}\setcounter{offset}{-1}\addtocounter{offset}{#2}
\ifthenelse{#2=0}{}{\ifthenelse{#2=1}{\stepcounter{legSteps}\leg{(v2)}{\zeroAngle}{\arabic{legSteps}};}{
\foreach\n in {1,...,#2}{\def\eph{\arabic{offset}}\def\angle{\zeroAngle-2*\n*\spread/\eph+#2*\spread/\eph+\spread/\eph}\stepcounter{legSteps}\leg{(v2)}{\angle}{\arabic{legSteps}}}}}

\def\zeroAngle{90}\def\spread{15}\setcounter{offset}{-1}\addtocounter{offset}{#3}
\ifthenelse{#3=0}{}{\ifthenelse{#3=1}{\stepcounter{legSteps}\leg{(v3)}{\zeroAngle}{\arabic{legSteps}};}{
\foreach\n in {1,...,#3}{\def\eph{\arabic{offset}}\def\angle{\zeroAngle-2*\n*\spread/\eph+#3*\spread/\eph+\spread/\eph}\stepcounter{legSteps}\leg{(v3)}{\angle}{\arabic{legSteps}}}}}

\def\zeroAngle{45}\def\spread{45}\setcounter{offset}{-1}\addtocounter{offset}{#4}
\ifthenelse{#4=0}{}{\ifthenelse{#4=1}{\stepcounter{legSteps}\leg{(v4)}{\zeroAngle}{\arabic{legSteps}};}{
\foreach\n in {1,...,#4}{\def\eph{\arabic{offset}}\def\angle{\zeroAngle-2*\n*\spread/\eph+#4*\spread/\eph+\spread/\eph}\stepcounter{legSteps}\leg{(v4)}{\angle}{\arabic{legSteps}}}}}

\def\zeroAngle{-45}\def\spread{45}\setcounter{offset}{-1}\addtocounter{offset}{#5}
\ifthenelse{#5=0}{}{\ifthenelse{#5=1}{\stepcounter{legSteps}\leg{(v5)}{\zeroAngle}{\arabic{legSteps}};}{
\foreach\n in {1,...,#5}{\def\eph{\arabic{offset}}\def\angle{\zeroAngle-2*\n*\spread/\eph+#5*\spread/\eph+\spread/\eph}\stepcounter{legSteps}\leg{(v5)}{\angle}{\arabic{legSteps}}}}}
}

\newcommand{\pBoxLegs}[7]{
\setcounter{legSteps}{0}
\def\zeroAngle{-108}\def\spread{45}\setcounter{offset}{-1}\addtocounter{offset}{#1}
\ifthenelse{#1=0}{}{\ifthenelse{#1=1}{\stepcounter{legSteps}\leg{(v1)}{\zeroAngle}{\arabic{legSteps}};}{
\foreach\n in {1,...,#1}{\def\eph{\arabic{offset}}\def\angle{\zeroAngle-2*\n*\spread/\eph+#1*\spread/\eph+\spread/\eph}\stepcounter{legSteps}\leg{(v1)}{\angle}{\arabic{legSteps}}}}}

\def\zeroAngle{180}\def\spread{45}\setcounter{offset}{-1}\addtocounter{offset}{#2}
\ifthenelse{#2=0}{}{\ifthenelse{#2=1}{\stepcounter{legSteps}\leg{(v2)}{\zeroAngle}{\arabic{legSteps}};}{
\foreach\n in {1,...,#2}{\def\eph{\arabic{offset}}\def\angle{\zeroAngle-2*\n*\spread/\eph+#2*\spread/\eph+\spread/\eph}\stepcounter{legSteps}\leg{(v2)}{\angle}{\arabic{legSteps}}}}}

\def\zeroAngle{108}\def\spread{45}\setcounter{offset}{-1}\addtocounter{offset}{#3}
\ifthenelse{#3=0}{}{\ifthenelse{#3=1}{\stepcounter{legSteps}\leg{(v3)}{\zeroAngle}{\arabic{legSteps}};}{
\foreach\n in {1,...,#3}{\def\eph{\arabic{offset}}\def\angle{\zeroAngle-2*\n*\spread/\eph+#3*\spread/\eph+\spread/\eph}\stepcounter{legSteps}\leg{(v3)}{\angle}{\arabic{legSteps}}}}}

\def\zeroAngle{81}\def\spread{20}\setcounter{offset}{-1}\addtocounter{offset}{#4}
\ifthenelse{#4=0}{}{\ifthenelse{#4=1}{\stepcounter{legSteps}\leg{(v4)}{\zeroAngle}{\arabic{legSteps}};}{
\foreach\n in {1,...,#4}{\def\eph{\arabic{offset}}\def\angle{\zeroAngle-2*\n*\spread/\eph+#4*\spread/\eph+\spread/\eph}\stepcounter{legSteps}\leg{(v4)}{\angle}{\arabic{legSteps}}}}}

\def\zeroAngle{45}\def\spread{45}\setcounter{offset}{-1}\addtocounter{offset}{#5}
\ifthenelse{#5=0}{}{\ifthenelse{#5=1}{\stepcounter{legSteps}\leg{(v5)}{\zeroAngle}{\arabic{legSteps}};}{
\foreach\n in {1,...,#5}{\def\eph{\arabic{offset}}\def\angle{\zeroAngle-2*\n*\spread/\eph+#5*\spread/\eph+\spread/\eph}\stepcounter{legSteps}\leg{(v5)}{\angle}{\arabic{legSteps}}}}}

\def\zeroAngle{-45}\def\spread{45}\setcounter{offset}{-1}\addtocounter{offset}{#6}
\ifthenelse{#6=0}{}{\ifthenelse{#6=1}{\stepcounter{legSteps}\leg{(v6)}{\zeroAngle}{\arabic{legSteps}};}{
\foreach\n in {1,...,#6}{\def\eph{\arabic{offset}}\def\angle{\zeroAngle-2*\n*\spread/\eph+#6*\spread/\eph+\spread/\eph}\stepcounter{legSteps}\leg{(v6)}{\angle}{\arabic{legSteps}}}}}

\def\zeroAngle{-81}\def\spread{20}\setcounter{offset}{-1}\addtocounter{offset}{#7}
\ifthenelse{#7=0}{}{\ifthenelse{#7=1}{\stepcounter{legSteps}\leg{(v7)}{\zeroAngle}{\arabic{legSteps}};}{
\foreach\n in {1,...,#7}{\def\eph{\arabic{offset}}\def\angle{\zeroAngle-2*\n*\spread/\eph+#7*\spread/\eph+\spread/\eph}\stepcounter{legSteps}\leg{(v7)}{\angle}{\arabic{legSteps}}}}}
}

\newcommand{\hBoxLegs}[7]{
\setcounter{legSteps}{0}
\def\zeroAngle{-115}\def\spread{45}\setcounter{offset}{-1}\addtocounter{offset}{#1}
\ifthenelse{#1=0}{}{\ifthenelse{#1=1}{\stepcounter{legSteps}\leg{(v1)}{\zeroAngle}{\arabic{legSteps}};}{
\foreach\n in {1,...,#1}{\def\eph{\arabic{offset}}\def\angle{\zeroAngle-2*\n*\spread/\eph+#1*\spread/\eph+\spread/\eph}\stepcounter{legSteps}\leg{(v1)}{\angle}{\arabic{legSteps}}}}}

\def\zeroAngle{180}\def\spread{45}\setcounter{offset}{-1}\addtocounter{offset}{#2}
\ifthenelse{#2=0}{}{\ifthenelse{#2=1}{\stepcounter{legSteps}\leg{(v2)}{\zeroAngle}{\arabic{legSteps}};}{
\foreach\n in {1,...,#2}{\def\eph{\arabic{offset}}\def\angle{\zeroAngle-2*\n*\spread/\eph+#2*\spread/\eph+\spread/\eph}\stepcounter{legSteps}\leg{(v2)}{\angle}{\arabic{legSteps}}}}}

\def\zeroAngle{115}\def\spread{45}\setcounter{offset}{-1}\addtocounter{offset}{#3}
\ifthenelse{#3=0}{}{\ifthenelse{#3=1}{\stepcounter{legSteps}\leg{(v3)}{\zeroAngle}{\arabic{legSteps}};}{
\foreach\n in {1,...,#3}{\def\eph{\arabic{offset}}\def\angle{\zeroAngle-2*\n*\spread/\eph+#3*\spread/\eph+\spread/\eph}\stepcounter{legSteps}\leg{(v3)}{\angle}{\arabic{legSteps}}}}}

\def\zeroAngle{61.5}\def\spread{30}\setcounter{offset}{-1}\addtocounter{offset}{#4}
\ifthenelse{#4=0}{}{\ifthenelse{#4=1}{\stepcounter{legSteps}\leg{(v4)}{\zeroAngle}{\arabic{legSteps}};}{
\foreach\n in {1,...,#4}{\def\eph{\arabic{offset}}\def\angle{\zeroAngle-2*\n*\spread/\eph+#4*\spread/\eph+\spread/\eph}\stepcounter{legSteps}\leg{(v4)}{\angle}{\arabic{legSteps}}}}}

\def\zeroAngle{0}\def\spread{45}\setcounter{offset}{-1}\addtocounter{offset}{#5}
\ifthenelse{#5=0}{}{\ifthenelse{#5=1}{\stepcounter{legSteps}\leg{(v5)}{\zeroAngle}{\arabic{legSteps}};}{
\foreach\n in {1,...,#5}{\def\eph{\arabic{offset}}\def\angle{\zeroAngle-2*\n*\spread/\eph+#5*\spread/\eph+\spread/\eph}\stepcounter{legSteps}\leg{(v5)}{\angle}{\arabic{legSteps}}}}}

\def\zeroAngle{-61.5}\def\spread{30}\setcounter{offset}{-1}\addtocounter{offset}{#6}
\ifthenelse{#6=0}{}{\ifthenelse{#6=1}{\stepcounter{legSteps}\leg{(v6)}{\zeroAngle}{\arabic{legSteps}};}{
\foreach\n in {1,...,#6}{\def\eph{\arabic{offset}}\def\angle{\zeroAngle-2*\n*\spread/\eph+#6*\spread/\eph+\spread/\eph}\stepcounter{legSteps}\leg{(v6)}{\angle}{\arabic{legSteps}}}}}

\def\zeroAngle{180}\def\spread{25}\setcounter{offset}{-1}\addtocounter{offset}{#7}
\ifthenelse{#7=0}{}{\ifthenelse{#7=1}{\stepcounter{legSteps}\leg{(v7)}{\zeroAngle}{\arabic{legSteps}};}{
\foreach\n in {1,...,#7}{\def\eph{\arabic{offset}}\def\angle{\zeroAngle-2*\n*\spread/\eph+#7*\spread/\eph+\spread/\eph}\stepcounter{legSteps}\leg{(v7)}{\angle}{\arabic{legSteps}}}}}
}

\newcommand{\npPboxLegs}[6]{
\setcounter{legSteps}{0}
\def\zeroAngle{-135}\def\spread{45}\setcounter{offset}{-1}\addtocounter{offset}{#1}
\ifthenelse{#1=0}{}{\ifthenelse{#1=1}{\stepcounter{legSteps}\leg{(v1)}{\zeroAngle}{{\color{labelcolor}\arabic{legSteps}}};}{
\foreach\n in {1,...,#1}{\def\eph{\arabic{offset}}\def\angle{\zeroAngle-2*\n*\spread/\eph+#1*\spread/\eph+\spread/\eph}\stepcounter{legSteps}\leg{(v1)}{\angle}{{\color{labelcolor}\arabic{legSteps}}}}}}

\def\zeroAngle{135}\def\spread{45}\setcounter{offset}{-1}\addtocounter{offset}{#2}
\ifthenelse{#2=0}{}{\ifthenelse{#2=1}{\stepcounter{legSteps}\leg{(v2)}{\zeroAngle}{{\color{labelcolor}\arabic{legSteps}}};}{
\foreach\n in {1,...,#2}{\def\eph{\arabic{offset}}\def\angle{\zeroAngle-2*\n*\spread/\eph+#2*\spread/\eph+\spread/\eph}\stepcounter{legSteps}\leg{(v2)}{\angle}{{\color{labelcolor}\arabic{legSteps}}}}}}

\def\zeroAngle{61.5}\def\spread{30}\setcounter{offset}{-1}\addtocounter{offset}{#3}
\ifthenelse{#3=0}{}{\ifthenelse{#3=1}{\stepcounter{legSteps}\leg{(v3)}{\zeroAngle}{{\color{labelcolor}\arabic{legSteps}}};}{
\foreach\n in {1,...,#3}{\def\eph{\arabic{offset}}\def\angle{\zeroAngle-2*\n*\spread/\eph+#3*\spread/\eph+\spread/\eph}\stepcounter{legSteps}\leg{(v3)}{\angle}{{\color{labelcolor}\arabic{legSteps}}}}}}

\def\zeroAngle{0}\def\spread{45}\setcounter{offset}{-1}\addtocounter{offset}{#4}
\ifthenelse{#4=0}{}{\ifthenelse{#4=1}{\stepcounter{legSteps}\leg{(v4)}{\zeroAngle}{{\color{labelcolor}\arabic{legSteps}}};}{
\foreach\n in {1,...,#4}{\def\eph{\arabic{offset}}\def\angle{\zeroAngle-2*\n*\spread/\eph+#4*\spread/\eph+\spread/\eph}\stepcounter{legSteps}\leg{(v4)}{\angle}{{\color{labelcolor}\arabic{legSteps}}}}}}

\def\zeroAngle{-61.5}\def\spread{30}\setcounter{offset}{-1}\addtocounter{offset}{#5}
\ifthenelse{#5=0}{}{\ifthenelse{#5=1}{\stepcounter{legSteps}\leg{(v5)}{\zeroAngle}{{\color{labelcolor}\arabic{legSteps}}};}{
\foreach\n in {1,...,#5}{\def\eph{\arabic{offset}}\def\angle{\zeroAngle-2*\n*\spread/\eph+#5*\spread/\eph+\spread/\eph}\stepcounter{legSteps}\leg{(v5)}{\angle}{{\color{labelcolor}\arabic{legSteps}}}}}}

\def\zeroAngle{180}\def\spread{25}\setcounter{offset}{-1}\addtocounter{offset}{#6}
\ifthenelse{#6=0}{}{\ifthenelse{#6=1}{\stepcounter{legSteps}\leg{(v6)}{\zeroAngle}{{\color{labelcolor}\arabic{legSteps}}};}{
\foreach\n in {1,...,#6}{\def\eph{\arabic{offset}}\def\angle{\zeroAngle-2*\n*\spread/\eph+#6*\spread/\eph+\spread/\eph}\stepcounter{legSteps}\leg{(v6)}{\angle}{{\color{labelcolor}\arabic{legSteps}}}}}}
}

\newcommand{\pTLegs}[6]{
\setcounter{legSteps}{0}
\def\zeroAngle{-115}\def\spread{45}\setcounter{offset}{-1}\addtocounter{offset}{#1}
\ifthenelse{#1=0}{}{\ifthenelse{#1=1}{\stepcounter{legSteps}\leg{(v1)}{\zeroAngle}{\arabic{legSteps}};}{
\foreach\n in {1,...,#1}{\def\eph{\arabic{offset}}\def\angle{\zeroAngle-2*\n*\spread/\eph+#1*\spread/\eph+\spread/\eph}\stepcounter{legSteps}\leg{(v1)}{\angle}{\arabic{legSteps}}}}}

\def\zeroAngle{180}\def\spread{45}\setcounter{offset}{-1}\addtocounter{offset}{#2}
\ifthenelse{#2=0}{}{\ifthenelse{#2=1}{\stepcounter{legSteps}\leg{(v2)}{\zeroAngle}{\arabic{legSteps}};}{
\foreach\n in {1,...,#2}{\def\eph{\arabic{offset}}\def\angle{\zeroAngle-2*\n*\spread/\eph+#2*\spread/\eph+\spread/\eph}\stepcounter{legSteps}\leg{(v2)}{\angle}{\arabic{legSteps}}}}}

\def\zeroAngle{115}\def\spread{45}\setcounter{offset}{-1}\addtocounter{offset}{#3}
\ifthenelse{#3=0}{}{\ifthenelse{#3=1}{\stepcounter{legSteps}\leg{(v3)}{\zeroAngle}{\arabic{legSteps}};}{
\foreach\n in {1,...,#3}{\def\eph{\arabic{offset}}\def\angle{\zeroAngle-2*\n*\spread/\eph+#3*\spread/\eph+\spread/\eph}\stepcounter{legSteps}\leg{(v3)}{\angle}{\arabic{legSteps}}}}}

\def\zeroAngle{61.5}\def\spread{30}\setcounter{offset}{-1}\addtocounter{offset}{#4}
\ifthenelse{#4=0}{}{\ifthenelse{#4=1}{\stepcounter{legSteps}\leg{(v4)}{\zeroAngle}{\arabic{legSteps}};}{
\foreach\n in {1,...,#4}{\def\eph{\arabic{offset}}\def\angle{\zeroAngle-2*\n*\spread/\eph+#4*\spread/\eph+\spread/\eph}\stepcounter{legSteps}\leg{(v4)}{\angle}{\arabic{legSteps}}}}}

\def\zeroAngle{0}\def\spread{45}\setcounter{offset}{-1}\addtocounter{offset}{#5}
\ifthenelse{#5=0}{}{\ifthenelse{#5=1}{\stepcounter{legSteps}\leg{(v5)}{\zeroAngle}{\arabic{legSteps}};}{
\foreach\n in {1,...,#5}{\def\eph{\arabic{offset}}\def\angle{\zeroAngle-2*\n*\spread/\eph+#5*\spread/\eph+\spread/\eph}\stepcounter{legSteps}\leg{(v5)}{\angle}{\arabic{legSteps}}}}}

\def\zeroAngle{-61.5}\def\spread{30}\setcounter{offset}{-1}\addtocounter{offset}{#6}
\ifthenelse{#6=0}{}{\ifthenelse{#6=1}{\stepcounter{legSteps}\leg{(v6)}{\zeroAngle}{\arabic{legSteps}};}{
\foreach\n in {1,...,#6}{\def\eph{\arabic{offset}}\def\angle{\zeroAngle-2*\n*\spread/\eph+#6*\spread/\eph+\spread/\eph}\stepcounter{legSteps}\leg{(v6)}{\angle}{\arabic{legSteps}}}}}
}

\newcommand{\dPentLegs}[7]{
\setcounter{legSteps}{0}
\def\zeroAngle{-90-45}\def\spread{45}\setcounter{offset}{-1}\addtocounter{offset}{#1}
\ifthenelse{#1=0}{}{\ifthenelse{#1=1}{\stepcounter{legSteps}\leg{(v1)}{\zeroAngle}{\arabic{legSteps}};}{
\foreach\n in {1,...,#1}{\def\eph{\arabic{offset}}\def\angle{\zeroAngle-2*\n*\spread/\eph+#1*\spread/\eph+\spread/\eph}\stepcounter{legSteps}\leg{(v1)}{\angle}{\arabic{legSteps}}}}}

\def\zeroAngle{90+45}\def\spread{45}\setcounter{offset}{-1}\addtocounter{offset}{#2}
\ifthenelse{#2=0}{}{\ifthenelse{#2=1}{\stepcounter{legSteps}\leg{(v2)}{\zeroAngle}{\arabic{legSteps}};}{
\foreach\n in {1,...,#2}{\def\eph{\arabic{offset}}\def\angle{\zeroAngle-2*\n*\spread/\eph+#2*\spread/\eph+\spread/\eph}\stepcounter{legSteps}\leg{(v2)}{\angle}{\arabic{legSteps}}}}}

\def\zeroAngle{90}\def\spread{25}\setcounter{offset}{-1}\addtocounter{offset}{#3}
\ifthenelse{#3=0}{}{\ifthenelse{#3=1}{\stepcounter{legSteps}\leg{(v3)}{\zeroAngle}{\arabic{legSteps}};}{
\foreach\n in {1,...,#3}{\def\eph{\arabic{offset}}\def\angle{\zeroAngle-2*\n*\spread/\eph+#3*\spread/\eph+\spread/\eph}\stepcounter{legSteps}\leg{(v3)}{\angle}{\arabic{legSteps}}}}}

\def\zeroAngle{45}\def\spread{45}\setcounter{offset}{-1}\addtocounter{offset}{#4}
\ifthenelse{#4=0}{}{\ifthenelse{#4=1}{\stepcounter{legSteps}\leg{(v4)}{\zeroAngle}{\arabic{legSteps}};}{
\foreach\n in {1,...,#4}{\def\eph{\arabic{offset}}\def\angle{\zeroAngle-2*\n*\spread/\eph+#4*\spread/\eph+\spread/\eph}\stepcounter{legSteps}\leg{(v4)}{\angle}{\arabic{legSteps}}}}}

\def\zeroAngle{-45}\def\spread{45}\setcounter{offset}{-1}\addtocounter{offset}{#5}
\ifthenelse{#5=0}{}{\ifthenelse{#5=1}{\stepcounter{legSteps}\leg{(v5)}{\zeroAngle}{\arabic{legSteps}};}{
\foreach\n in {1,...,#5}{\def\eph{\arabic{offset}}\def\angle{\zeroAngle-2*\n*\spread/\eph+#5*\spread/\eph+\spread/\eph}\stepcounter{legSteps}\leg{(v5)}{\angle}{\arabic{legSteps}}}}}

\def\zeroAngle{-90}\def\spread{25}\setcounter{offset}{-1}\addtocounter{offset}{#6}
\ifthenelse{#6=0}{}{\ifthenelse{#6=1}{\stepcounter{legSteps}\leg{(v6)}{\zeroAngle}{\arabic{legSteps}};}{
\foreach\n in {1,...,#6}{\def\eph{\arabic{offset}}\def\angle{\zeroAngle-2*\n*\spread/\eph+#6*\spread/\eph+\spread/\eph}\stepcounter{legSteps}\leg{(v6)}{\angle}{\arabic{legSteps}}}}}

\def\zeroAngle{180}\def\spread{20}\setcounter{offset}{-1}\addtocounter{offset}{#7}
\ifthenelse{#7=0}{}{\ifthenelse{#7=1}{\stepcounter{legSteps}\leg{(v7)}{\zeroAngle}{\arabic{legSteps}};}{
\foreach\n in {1,...,#7}{\def\eph{\arabic{offset}}\def\angle{\zeroAngle-2*\n*\spread/\eph+#7*\spread/\eph+\spread/\eph}\stepcounter{legSteps}\leg{(v7)}{\angle}{\arabic{legSteps}}}}}
}

\newcommand{\dBoxLegs}[6]{
\setcounter{legSteps}{0}
\def\zeroAngle{-90-45}\def\spread{45}\setcounter{offset}{-1}\addtocounter{offset}{#1}
\ifthenelse{#1=0}{}{\ifthenelse{#1=1}{\stepcounter{legSteps}\leg{(v1)}{\zeroAngle}{\arabic{legSteps}};}{
\foreach\n in {1,...,#1}{\def\eph{\arabic{offset}}\def\angle{\zeroAngle-2*\n*\spread/\eph+#1*\spread/\eph+\spread/\eph}\stepcounter{legSteps}\leg{(v1)}{\angle}{\arabic{legSteps}}}}}

\def\zeroAngle{90+45}\def\spread{45}\setcounter{offset}{-1}\addtocounter{offset}{#2}
\ifthenelse{#2=0}{}{\ifthenelse{#2=1}{\stepcounter{legSteps}\leg{(v2)}{\zeroAngle}{\arabic{legSteps}};}{
\foreach\n in {1,...,#2}{\def\eph{\arabic{offset}}\def\angle{\zeroAngle-2*\n*\spread/\eph+#2*\spread/\eph+\spread/\eph}\stepcounter{legSteps}\leg{(v2)}{\angle}{\arabic{legSteps}}}}}

\def\zeroAngle{90}\def\spread{25}\setcounter{offset}{-1}\addtocounter{offset}{#3}
\ifthenelse{#3=0}{}{\ifthenelse{#3=1}{\stepcounter{legSteps}\leg{(v3)}{\zeroAngle}{\arabic{legSteps}};}{
\foreach\n in {1,...,#3}{\def\eph{\arabic{offset}}\def\angle{\zeroAngle-2*\n*\spread/\eph+#3*\spread/\eph+\spread/\eph}\stepcounter{legSteps}\leg{(v3)}{\angle}{\arabic{legSteps}}}}}

\def\zeroAngle{45}\def\spread{45}\setcounter{offset}{-1}\addtocounter{offset}{#4}
\ifthenelse{#4=0}{}{\ifthenelse{#4=1}{\stepcounter{legSteps}\leg{(v4)}{\zeroAngle}{\arabic{legSteps}};}{
\foreach\n in {1,...,#4}{\def\eph{\arabic{offset}}\def\angle{\zeroAngle-2*\n*\spread/\eph+#4*\spread/\eph+\spread/\eph}\stepcounter{legSteps}\leg{(v4)}{\angle}{\arabic{legSteps}}}}}

\def\zeroAngle{-45}\def\spread{45}\setcounter{offset}{-1}\addtocounter{offset}{#5}
\ifthenelse{#5=0}{}{\ifthenelse{#5=1}{\stepcounter{legSteps}\leg{(v5)}{\zeroAngle}{\arabic{legSteps}};}{
\foreach\n in {1,...,#5}{\def\eph{\arabic{offset}}\def\angle{\zeroAngle-2*\n*\spread/\eph+#5*\spread/\eph+\spread/\eph}\stepcounter{legSteps}\leg{(v5)}{\angle}{\arabic{legSteps}}}}}

\def\zeroAngle{-90}\def\spread{25}\setcounter{offset}{-1}\addtocounter{offset}{#6}
\ifthenelse{#6=0}{}{\ifthenelse{#6=1}{\stepcounter{legSteps}\leg{(v6)}{\zeroAngle}{\arabic{legSteps}};}{
\foreach\n in {1,...,#6}{\def\eph{\arabic{offset}}\def\angle{\zeroAngle-2*\n*\spread/\eph+#6*\spread/\eph+\spread/\eph}\stepcounter{legSteps}\leg{(v6)}{\angle}{\arabic{legSteps}}}}}
}

\newcommand{\bTLegs}[5]{
\setcounter{legSteps}{0}
\def\zeroAngle{-90-45}\def\spread{45}\setcounter{offset}{-1}\addtocounter{offset}{#1}
\ifthenelse{#1=0}{}{\ifthenelse{#1=1}{\stepcounter{legSteps}\leg{(v1)}{\zeroAngle}{{\color{labelcolor}\arabic{legSteps}}};}{
\foreach\n in {1,...,#1}{\def\eph{\arabic{offset}}\def\angle{\zeroAngle-2*\n*\spread/\eph+#1*\spread/\eph+\spread/\eph}\stepcounter{legSteps}\leg{(v1)}{\angle}{{\color{labelcolor}\arabic{legSteps}}}}}}

\def\zeroAngle{90+45}\def\spread{45}\setcounter{offset}{-1}\addtocounter{offset}{#2}
\ifthenelse{#2=0}{}{\ifthenelse{#2=1}{\stepcounter{legSteps}\leg{(v2)}{\zeroAngle}{{\color{labelcolor}\arabic{legSteps}}};}{
\foreach\n in {1,...,#2}{\def\eph{\arabic{offset}}\def\angle{\zeroAngle-2*\n*\spread/\eph+#2*\spread/\eph+\spread/\eph}\stepcounter{legSteps}\leg{(v2)}{\angle}{{\color{labelcolor}\arabic{legSteps}}}}}}

\def\zeroAngle{70}\def\spread{25}\setcounter{offset}{-1}\addtocounter{offset}{#3}
\ifthenelse{#3=0}{}{\ifthenelse{#3=1}{\stepcounter{legSteps}\leg{(v3)}{\zeroAngle}{{\color{labelcolor}\arabic{legSteps}}};}{
\foreach\n in {1,...,#3}{\def\eph{\arabic{offset}}\def\angle{\zeroAngle-2*\n*\spread/\eph+#3*\spread/\eph+\spread/\eph}\stepcounter{legSteps}\leg{(v3)}{\angle}{{\color{labelcolor}\arabic{legSteps}}}}}}

\def\zeroAngle{0}\def\spread{45}\setcounter{offset}{-1}\addtocounter{offset}{#4}
\ifthenelse{#4=0}{}{\ifthenelse{#4=1}{\stepcounter{legSteps}\leg{(v4)}{\zeroAngle}{{\color{labelcolor}\arabic{legSteps}}};}{
\foreach\n in {1,...,#4}{\def\eph{\arabic{offset}}\def\angle{\zeroAngle-2*\n*\spread/\eph+#4*\spread/\eph+\spread/\eph}\stepcounter{legSteps}\leg{(v4)}{\angle}{{\color{labelcolor}\arabic{legSteps}}}}}}

\def\zeroAngle{-70}\def\spread{25}\setcounter{offset}{-1}\addtocounter{offset}{#5}
\ifthenelse{#5=0}{}{\ifthenelse{#5=1}{\stepcounter{legSteps}\leg{(v5)}{\zeroAngle}{{\color{labelcolor}\arabic{legSteps}}};}{
\foreach\n in {1,...,#5}{\def\eph{\arabic{offset}}\def\angle{\zeroAngle-2*\n*\spread/\eph+#5*\spread/\eph+\spread/\eph}\stepcounter{legSteps}\leg{(v5)}{\angle}{{\color{labelcolor}\arabic{legSteps}}}}}}
}

\newcommand{\dTLegs}[4]{
\setcounter{legSteps}{0}
\def\zeroAngle{180}\def\spread{45}\setcounter{offset}{-1}\addtocounter{offset}{#1}
\ifthenelse{#1=0}{}{\ifthenelse{#1=1}{\stepcounter{legSteps}\leg{(v1)}{\zeroAngle}{{\color{labelcolor}\arabic{legSteps}}};}{
\foreach\n in {1,...,#1}{\def\eph{\arabic{offset}}\def\angle{\zeroAngle-2*\n*\spread/\eph+#1*\spread/\eph+\spread/\eph}\stepcounter{legSteps}\leg{(v1)}{\angle}{{\color{labelcolor}\arabic{legSteps}}}}}}

\def\zeroAngle{90}\def\spread{55}\setcounter{offset}{-1}\addtocounter{offset}{#2}
\ifthenelse{#2=0}{}{\ifthenelse{#2=1}{\stepcounter{legSteps}\leg{(v2)}{\zeroAngle}{{\color{labelcolor}\arabic{legSteps}}};}{
\foreach\n in {1,...,#2}{\def\eph{\arabic{offset}}\def\angle{\zeroAngle-2*\n*\spread/\eph+#2*\spread/\eph+\spread/\eph}\stepcounter{legSteps}\leg{(v2)}{\angle}{{\color{labelcolor}\arabic{legSteps}}}}}}

\def\zeroAngle{0}\def\spread{60}\setcounter{offset}{-1}\addtocounter{offset}{#3}
\ifthenelse{#3=0}{}{\ifthenelse{#3=1}{\stepcounter{legSteps}\leg{(v3)}{\zeroAngle}{{\color{labelcolor}\arabic{legSteps}}};}{
\foreach\n in {1,...,#3}{\def\eph{\arabic{offset}}\def\angle{\zeroAngle-2*\n*\spread/\eph+#3*\spread/\eph+\spread/\eph}\stepcounter{legSteps}\leg{(v3)}{\angle}{{\color{labelcolor}\arabic{legSteps}}}}}}

\def\zeroAngle{-90}\def\spread{55}\setcounter{offset}{-1}\addtocounter{offset}{#4}
\ifthenelse{#4=0}{}{\ifthenelse{#4=1}{\stepcounter{legSteps}\leg{(v4)}{\zeroAngle}{{\color{labelcolor}\arabic{legSteps}}};}{
\foreach\n in {1,...,#4}{\def\eph{\arabic{offset}}\def\angle{\zeroAngle-2*\n*\spread/\eph+#4*\spread/\eph+\spread/\eph}\stepcounter{legSteps}\leg{(v4)}{\angle}{{\color{labelcolor}\arabic{legSteps}}}}}}
}

\newcommand{\tardiLegs}[5]{
\setcounter{legSteps}{0}
\def\zeroAngle{180}\def\spread{45}\setcounter{offset}{-1}\addtocounter{offset}{#1}
\ifthenelse{#1=0}{}{\ifthenelse{#1=1}{\stepcounter{legSteps}\leg{(v1)}{\zeroAngle}{{\color{labelcolor}\arabic{legSteps}}};}{
\foreach\n in {1,...,#1}{\def\eph{\arabic{offset}}\def\angle{\zeroAngle-2*\n*\spread/\eph+#1*\spread/\eph+\spread/\eph}\stepcounter{legSteps}\leg{(v1)}{\angle}{{\color{labelcolor}\arabic{legSteps}}}}}}

\def\zeroAngle{90}\def\spread{55}\setcounter{offset}{-1}\addtocounter{offset}{#2}
\ifthenelse{#2=0}{}{\ifthenelse{#2=1}{\stepcounter{legSteps}\leg{(v2)}{\zeroAngle}{{\color{labelcolor}\arabic{legSteps}}};}{
\foreach\n in {1,...,#2}{\def\eph{\arabic{offset}}\def\angle{\zeroAngle-2*\n*\spread/\eph+#2*\spread/\eph+\spread/\eph}\stepcounter{legSteps}\leg{(v2)}{\angle}{{\color{labelcolor}\arabic{legSteps}}}}}}

\def\zeroAngle{0}\def\spread{45}\setcounter{offset}{-1}\addtocounter{offset}{#3}
\ifthenelse{#3=0}{}{\ifthenelse{#3=1}{\stepcounter{legSteps}\leg{(v3)}{\zeroAngle}{{\color{labelcolor}\arabic{legSteps}}};}{
\foreach\n in {1,...,#3}{\def\eph{\arabic{offset}}\def\angle{\zeroAngle-2*\n*\spread/\eph+#3*\spread/\eph+\spread/\eph}\stepcounter{legSteps}\leg{(v3)}{\angle}{{\color{labelcolor}\arabic{legSteps}}}}}}

\def\zeroAngle{-90}\def\spread{55}\setcounter{offset}{-1}\addtocounter{offset}{#4}
\ifthenelse{#4=0}{}{\ifthenelse{#4=1}{\stepcounter{legSteps}\leg{(v4)}{\zeroAngle}{{\color{labelcolor}\arabic{legSteps}}};}{
\foreach\n in {1,...,#4}{\def\eph{\arabic{offset}}\def\angle{\zeroAngle-2*\n*\spread/\eph+#4*\spread/\eph+\spread/\eph}\stepcounter{legSteps}\leg{(v4)}{\angle}{{\color{labelcolor}\arabic{legSteps}}}}}}

\def\zeroAngle{180}\def\spread{35}\setcounter{offset}{-1}\addtocounter{offset}{#5}
\ifthenelse{#5=0}{}{\ifthenelse{#5=1}{\stepcounter{legSteps}\leg{(v5)}{\zeroAngle}{{\color{labelcolor}\arabic{legSteps}}};}{\ifthenelse{#5=2}{\stepcounter{legSteps}\leg{(v5)}{180}{{\color{labelcolor}\arabic{legSteps}}};\stepcounter{legSteps}\leg{(v5)}{0}{{\color{labelcolor}\arabic{legSteps}}};}{
\foreach\n in {1,...,#5}{\def\eph{\arabic{offset}}\def\angle{\zeroAngle-2*\n*\spread/\eph+#5*\spread/\eph+\spread/\eph}\stepcounter{legSteps}\leg{(v5)}{\angle}{{\color{labelcolor}\arabic{legSteps}}}}}}}

}

\newcommand{\pBoxCoords}{\coordinate(v2)at($\figScale*(-1.33125,0)$);\coordinate(v1)at($(v2)+(-54:\figScale*0.919)$);\coordinate(v3)at($(v2)+(54:\figScale*0.919)$);\coordinate(v4)at($(v3)+(54-72:\figScale*0.919)$);\coordinate(v5)at($(v4)+(0:\figScale*0.919)$);\coordinate(v6)at($(v5)+(-90:\figScale*0.919)$);\coordinate(v7)at($(v4)+(-90:\figScale*0.919)$);}

\let\olditemize\itemize\renewcommand{\itemize}{\vspace{-2pt}\olditemize\setlength{\itemsep}{1pt}\setlength{\parskip}{0pt}\setlength{\parsep}{-0pt}}
\let\oldenumerate\enumerate\renewcommand{\enumerate}{\vspace{-4pt}\oldenumerate\setlength{\itemsep}{1pt}\setlength{\parskip}{0pt}\setlength{\parsep}{0pt}}
\makeatletter\renewcommand\section{\addtocontents{toc}{\protect\addvspace{-2.25\p@}}\@startsection {section}{1}{\z@}{-0.0ex \@plus .2ex \@minus 0.2ex}{1ex \@plus.1ex\@minus .5ex}{\normalfont\large\bfseries}}
\renewcommand\subsection{\addtocontents{toc}{\protect\addvspace{-2.5\p@}}\@startsection {subsection}{1}{\z@}{0.5ex \@plus .2ex \@minus 0.2ex}{0.75ex \@plus.1ex\@minus .5ex}{\normalfont\bfseries}}
\renewcommand\subsubsection{\addtocontents{toc}{\protect\addvspace{-2.5\p@}}\@startsection {subsubsection}{1}{\z@}{0.5ex \@plus .2ex \@minus 0.2ex}{0.75ex \@plus.1ex\@minus .5ex}{\normalfont\bfseries}}
\newcommand{\eq}[1]{\vspace{-0.5pt}\begin{equation}#1\vspace{-0.5pt}\end{equation}}

\newcommand{\fwbox}[2]{\text{\makebox[#1][c]{$\hspace{-150pt}\displaystyle#2\hspace{-150pt}$}}}
\newcommand{\fwboxL}[2]{\text{\makebox[#1][l]{$#2$}}}
\newcommand{\fwboxR}[2]{\text{\makebox[#1][r]{$#2$}}}
\newcommand{\equivR}{\fwbox{14.5pt}{\hspace{-0pt}\fwboxR{0pt}{\raisebox{0.47pt}{\hspace{1.25pt}:\hspace{-4pt}}}=\fwboxL{0pt}{}}}
\newcommand{\equivL}{\fwbox{14.5pt}{\fwboxR{0pt}{}=\fwboxL{0pt}{\raisebox{0.47pt}{\hspace{-4pt}:\hspace{1.25pt}}}}}

\newcommand{\bigger}[1]{\raisebox{-0.95pt}{\scalebox{1.25}{$#1$}}}

\renewcommand{\phi}{\varphi}

\renewcommand{\hat}{\widehat}

\renewcommand{\tilde}{\widetilde}
\newcommand{\ab}[1]{\langle #1\rangle}

\newcommand{\newcap}{\mathrm{\raisebox{0.75pt}{{$\,\bigcap\,$}}}}
\newcommand{\tcap}{\scalebox{0.9}{$\!\newcap\!$}}
\newcommand{\tncap}{\scalebox{0.8}{$\!\newcap\!$}}
\newcommand{\x}[2]{{\color{black}(}\hspace{-0.85pt}{\color{black}#1}\hspace{-0.25pt}{\color{black}|}\hspace{-0.25pt}{\color{black}#2}\hspace{-0.85pt}{\color{black})}}
\newcommand{\dbar}{\fwboxL{7.2pt}{\raisebox{4.5pt}{\fwboxL{0pt}{\scalebox{1.5}[0.75]{\hspace{1.25pt}\text{-}}}}d}}

\definecolor{varcolor}{rgb}{0.08,0.44,0.2}
\definecolor{functioncolor}{rgb}{0.08,0.28,0.6}

\newcommand{\uscore}{\rule[-1.05pt]{7.5pt}{.75pt}}

\definecolor{rindou1}{rgb}{0.4431,0.2862,0.7960}
\definecolor{rindou2}{rgb}{0.0078,0.1215,0.4392}
\definecolor{lapis}{rgb}{0.0.0470,0.2941,0.5568}
\definecolor{emerald}{rgb}{0.31, 0.78, 0.47}
\definecolor{pinegreen}{rgb}{0.0, 0.47, 0.44}
\definecolor{jade}{rgb}{0.0, 0.66, 0.42}
\definecolor{teal}{rgb}{0.0, 0.5, 0.5}
\definecolor{totalCount}{rgb}{0,0,0.575}
\definecolor{topCount}{rgb}{0.575,0.0,0.225}
\definecolor{dim}{rgb}{0.55,0.55,0.55}
\definecolor{deemph}{rgb}{0.25,0.25,0.25}
\definecolor{hblue}{rgb}{0,0,0.575}
\definecolor{hred}{rgb}{0.575,0.0,0.225}
\definecolor{hgreen}{rgb}{0.0,0.4,0.2}
\definecolor{hteal}{rgb}{0.0,0.445,0.6451}
\renewcommand{\r}[1]{{\color{hred}#1}}
\renewcommand{\b}[1]{{\color{hblue}#1}}
\newcommand{\g}[1]{{\color{hteal}#1}}

\newcommand{\biketB}[1]{|\hspace{-0.85pt}\fwbox{14pt}{{\color{hred}#1}}\hspace{-0.85pt})}
\newcommand{\biketC}[1]{|\hspace{-0.85pt}\fwbox{10pt}{{\color{hred}#1}}\hspace{-0.85pt})}
\newcommand{\infX}{{\color{hblue}X}}

\title{\texorpdfstring{{\huge \mbox{The Stratification of Rigidity\hspace{-10pt}}}\\[-6pt]}{Stratifications of Rigidity}}
\author[a,b]{\vspace{-24pt}Jacob~L.~Bourjaily,}\emailAdd{bourjaily@psu.edu}
\author[a]{Nikhil~Kalyanapuram}\emailAdd{nkalyanapuram@psu.edu}
\affiliation[a]{Institute for Gravitation and the Cosmos, Department of Physics,\\Pennsylvania State University, University Park, PA 16802, USA}
\affiliation[b]{Niels Bohr International Academy and Discovery Center, Niels Bohr Institute,\\University of Copenhagen, Blegdamsvej 17, DK-2100, Copenhagen \O, Denmark}
\abstract{%
We show that a master integrand basis exists for all planar, two-loop amplitudes in massless four-dimensional theories which is fully stratified by rigidity---with each integrand being either pure and strictly polylogarithmic or (pure and) strictly elliptic-polylogarithmic, with each of the later involving a single elliptic curve. Such integrands can be said to have \emph{definite} rigidity.
}
\preprint{}

\begin{document}

\maketitle\thispagestyle{empty}
\pagenumbering{roman}\clearpage

\setcounter{section}{0}

\newpage
\pagenumbering{arabic}
\vspace{0pt}%
\section{Introduction and Overview}\label{sec:introduction}\vspace{0pt}
The appearance of non-polylogarithmic contributions to scattering amplitudes beyond one loop has been the source of much interest and development in recent years (see \emph{e.g.} ~\cite{Adams:2013nia,Bloch:2013tra,Remiddi:2013joa,Adams:2016xah,Broedel:2017kkb,Broedel:2017siw,Bourjaily:2017bsb,Bourjaily:2018ycu,Bourjaily:2018yfy,Bourjaily:2019hmc,Bourjaily:2020hjv,Bourjaily:2021vyj,Bourjaily:2021iyq}). In the case of massless four-dimensional theories, the simplest such example arises in the case of the elliptic double-box,
\vspace{-5pt}\eq{\dBox{\draw[int](v5)--(v1);\draw[int](v1)--(v2);\draw[int](v2)--(v3);\draw[int](v3)--(v4);\draw[int](v4)--(v5);\draw[int](v6)--(v3);\legMassive{(v1)}{-135}{$$}\legMassive{(v2)}{135}{$$}\legMassive{(v3)}{90}{$$}\legMassive{(v4)}{45}{$$}\legMassive{(v5)}{-45}{$$}\legMassive{(v6)}{-90}{$$}
\coordinate (ella) at ($(v6)!.5!(v1)+(90:\figScale*0.65)$);
\coordinate (ellb) at ($(v6)!.5!(v5)+(90:\figScale*0.65)$);
\coordinate (a1) at ($(v6)!.55!(v1)+(-90:0.25)$);
\coordinate (a2) at ($(v1)!.5!(v2)+(180:0.25)$);
\coordinate (a3) at ($(v3)!.55!(v2)+(90:0.25)$);
\coordinate (b1) at ($(v3)!.55!(v4)+(90:0.25)$);
\coordinate (b2) at ($(v4)!.5!(v5)+(0:0.25)$);
\coordinate (b3) at ($(v6)!.55!(v5)+(-90:0.25)$);
\node[rdot] at (a1){};
\node[rdot] at (a2){};
\node[rdot] at (a3){};
\node[rdot] at (b1){};
\node[rdot] at (b2){};
\node[rdot] at (b3){};
\node at ($(ella)$) {\text{\normalsize{$\ell_1$}}};
\node at ($(ellb)$) {\text{\normalsize{$\ell_2$}}};
\node at ($(a1)+(-90:0.25)$) {\text{{\normalsize$\r{a_1}$}}};
\node at ($(a2)+(180:0.275)+(90:0.0)$) {\text{{\normalsize$\r{a_2}$}}};
\node at ($(a3)+(90:0.25)$) {\text{{\normalsize$\r{a_3}$}}};
\node at ($(b1)+(0:0.05)+(90:0.25)$) {\text{{\normalsize$\r{b_1}$}}};
\node at ($(b2)+(0:0.275)+(90:0.05)$) {\text{{\normalsize$\r{b_2}$}}};
\node at ($(b3)+(-90:0.25)$) {\text{{\normalsize$\r{b_3}$}}};
}\label{scalar_double_box}\vspace{-5pt}}
which can be normalized (by a function of the external momenta) so that it is dual-conformally invariant  \cite{Drummond:2006rz,Alday:2007hr,Drummond:2008vq} and is known to integrate to an elliptic-multiple-polylogarithm \cite{Bourjaily:2017bsb}. Here, we have used wedges to indicate massive external momenta flowing into the graph. (Massless momenta will be indicated by thin-stroke lines.) In dual-momentum coordinates appropriate for planar integrands, this is simply a contribution to the tree-level 6-point correlation function in massless $\phi^4$-theory in \emph{position-space}:
\vspace{-5pt}\eq{\dBox{\dimLines
\draw[int](v5)--(v1);\draw[int](v1)--(v2);\draw[int](v2)--(v3);\draw[int](v3)--(v4);\draw[int](v4)--(v5);\draw[int](v6)--(v3);\legMassive{(v1)}{-135}{$$}\legMassive{(v2)}{135}{$$}\legMassive{(v3)}{90}{$$}\legMassive{(v4)}{45}{$$}\legMassive{(v5)}{-45}{$$}\legMassive{(v6)}{-90}{$$}
\restoreDark
\coordinate (ella) at ($(v6)!.5!(v1)+(90:\figScale*0.65)$);
\coordinate (ellb) at ($(v6)!.5!(v5)+(90:\figScale*0.65)$);
\coordinate (a1) at ($(v6)!.55!(v1)+(-90:0.25)$);
\coordinate (a2) at ($(v1)!.5!(v2)+(180:0.25)$);
\coordinate (a3) at ($(v3)!.55!(v2)+(90:0.25)$);
\coordinate (b1) at ($(v3)!.55!(v4)+(90:0.25)$);
\coordinate (b2) at ($(v4)!.5!(v5)+(0:0.25)$);
\coordinate (b3) at ($(v6)!.55!(v5)+(-90:0.25)$);
\draw[int] (ella)--(a1);
\draw[int] (ella)--(a2);
\draw[int] (ella)--(a3);
\draw[int] (ella)--(ellb);
\draw[int] (ellb)--(b1);
\draw[int] (ellb)--(b2);
\draw[int] (ellb)--(b3);
\node[rdot] at (a1){};
\node[rdot] at (a2){};
\node[rdot] at (a3){};
\node[rdot] at (b1){};
\node[rdot] at (b2){};
\node[rdot] at (b3){};
\node[ddot] at (ella){};
\node[ddot] at (ellb){};
\node at ($(ella)+(-130:0.305)$) {\text{\normalsize{$\ell_1$}}};
\node at ($(ellb)+(-130:0.305)$) {\text{\normalsize{$\ell_2$}}};
\node at ($(a1)+(-90:0.25)$) {\text{{\normalsize$\r{a_1}$}}};
\node at ($(a2)+(180:0.275)+(90:0.0)$) {\text{{\normalsize$\r{a_2}$}}};
\node at ($(a3)+(90:0.25)$) {\text{{\normalsize$\r{a_3}$}}};
\node at ($(b1)+(0:0.05)+(90:0.25)$) {\text{{\normalsize$\r{b_1}$}}};
\node at ($(b2)+(0:0.275)+(90:0.05)$) {\text{{\normalsize$\r{b_2}$}}};
\node at ($(b3)+(-90:0.20)$) {\text{{\normalsize$\r{b_3}$}}};
}\bigger{\Leftrightarrow\;}\int\!\!\dbar^4\!\ell_1\,\dbar^4\!\ell_2\frac{1}{\x{\ell_1}{\r{a_1}}\x{\ell_1}{\r{a_2}}\x{\ell_1}{\r{a_3}}\x{\ell_1}{\ell_2}\x{\ell_2}{\r{b_1}}\x{\ell_2}{\r{b_2}}\x{\ell_2}{\r{b_3}}}
\label{double_box_in_dual_coordaintes}\vspace{-5pt}}
where  $\dbar\ell_i^\mu\equivR d\ell_i^\mu/(2\pi)$ and $\x{\r{a}}{\r{b}}\equivR(x_{\r{a}}{-}x_{\r{b}})^2$ and we have identified $\ell_i$ with $x_{\ell_i}$ (with apologies for the slight abuse of notation). 

Viewed as a two-loop integrand in momentum space, this integral (\ref{double_box_in_dual_coordaintes}) is needed for scattering amplitudes in massless $\phi^4$-theory \cite{Bourjaily:2017bsb}, integrable fishnet theories \mbox{\cite{Gurdogan:2015csr,Sieg:2016vap,Grabner:2017pgm}}, maximally supersymmetric $(\mathcal{N}\!=\!4)$ Yang-Mills theory (sYM) \cite{Caron-Huot:2012awx,Bourjaily:2015jna,Bourjaily:2015bpz,Bourjaily:2020hjv}, and therefore also pure Yang-Mills and the Standard Model. It is a necessary element of virtually any master integrand basis, and its non-polylogarithmicity (`rigidity' \cite{Bourjaily:2018yfy}) has been the source of much research in recent years (see \emph{e.g.}~\cite{Bloch:2013tra,Bloch:2014qca,Bloch:2016izu,Broadhurst:2016myo,Bogner:2017vim}).

By itself, an integrand such as (\ref{double_box_in_dual_coordaintes}) may appear somewhat unusual, but hardly \emph{problematic}: upon normalizing it appropriately, it can be seen to satisfy canonical differential equations and therefore considered a convenient choice of basis element among master integrands. 

However, consider the case of a `scalar' pentabox integrand\footnote{Here, the numerator `$1$' may be viewed as the inverse propagator $\x{\ell}{\infX}$ involving the `point at infinity' $\infX$ in embedding space; this view will prove useful to us later.} where all inflowing momenta are taken to be massive:
\vspace{-3pt}\eq{\fwbox{0pt}{\hspace{-30pt}\pBox{\pBoxPlainEdges
\legMassive{(v1)}{-110}{$$}\legMassive{(v2)}{180}{$$}\legMassive{(v3)}{110}{$$}\legMassive{(v4)}{81}{$$}\legMassive{(v5)}{45}{$$}\legMassive{(v6)}{-45}{$$}\legMassive{(v7)}{-81}{$$}
\coordinate (ella) at ($(v7)!.5!(v4)+(180:0.575)$);
\coordinate (ellb) at ($(v7)!.5!(v4)+(0:\figScale*0.45)$);
\coordinate (a1) at ($(v7)!.5!(v1)+(-72:0.25)$);
\coordinate (a2) at ($(v1)!.5!(v2)+(-144:0.25)$);
\coordinate (a3) at ($(v2)!.5!(v3)+(144:0.25)$);
\coordinate (a4) at ($(v3)!.5!(v4)+(72:0.25)$);
\coordinate (b1) at ($(v4)!.5!(v5)+(90:0.205)$);
\coordinate (b2) at ($(v5)!.5!(v6)+(0:0.25)$);
\coordinate (b3) at ($(v6)!.5!(v7)-(0:0.25)$);
\coordinate (b3) at ($(v6)!.5!(v7)-(90:0.205)$);
\node[rdot] at (a1){};
\node[rdot] at (a2){};
\node[rdot] at (a3){};
\node[rdot] at (a4){};
\node[rdot] at (b1){};
\node[rdot] at (b2){};
\node[rdot] at (b3){};
\node at ($(ella)$) {\text{{\normalsize${\ell_1}$}}};
\node at ($(ellb)$) {\text{{\normalsize${\ell_2}$}}};
\node at ($(a1)+(-72:0.195)+(-90:0.05)+(0:0.0)$) {\text{{\normalsize$\r{a_1}$}}};
\node at ($(a2)+(-144:0.195)+(-90:0.1)+(0:0.2)$) {\text{{\normalsize$\r{a_2}$}}};
\node at ($(a3)+(144:0.195)+(90:0.05)+(0:0.2)$) {\text{{\normalsize$\r{a_3}$}}};
\node at ($(a4)+(72:0.195)$) {\text{{\normalsize$\r{a_4}$}}};
\node at ($(b1)+(80:0.195)+(0:0.1)+(80:0.05)$) {\text{{\normalsize$\r{b_1}$}}};
\node at ($(b2)+(0:0.25)$) {\text{{\normalsize$\r{b_2}$}}};
\node at ($(b3)+(-80:0.195)+(0:0.2)+(90:0.05)$) {\text{{\normalsize$\r{b_3}$}}};
}\bigger{\Leftrightarrow}\;\dbar^4\!\ell_1\,\dbar^4\!\ell_2\frac{1}{\x{\ell_1}{\r{a_1}}\x{\ell_1}{\r{a_2}}\x{\ell_1}{\r{a_3}}\x{\ell_1}{\r{a_4}}\x{\ell_1}{\ell_2}\x{\ell_2}{\r{b_1}}\x{\ell_2}{\r{b_2}}\x{\ell_2}{\r{b_3}}}.}
\label{scalar_pentabox_integrand}\vspace{-6pt}}
This integrand does not need to be regulated in four dimensions, and may be considered a viable element of a basis of master integrands. Indeed, most software used to construct master integrand bases (see \emph{e.g.}~\cite{Smirnov:2008iw}) would seem to prefer it. However, not only is this integrand neither pure (as evidenced by the fact that its maximal co-dimension residues are distinct) nor dual-conformal, but worse: it would integrate to a \emph{mixture} of polylogarithmic and elliptic-polylogarithmic pieces involving \emph{four distinct elliptic curves}---those associated with its double-box contact terms arising from the collapse of any one of the external edges of the pentagon loop. An integrand which integrates to combinations of functions of different rigidity is said to have \emph{indefinite rigidity}. Clearly, a basis of master integrands with definite rigidity would be preferential. 

The unpleasantness of this integrand would be slightly remedied by adding some loop-dependent numerator to the integrand, by which it can be rendered at least dual-conformally invariant. There is a six-dimensional space of numerators of the form $\x{\ell_1}{\r{N}}$---four of which are naturally identified with its double-box contact terms. In all, we may express any such numerator as being in the space `$[\ell_1]$' defined by
\eq{
[\ell_1]\equivR\underset{\b{N}\in\mathbb{R}^{4}
}{\mathrm{span}}\big\{\x{\ell_1}{\b{N}}\big\}\simeq\mathrm{span}\Big\{\!\underbrace{\x{\ell_1}{\b{Q}^{\b{1}}_{\r{\vec{a}}}},\x{\ell_1}{\b{Q}^{\b{2}}_{\r{\vec{a}}}}}_{\text{`top-level'}},
\underbrace{\x{\ell_1}{\r{a_1}},\x{\ell_1}{\r{a_2}},\x{\ell_1}{\r{a_3}},\x{\ell_1}{\r{a_4}}}_{\text{`contact-terms'}}\!\Big\}\label{basis_for_conformal_pentaboxes}}
where the two `top-level' numerators (spanning the space orthogonal to that spanned by the contact terms) involve the two solutions to the \emph{cut equations} $\x{\b{Q}^{\b{i}}_{\r{\vec{\r{a}}}}}{\r{a_1}}\!=\!\x{\b{Q}^{\b{i}}_{\r{\vec{\r{a}}}}}{\r{a_2}}\!=\!\x{\b{Q}^{\b{i}}_{\r{\vec{\r{a}}}}}{\r{a_3}}\!=\!\x{\b{Q}^{\b{i}}_{\r{\vec{\r{a}}}}}{\r{a_4}}\!=\!0$. It is worth remarking that the loop-independent monomial `$1$'$\in\![\ell_1]$; as such, the so-called `scalar' pentabox in (\ref{scalar_pentabox_integrand}) can be viewed as a linear combination of the six master integrands involving numerators chosen from (\ref{basis_for_conformal_pentaboxes}). (This fact is most obvious in embedding space where `$1$'$\Leftrightarrow\!\x{\ell}{\infX}$.)

In this new basis, four of the integrands---those directly associated with double-box contact-terms---are simply instances of (\ref{scalar_double_box}), and may be viewed as adequate master integrands of definite rigidity; but what about the remaining two, `top-level' integrands? These are little better than the original, scalar pentabox, as they can be normalized so that their \emph{polylogarithmic parts} are pure (have coefficients independent of kinematics)---and such a normalization, incidentally, ensures that the integral is dual-conformally invariant; however, they involve both polylogarithmic \emph{and} elliptic-polylogarithmic contributions involving all four distinct elliptic curves. (The best one can do is construct one linear combination of the two that vanishes upon integration.) It is not hard to see that no linear combination of top-level and contact-term numerators can generate a differential form which would be cohomologous to zero on each of the elliptic curves. We discuss this example in much more detail below, and show how increasing the space of numerators allows these problems to be solved.

\newpage
In this work, we show that this problem can be fully remedied by considering a larger space master loop integrands---specifically, those with so-called `triangle' power-counting (or worse) in four dimensions \cite{Bourjaily:2020qca}. In this space, a basis of masters can be chosen for which every integrand is either pure and polylogarithmic or (pure and) elliptic-polylogarithmic---with each of the latter depending on a single elliptic curve. We may describe a basis of master integrands with this property as being \emph{stratified in rigidity}---where `rigidity' is defined as an integral's degree of \emph{non}-polylogarithmicity as defined in \cite{Bourjaily:2018yfy}. For planar integrands at two loops, the only degrees of rigidity that arise are $0$ (polylogarithms) and $1$ (elliptic).

As two-loop integrands with triangle power-counting in massless theories are known to span all those with maximal transcendental weight and the only ones involving non-polylogarithmic contributions, this suffices to demonstrate that the space of all planar integrands relevant for massless four-dimensional theories can be stratified by rigidity.

\subsection{Organization and Outline}\label{subsec:outline}\vspace{0pt}
This work is organized as follows. In \mbox{section \ref{sec:power_counting}}, we review how master integrand bases can be defined and organized by their `power-counting'---a notion defined precisely in \cite{Bourjaily:2020qca}, but which matches a more na\"ive sense for planar integrands when expressed in dual-momentum coordinates; in \mbox{section \ref{subsec:prescriptivity_review}} we review the notion of \emph{prescriptivity} for integrand bases (relative to choice of a spanning set of integration cycles) and its relation to integrand/integral purity and the definiteness of rigidity. In \mbox{section \ref{sec:stratifying_rigidity}} we describe our main results: in \mbox{section \ref{subsec:source_of_rigidity}} we review the appearance of rigid integrands at two loops, and discuss the role of prescriptivity for elliptic integrands in \mbox{section \ref{subsec:prescriptivity_and_ellipticity}}; in \mbox{section \ref{subsec:diagonalizing_rigidity_box_triangle_basis}} we show that master integrand bases with triangle power-counting (or worse) in four dimensions \emph{can} be constructed which are individually of definite rigidity; and in \mbox{section \ref{subsec:mixed_rigidity}} we show that such bases do not exist for integrands with box power-counting (or better). In \mbox{section \ref{sec:conclusions}} we discuss other instances where integrands with indefinite rigidity have appeared in \mbox{section \ref{subsec:multiple_curves}}, and the case of higher (than elliptic) rigidity in \mbox{sections \ref{subsec:non_planar_two_loops}--\ref{subsec:higher_loops}}.

Many of the specific examples discussed in this work are described in terms of embedding space and momentum-twistors (when needed). For the sake of completeness, we have included a review of these ideas in \mbox{appendix \ref{appendix:embedding_and_momentum_twistors}}. Specifically, we review the embedding-space formalism in \mbox{appendix \ref{appendix_subsec:embedding_space}}, and the relationship between embedding space and momentum-twistor space in \mbox{append \ref{appendix_subsec:momentum_twistors_review}}; several notational technicalities of the momentum-twistor formalism are reviewed in \mbox{appendix \ref{appendix_subsec:momentum_twistor_notation}}. 

We give complete details for the stratification of rigidity for the box-triangle integrands in \mbox{appendix \ref{appendix:box_triangle_masters}}. These results have also been prepared as part of ancillary files for this work's submission to the \texttt{arXiv}, the contents of which are summarized in \mbox{appendix \ref{appendix:ancillary_files}}. These ancillary files have been prepared in \textsc{Mathematica} and are fully self-contained---but make use of code developed as part of the works \cite{Bourjaily:2010wh,Bourjaily:2012gy,Bourjaily:2013mma,Bourjaily:2015jna}. 

\newpage
\section[Review of Power-Counting Stratification of Master Integrand Bases]{Power-Counting Stratification of Master Integrand Bases}\label{sec:power_counting}\vspace{0pt}
In this section we review the organization of master loop integrand bases according to their `power-counting' as described in \cite{Bourjaily:2020qca}. In the case of planar integrands at two loops, this notion of power-counting exactly matches a more na\"ive definition appropriate for loop integrands expressed in dual-momentum coordinates. In particular, a loop integrand is said to behave like a `scalar $p$-gon' (at infinity)---that is, to have `$p$-gon power-counting'---if for each $\ell_i$, 
\eq{\lim_{\ell_i\to\infty}\Big[\mathcal{I}(\ell_1,\ell_2)\Big]=\frac{1}{(\ell_i^2)^{q\geq p}}\Big[1+\mathcal{O}(1/\ell_i^2)\Big]\,.\label{naive_power_counting}}
Notice that an integrand which vanishes \emph{faster} at infinity than $p$ propagators is also said to have $p$-gon power-counting. 

Even for planar integrands, however, this definition is more subtle than it may at first appear: it depends upon how the undetermined loop momenta are `routed' through the Feynman graph---that is, how momentum conservation is solved at each vertex in the graph in order to write the integrand as a rational expression involving exactly $L$ specific loop momenta to be integrated. For example, consider the following integrand with a somewhat peculiar choice of loop momentum routing:
\vspace{-10pt}\eq{\begin{tikzpicture}[scale=\figScale,baseline=-3.05]\useasboundingbox ($(-\pageW/9,-1.4)$) rectangle ($(\pageW/9,1.4)$);
\coordinate(v1) at (.5,.5);
\coordinate(v2) at (.5,-.5);
\coordinate(v3) at (-.5,-.5);
\coordinate(v4) at (-.5,.5);
\draw[int](v2)--(v3);
\draw[int](v3)--(v4);
\draw[int](v4)--(v1);
\draw[int](v1) arc (90:270:0.2 and 0.5) (0,0);
\draw[int](v1) arc (90:-90:0.2 and 0.5) (0,0);
\draw[directedEdge](v1) arc (90:270:0.2 and 0.5) (0,0);
\draw[directedEdge](v1) arc (90:-90:0.2 and 0.5) (0,0);
\legMassive{(v1)}{50}{$$}\legMassive{(v2)}{-50}{$$}\legMassive{(v3)}{-135}{$$}\legMassive{(v4)}{135}{$$}\node at (0.05,0) [] {$\ell_1$};\node at (1.05,0) [] {$\ell_2$};
\end{tikzpicture}\hspace{-10pt}\bigger{\Leftrightarrow}\,\,\dbar^4\!\ell_1\,\dbar^4\!\ell_2\frac{1}{\x{\ell_1}{\r{a_1}}\x{\ell_2}{\r{a_1}}\x{\ell_1{+}\ell_2}{\b{a_2}}\x{\ell_1{+}\ell_2}{\b{a_3}}\x{\ell_1{+}\ell_2}{\b{a_4}}}
\fwboxL{0pt}{\hspace{-0pt}.}
\vspace{-10pt}\label{example_on_routing}}
There is nothing preventing us from using this choice for the undetermined loop momenta, and our conclusion (according to the definition of (\ref{naive_power_counting})) would be that the integrand (\ref{example_on_routing}) behaves like a scalar `box' at infinity as each $\ell_i$ appears in exactly four propagators. Obviously, this conclusion would be misleading, as a simple change of integration variables would expose the obvious bubble in (\ref{example_on_routing}). Thus, any definition of `power-counting' analogous to (\ref{naive_power_counting}) must be used with some care. 

To avoid the need for any reference to loop momentum routing in the notion of `power-counting', the authors of \cite{Bourjaily:2020qca} proposed an intrinsically graph-theoretic definition. In the case of planar integrands at two loops, this graph-theoretic notion matches the na\"ive definition (\ref{naive_power_counting}) \emph{provided that loop momenta are expressed in dual-momentum coordinates} defined the graph's planar-dual and taking the external momenta to be defined by $p_\r{a}\equivL(x_{\r{a+1}}{-}x_{\r{a}})$ with each $x_{\r{a}}$ labeling the external pairs of faces of the graph.\\ 

In these dual-momentum coordinates, we may construct sets of loop integrands by starting with some Feynman graph of scalar propagators and considering the vector space of loop-dependent polynomials in their numerators. Taking inverse-propagators as natural Lorentz-invariant factors to include in the numerators of loop integrands, we consider the space of numerators of the form\footnote{More generally, we could consider numerators of the form $[\ell_1]^a[\ell_1{-}\ell_2]^b[\ell_2]^c$ or even linear combinations of such spaces of numerators; however, we will not require such numerators in this work.} $[\ell_1]^a[\ell_2]^b$ where 
\eq{[\ell_i]^k\equivR\underset{\b{N_i}\in\mathbb{R}^{4}}{\mathrm{span}}\Big[\x{\ell_i}{\b{N_1}}\cdots\x{\ell_i}{\b{N_k}}\Big]\,.}
We remind the reader that $\x{\ell_i}{\r{Q}}\equivR(\ell_i{-}\r{Q})^2$, an inverse propagator involving a \emph{translate} of the loop momentum $\ell_i$; as such, $[\ell_i]\!\simeq\![\ell_i{+}\r{Q}]$ for any constant $\r{Q}\!\in\!\mathbb{R}^4$. As inverse propagators can be represented in embedding formalism by inner products in a $(d{+}2)$-dimensional space, it is easy to see that $\mathrm{rank}\big([\ell_i]\big)\!=\!d{+}2$. Similarly, a $k$-fold product of inverse-propagators encodes a symmetric, traceless product of $(d{+}2)$-dimensional representations of $SO(d{+}2)$; as such, we can readily see that
\eq{\mathrm{rank}\big([\ell_i]^k\big)=\binom{d{+}k}{d}{+}\binom{d{+}k{-}1}{d}\,.}
For our purposes, it is useful to note that in four dimensions, $\mathrm{rank}([\ell])\!=\!6$, \mbox{$\mathrm{rank}([\ell]^2)\!=\!20$}, $\mathrm{rank}([\ell]^3)\!=\!50$, and so-on. It is not hard to see that the product space $[\ell_1]^a[\ell_2]^b$ has a total rank given by the products of the two spaces' ranks.  
%
%

Graphically, we may denote a space of integrands in this way by defining\footnote{Here, `$1$' simply denotes a loop-independent monomial; obviously, it would necessarily carry some scaling dimension.}
\eq{\fwbox{65pt}{\tikzBox{\draw[int](0,0)--(1,0);\draw[markedEdge](0,0)--(1,0);\node[anchor=north] at (0.5,0) {$\vec{\ell}$};}}\equivR\frac{[\ell]}{\ell^2}=\mathrm{span}\left\{1,\frac{\ell\!\cdot\!e_1}{\ell^2},\frac{\ell\!\cdot\!e_2}{\ell^2},\frac{\ell\!\cdot\!e_3}{\ell^2},\frac{\ell\!\cdot\!e_4}{\ell^2},\frac{1}{\ell^2}\right\}\label{vector_space_of_decorated_edge}}
where $e_\mu$ form a basis for $\mathbb{R}^4$. Thus, we may denote a space of a pentabox integrands (\ref{scalar_pentabox_integrand}) endowed with a loop-dependent numerator chosen from the space `$[\ell_1]$' by any of the following figures:
\eq{\pBox{\pBoxPlainEdges
\legMassive{(v1)}{-110}{$$}\legMassive{(v2)}{180}{$$}\legMassive{(v3)}{110}{$$}\legMassive{(v4)}{80}{$$}\legMassive{(v5)}{45}{$$}\legMassive{(v6)}{-45}{$$}\legMassive{(v7)}{-80}{$$}
\draw[markedEdge] (v3)--(v4);
}\hspace{-14pt}\simeq\hspace{-5pt}\pBox{\pBoxPlainEdges
\legMassive{(v1)}{-110}{$$}\legMassive{(v2)}{180}{$$}\legMassive{(v3)}{110}{$$}\legMassive{(v4)}{80}{$$}\legMassive{(v5)}{45}{$$}\legMassive{(v6)}{-45}{$$}\legMassive{(v7)}{-80}{$$}
\draw[markedEdge] (v2)--(v3);
}\hspace{-14pt}\simeq\hspace{-5pt}\pBox{\pBoxPlainEdges
\legMassive{(v1)}{-110}{$$}\legMassive{(v2)}{180}{$$}\legMassive{(v3)}{110}{$$}\legMassive{(v4)}{80}{$$}\legMassive{(v5)}{45}{$$}\legMassive{(v6)}{-45}{$$}\legMassive{(v7)}{-80}{$$}
\draw[markedEdge] (v1)--(v2);
}\hspace{-14pt}\simeq\hspace{-5pt}\pBox{\pBoxPlainEdges
\legMassive{(v1)}{-110}{$$}\legMassive{(v2)}{180}{$$}\legMassive{(v3)}{110}{$$}\legMassive{(v4)}{80}{$$}\legMassive{(v5)}{45}{$$}\legMassive{(v6)}{-45}{$$}\legMassive{(v7)}{-80}{$$}
\draw[markedEdge] (v7)--(v1);
}}
where their equivalence follows from the fact that $[\ell_i]\!\simeq\![\ell_i{+}\r{Q}]$ for any $\r{Q}\!\in\!\mathbb{R}^4$. 

Using such notation, we may easily discuss increasingly larger spaces of master integrands by endowing a scalar integrand with increasingly large vector-spaces of loop-dependent numerators. This is illustrated in \mbox{Table \ref{pentabox_power_counting_strata_table}}, where we have indicated the integrand numerator spaces that would be defined as part of bases defined with box power-counting through `0-gon' power-counting, denoted $\mathfrak{B}_4$, \ldots, $\mathfrak{B}_0$ in \cite{Bourjaily:2020qca}, respectively. It is important to appreciate that these master integrand bases are strictly ordered by their spans, with $\mathfrak{B}_4\!\subsetneq\!\mathfrak{B}_3\!\subsetneq\!\cdots\!\subsetneq\mathfrak{B}_0$.

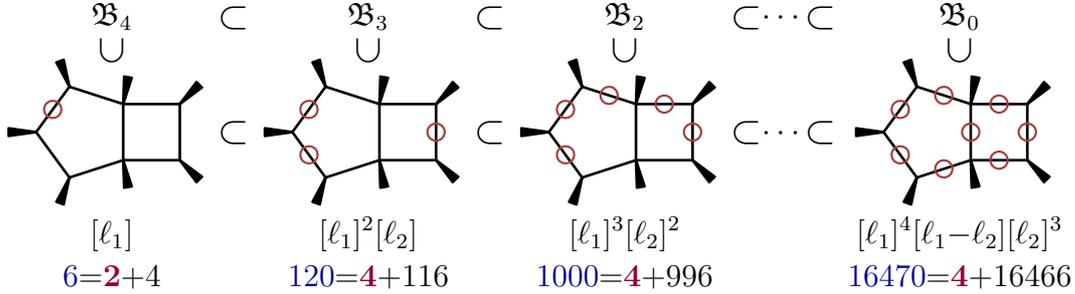
\begin{table}[t]\caption{Master integrand spaces for pentaboxes arranged by `power-counting' as defined in \cite{Bourjaily:2020qca}. For each, we indicate the {\color{totalCount}total rank} of the space of masters, and the decomposition of this space into {\color{topCount}top-level} numerators and those spanned by `contact-terms'.\label{pentabox_power_counting_strata_table}}$$\fwbox{0pt}{\hspace{-20pt}\begin{array}{@{}c@{}c@{}c@{}c@{}c@{}c@{}c@{}c@{}c@{}c@{}}\mathfrak{B}_4&\fwbox{0pt}{\hspace{-5pt}\bigger{\subset}}&\mathfrak{B}_3&\fwbox{0pt}{\hspace{-5pt}\bigger{\subset}}&\mathfrak{B}_2&\fwbox{0pt}{\hspace{-5pt}\bigger{\subset}}\;&\cdots&\fwboxL{10pt}{\hspace{0pt}\bigger{\subset}}&\mathfrak{B}_0\\[-12pt]
\rotatebox{-90}{\bigger{\supset}}&&\rotatebox{-90}{\bigger{\supset}}&&\rotatebox{-90}{\bigger{\supset}}&&&&\rotatebox{-90}{\bigger{\supset}}\\[-11pt]
\pBox{\pBoxPlainEdges
\legMassive{(v1)}{-110}{$$}\legMassive{(v2)}{180}{$$}\legMassive{(v3)}{110}{$$}\legMassive{(v4)}{80}{$$}\legMassive{(v5)}{45}{$$}\legMassive{(v6)}{-45}{$$}\legMassive{(v7)}{-80}{$$}
\draw[markedEdge] (v2)--(v3);
}&\fwbox{0pt}{\hspace{-5pt}\bigger{\subset}}&
\pBox{\pBoxPlainEdges
\legMassive{(v1)}{-110}{$$}\legMassive{(v2)}{180}{$$}\legMassive{(v3)}{110}{$$}\legMassive{(v4)}{80}{$$}\legMassive{(v5)}{45}{$$}\legMassive{(v6)}{-45}{$$}\legMassive{(v7)}{-80}{$$}
\draw[markedEdge] (v2)--(v3);\draw[markedEdge] (v1)--(v2);\draw[markedEdge] (v5)--(v6);
}&\fwbox{0pt}{\hspace{-5pt}\bigger{\subset}}&
\pBox{\pBoxPlainEdges
\legMassive{(v1)}{-110}{$$}\legMassive{(v2)}{180}{$$}\legMassive{(v3)}{110}{$$}\legMassive{(v4)}{80}{$$}\legMassive{(v5)}{45}{$$}\legMassive{(v6)}{-45}{$$}\legMassive{(v7)}{-80}{$$}
\draw[markedEdge] (v2)--(v3);\draw[markedEdge] (v1)--(v2);\draw[markedEdge] (v3)--(v4);\draw[markedEdge] (v4)--(v5);\draw[markedEdge] (v5)--(v6);
}&\fwbox{0pt}{\hspace{-5pt}\bigger{\subset}}\;&\cdots&\fwboxL{10pt}{\hspace{0pt}\bigger{\subset}}&
\pBox{\pBoxPlainEdges
\legMassive{(v1)}{-110}{$$}\legMassive{(v2)}{180}{$$}\legMassive{(v3)}{110}{$$}\legMassive{(v4)}{80}{$$}\legMassive{(v5)}{45}{$$}\legMassive{(v6)}{-45}{$$}\legMassive{(v7)}{-80}{$$}
\draw[markedEdge] (v2)--(v3);\draw[markedEdge] (v1)--(v2);\draw[markedEdge] (v7)--(v1);\draw[markedEdge] (v3)--(v4);\draw[markedEdge] (v4)--(v5);\draw[markedEdge] (v5)--(v6);\draw[markedEdge] (v6)--(v7);\draw[markedEdge] (v7)--(v4);
}\\[-5pt]
\text{$[\ell_1]$}&&[\ell_1]^2[\ell_2]&&[\ell_1]^3[\ell_2]^2&&&&[\ell_1]^4[\ell_1{-}\ell_2][\ell_2]^3\\
{\color{totalCount}{6}}{=}{\color{topCount}\mathbf{2}}{\color{black}{+}{4}}&&{\color{totalCount}{120}}{=}{\color{topCount}\mathbf{4}}{\color{black}{+}{116}}&&{\color{totalCount}{1000}}{=}{\color{topCount}\mathbf{4}}{\color{black}{+}{996}}&&&&{\color{totalCount}{16470}}{=}{\color{topCount}\mathbf{4}}{\color{black}{+}{16466}}
\end{array}}
\vspace{-5pt}$$\end{table}

\subsection{Review of Master Integrand Bases at Two Loops in Four Dimensions}\label{subsec:two_loop_bases_review}\vspace{0pt}
In four dimensions, there are two seed topologies that generate (together with all their contact terms) the full basis of integrands at two loops with triangle power-counting (or worse) as defined by \cite{Bourjaily:2020qca}. These are the pentabox and kissing box topologies, with sets of loop-dependent numerators determined according to
\eq{\fwboxR{0pt}{\mathfrak{B}_3^{(4)}\equivR}\left\{\rule{0pt}{35pt}\right.\pBox{\pBoxPlainEdges
\legMassive{(v1)}{-110}{$$}\legMassive{(v2)}{180}{$$}\legMassive{(v3)}{110}{$$}\legMassive{(v4)}{80}{$$}\legMassive{(v5)}{45}{$$}\legMassive{(v6)}{-45}{$$}\legMassive{(v7)}{-80}{$$}
\draw[markedEdge] (v2)--(v3);\draw[markedEdge] (v1)--(v2);\draw[markedEdge] (v5)--(v6);
},
\kBox{\kBoxPlainEdges\legMassive{(v1)}{-90}{$$}\legMassive{(v2)}{-180}{$$}\legMassive{(v3)}{90}{$$}\legMassive{(v4)}{90}{$$}\legMassive{(v5)}{90}{$$}\legMassive{(v6)}{0}{$$}\legMassive{(v7)}{-90}{$$}\draw[markedEdge] (v2)--(v3);\draw[markedEdge] (v5)--(v6);}\left.\rule{0pt}{32pt}\right\}\fwboxL{0pt}{.}\label{seed_topologies_for_b3}}
These spaces of master integrands have numerators given by $[\ell_1]^2[\ell_2]$ and $[\ell_1][\ell_2]$, as indicated graphically above, spanning total spaces of dimension ${\color{totalCount}120}$ and ${\color{totalCount}36}$, respectively (for four-dimensional loop momenta). 

Each of these spaces can be divided into sub-spaces spanned by contact terms---those numerators directly proportional to some inverses of propagators directly appearing in the original Feynman integrand---and top-level numerators (defined as the complement of the space spanned by contact terms). For example, the ${\color{totalCount}36}$-dimensional space of master integrands for the kissing boxes can be represented by
\eq{\mathfrak{n}(\ell_1,\ell_2)\in[\ell_1][\ell_2]=\mathrm{span}\big\{\underbrace{\x{\ell_1}{\b{N^1_i}}\x{\ell_2}{\b{N^2_j}}}_{\text{`{\color{topCount}top-level}' (${\color{topCount}4}\times$)}},\underbrace{\x{\ell_1}{\b{N^1_i}}\x{\ell_2}{\r{b_j}},\x{\ell_1}{\r{a_i}}\x{\ell_2}{\b{N^2_j}},\x{\ell_1}{\r{a_i}}\x{\ell_2}{\r{b_j}}}_{\text{`contact-terms'}~(4\times{\color{topCount}2}{+}4\times{\color{topCount}2}{+}4\times4\times{\color{topCount}1})=(32\times)}\big\}\,,\vspace{-5pt}}
where the $\b{N^i_j}$'s represent the complementary ${\color{topCount}2}$-dimensional space of numerators \emph{not spanned} by contact terms for each box (and can be taken, for example, to be given by those analogous to the example in (\ref{basis_for_conformal_pentaboxes})). The contact terms span a 32-dimensional space of master integrands, with $16$ of them representing the two `top-level' integrands for each of the 8 kissing box-triangles subtopologies, and another $16(=4\times4)$ representing the kissing-triangle integrands with loop-independent numerators (one-dimensional master integrand spaces each). 

The complete list of master integrand topologies and numerator spaces is given in \mbox{Table \ref{triangle_power_counting_basis_topologies}}. This Table illustrates how all the integrands are generated within the space of master integrands defining the pentabox and kissing boxes as in (\ref{seed_topologies_for_b3}). 
\begin{table}[t!]\caption{Master integrands defining the space of two-loop, planar integrands with triangle power-counting in four spacetime dimensions. For each topology, we indicate the corresponding vector space of loop-dependent numerators, the {\color{totalCount}total rank} of the space of these numerators, and the breakdown of these ranks into {\color{topCount}top-level} and contact-terms. \label{triangle_power_counting_basis_topologies}}\vspace{-5pt}$$\fwbox{0pt}{\begin{tikzpicture}[scale=\figScale,baseline=-3.05]\useasboundingbox ($(-\pageW/2,-4.2)$) rectangle ($(\pageW/2,4.2)$);\draw[int,line width=0.1,red,draw=\boundingDraw] ($(-\pageW/2,-4.2)$) rectangle ($(\pageW/2,4.2)$);
\node[] at (-6.5,2.9) {$\begin{array}{@{}c@{}}~\\[-8pt]\kBox{\kBoxPlainEdges\legMassive{(v1)}{-90}{$$}\legMassive{(v2)}{-180}{$$}\legMassive{(v3)}{90}{$$}\legMassive{(v4)}{90}{$$}\legMassive{(v5)}{90}{$$}\legMassive{(v6)}{0}{$$}\legMassive{(v7)}{-90}{$$}\draw[markedEdge] (v2)--(v3);\draw[markedEdge] (v5)--(v6);}\\[-8pt]{\color{totalCount}36}{=}{\color{topCount}\mathbf{4}}{+}32\end{array}$};
\node[] at (-6.5,-0.2) {$\begin{array}{@{}c@{}}~\\[-8pt]\pBoxB{\pBoxPlainEdges
\legMassive{(v1)}{-110}{$$}\legMassive{(v2)}{180}{$$}\legMassive{(v3)}{110}{$$}\legMassive{(v4)}{80}{$$}\legMassive{(v5)}{45}{$$}\legMassive{(v6)}{-45}{$$}\legMassive{(v7)}{-80}{$$}\draw[markedEdge] (v2)--(v3);\draw[markedEdge] (v1)--(v2);\draw[markedEdge] (v5)--(v6);
}\\[-8pt]{\color{totalCount}120}{=}{\color{topCount}\mathbf{4}}{+}116\end{array}$};
\node[] at (-1.8,3.2) {$\begin{array}{@{}c@{}}~\\[-8pt]\fwbox{0pt}{\rotatebox{180}{\hspace{0pt}\kTbox{\kTboxPlainEdges
\legMassive{(v1)}{-120}{$$}\legMassive{(v2)}{120}{$$}\legMassive{(v3)}{-90}{$$}\legMassive{(v4)}{90}{$$}\legMassive{(v5)}{0}{$$}\legMassive{(v6)}{-90}{$$}
\draw[markedEdge] (v5)--(v6);}}}\\[-8pt]{\color{totalCount}6}{=}{\color{topCount}\mathbf{2}}{+}4\end{array}$};
\node[] at (-1.8,0.45) {$\begin{array}{@{}c@{}}~\\[-9pt]\dBox{\dBoxPlainEdges
\legMassive{(v1)}{-135}{$$}\legMassive{(v2)}{135}{$$}\legMassive{(v3)}{90}{$$}\legMassive{(v4)}{45}{$$}\legMassive{(v5)}{-45}{$$}\legMassive{(v6)}{-90}{$$}
\draw[markedEdge] (v4)--(v5);\draw[markedEdge] (v1)--(v2);}\\[-9pt]{\color{totalCount}36}{=}{\color{topCount}\mathbf{8}}{+}28\end{array}$};
\node[] at (-1.8,-2.4) {$\begin{array}{@{}c@{}}~\\[-6pt]\pT{\pTPlainEdges
\legMassive{(v1)}{-110}{$$}\legMassive{(v2)}{180}{$$}\legMassive{(v3)}{110}{$$}\legMassive{(v4)}{63.5}{$$}\legMassive{(v5)}{0}{$$}\legMassive{(v6)}{-63.5}{$$}
\draw[markedEdge] (v2)--(v3);\draw[markedEdge] (v1)--(v2);}\\[-6pt]{\color{totalCount}20}{=}{\color{topCount}\mathbf{2}}{+}18\end{array}$};
\node[] at (2.3,2.5) {$\begin{array}{@{}c@{}}~\\[-9pt]\\\rotatebox{0}{\kT{\draw[int](v3)--(v1);\draw[int](v1)--(v2);\draw[int](v2)--(v3);\draw[int](v3)--(v4);\draw[int](v4)--(v5);\draw[int](v5)--(v3);
\legMassive{(v1)}{-130}{$$}\legMassive{(v2)}{130}{$$}\legMassive{(v3)}{90}{$$}\legMassive{(v4)}{50}{$$}\legMassive{(v5)}{-50}{$$}
}}\\[-9pt]{\color{totalCount}1}{=}{\color{topCount}\mathbf{1}}{+}0\end{array}$};
\node[] at (2.3,-0.55) {$\begin{array}{@{}c@{}}~\\[-6pt]\\\fwbox{0pt}{\hspace{10pt}\rotatebox{0}{\bT{\draw[int](v5)--(v1);\draw[int](v1)--(v2);\draw[int](v2)--(v3);\draw[int](v3)--(v4);\draw[int](v4)--(v5);\draw[int](v5)--(v3);
\legMassive{(v1)}{-135}{$$}\legMassive{(v2)}{135}{$$}\legMassive{(v3)}{71.5}{$$}\legMassive{(v4)}{0}{$$}\legMassive{(v5)}{-71.5}{$$}
\draw[markedEdge] (v1)--(v2);}}}\\[-6pt]{\color{totalCount}6}{=}{\color{topCount}\mathbf{3}}{+}3\end{array}$};
\node[] at (6.5,0.7) {$\begin{array}{@{}c@{}}~\\[-6pt]\\\dT{\draw[int](v4)--(v1);\draw[int](v1)--(v2);\draw[int](v2)--(v3);\draw[int](v3)--(v4);\draw[int](v4)--(v2);\legMassive{(v1)}{180}{$$}\legMassive{(v2)}{90}{$$}\legMassive{(v3)}{0}{$$}\legMassive{(v4)}{-90}{$$}}\\[-6pt]{\color{totalCount}1}{=}{\color{topCount}\mathbf{1}}{+}0\end{array}$};
\draw[int,->] (-4.8,2.9)--(-3.45,3.3) node[above,align=center,midway] {};\draw[int,->] (-4.8,-0.1)--(-3.45,3.1) node[above,align=center,midway] {};\draw[int,->] (-4.8,-0.2)--(-3.45,0.4) node[above,align=center,midway] {};
\draw[int,->] (-4.8,-0.3)--(-3.45,-2.4) node[above,align=center,midway] {};
\draw[int,->] (-0.1,-2.4)--(1.0,-1.05) node[above,align=center,midway] {};
\draw[int,->] (-0.1,0.4)--(1.0,-0.95) node[above,align=center,midway] {};
\draw[int,->] (-0.1,0.5)--(1.0,1.9) node[above,align=center,midway] {};
\draw[int,->] (-0.1,3.2)--(1.0,2.1) node[above,align=center,midway] {};
\draw[int,->] (4.1,-0.85)--(5.0,0.45) node[above,align=center,midway] {};
\end{tikzpicture}}$$\vspace{-20pt}\end{table}

Although we speak of top-level and contact-term subspaces of master integrands, the \emph{particular} choice of basis for master integrands may involve any linear combination of these subspaces. While the first step in constructing a master integrand basis involves identifying \emph{some} choice for master integrand which span the space, the particular choice for a `good' basis of master integrands often requires rotations (and rescalings) relative to whatever initial choice was made for the basis.

One particularly important mechanism to find good, particular sets of master integrands to serve as a basis is the requirement of \emph{prescriptivity}, which results in a unique set of integrands cohomologically-dual to some choice of integration contours. Such integrands turn out to have many important and desirable features (such as purity). We review how this works---and its limitations---in the following \mbox{section \ref{subsec:prescriptivity_review}}.

\newpage 
\subsection{Review of Prescriptivity for Choosing Master Integrand Bases}\label{subsec:prescriptivity_review}\vspace{0pt}
\subsubsection{Generalized and \emph{Prescriptive} Unitarity}\label{subsubsec:general_and_prescriptive_unitarity}
Generalized unitarity is usually described in terms of rational differential forms on the space of loop momenta---the space on which any Feynman diagram (or loop amplitude integrand) must be defined prior to loop integration. This description misses a key fact about Feynman loop integrands: that they are only defined as \emph{elements of cohomology classes} on the space of internal loop momenta. That is, integrands which differ by total derivatives cannot be distinguished, and there is no meaningful sense in which a particular loop integrand is preferred. As such, any basis of \emph{master} integrands must represent cohomologically-distinct differential forms on the space of loop momenta.  

Given any sufficiently large basis $\mathfrak{B}$ of master loop integrands, any loop amplitude integrand may be expressed as some linear combination of these basis integrands (with loop-momentum-independent coefficients) according to
\eq{\mathcal{A}=\sum_{\mathcal{I}^0_i\in\mathfrak{B}}\mathfrak{c}_i\,\mathcal{I}^0_i\,.\label{general_amplitude_expansion}}
The principle of (generalized) unitarity states that coefficients $\mathfrak{c}_i$ appearing above may be determined by `[unitarity] cuts'. What this means in practice is that one may choose some (any) \emph{spanning set} of maximal-dimensional, compact contours $\{\Omega_j\}$---the number of which equals the number of master integrands in the basis---each of which encloses the poles associated with least $(L{+}1)$ propagators (hence the name `unitarity cuts')\footnote{The fact that a spanning set of cycles always exists which cuts a sufficient number of propagators was recently made more rigorous in \cite{Caron-Huot:2021xqj,Caron-Huot:2021iev}.} such that the \emph{period matrix} $\mathbf{M}^0_{i\,j}$
\eq{\mathbf{M}^0_{i\,j}\equivR\oint\limits_{\Omega_j}\!\mathcal{I}_i\,\label{general_period_matrix}}
will be of full rank. (The existence of such a period matrix in fact guarantees that the set of master integrands represent distinct cohomology classes.) The principle of generalized unitarity follows from the fact that when any scattering amplitude integrand is integrated over such a unitarity cut, its `periods'
\eq{\mathfrak{a}_j\equivR\oint\limits_{\Omega_j}\!\mathcal{A}\label{on_shell_functions_as_periods}}
are determined by lower-loop (and, ultimately (via recursion) \emph{on-shell} tree-level) scattering data. As such, the \emph{periods} $\mathfrak{a}_j$ of amplitudes (\ref{on_shell_functions_as_periods}) correspond to particular \emph{on-shell functions} (see e.g.~\cite{Arkani-Hamed:2016byb}). Such on-shell functions are sometimes called `leading singularities' (although historically this term was reserved for periods of amplitudes associated with strictly polylogarithmic contours of integration---those encircling, iteratively, simple poles).

To understand how the on-shell data encoded in the periods $\mathfrak{a}_{i}$ determines the loop-momentum-independent coefficients $\mathfrak{c}_i$ in (\ref{general_amplitude_expansion}), one merely needs to consider the integration of both sides of (\ref{general_amplitude_expansion}) on each contour $\{\Omega_j\}$:
\eq{(\mathfrak{a}_j\equivR)\oint\limits_{\Omega_j}\!\mathcal{A}=\oint\limits_{\Omega_j}\Bigg[\sum_{\mathcal{I}^0_i\in\mathfrak{B}}\mathfrak{c}_i\,\mathcal{I}^0_i\Bigg]=\sum_{\mathcal{I}^0_i\in\mathfrak{B}}\mathfrak{c}_i\,\oint\limits_{\Omega_j}\!\mathcal{I}^0_i=\sum_{\mathcal{I}^0_i\in\mathfrak{B}}\mathfrak{c}_i\,\mathbf{M}^0_{i\,j}\Rightarrow \mathfrak{c}_i=\sum_{j}\Big(\mathbf{M}^0\Big)^{-1}_{j\,i}\mathfrak{a}_j\,.\label{general_solution_to_unitarity_problem_via_periods}}
To be clear, this procedure has nothing to do with planarity or even the routing of loop momentum variables---as all internal degrees of freedom are fully integrated-out in (\ref{general_solution_to_unitarity_problem_via_periods}). Moreover, this strategy works for \emph{any} master integrand basis given \emph{any} choice of a spanning set of contours. However, it should be clear that the linear algebra involved above would be simplified considerably if the master integrand basis were chosen so that its period matrix were simpler. 

A \emph{prescriptive} basis of master integrands $\{\mathcal{I}_i\}$ is defined as the cohomological-dual to \emph{some choice} of a spanning set of integration contours (viewed as elements of homology on the space defining the integrands). That is, relative to a particular choice of integration contours $\{\Omega_{j}\}$, an integrand basis $\{\mathcal{I}_{i}\}$ is said to be prescriptive (relative to these contours) if it has a unit period matrix:
\eq{\oint\limits_{\Omega_j}\mathcal{I}_i=\delta_{i\,j}\,.}
In such a basis, the coefficients $\mathfrak{c}_i$ in the expansion of an amplitude integrand (\ref{general_amplitude_expansion}) are simply on-shell functions $\mathfrak{a}_i$: $\mathcal{A}=\sum_i\,\mathfrak{a}_i\mathcal{I}_i$.

To be clear, given \emph{any} initial choice of master integrands $\{\mathcal{I}_i^0\}$ and \emph{any} choice of contours, a prescriptive basis may be constructed by simple linear algebra: 
\eq{\mathcal{I}_i\equivR\sum_{j}\Big(\!\mathbf{M}^0\!\Big)^{-1}_{i\,j}\mathcal{I}_j^0\,,}
where $\mathbf{M}^0$ is the period matrix for any initial choice of master integrands (\ref{general_period_matrix}) relative to this choice of contours. And the prescriptive basis (relative to this choice of contours) will be unique. \\

Suppose that for each Feynman integrand topology of a basis, there could be found a spanning set of of contours for the top-level degrees of freedom for the graph such that each contour is defined (in part, at least) by enclosing every one of the Feynman propagators of that integrand---a so-called `maximal cut' of the Feynman graph. Then any master integrand involving a strict subset of Feynman propagators (contact-terms) would necessarily vanish on the contours defining the top-level master integrands. Moreover, two graphs with different subsets of propagators would automatically vanish on the contours defining each other's numerators. This results in a `triangular' period matrix for integrands divided into manifest top-level subspaces and spaces spanned by contact terms. This structure was illustrated with examples in \cite{Bourjaily:2021hcp}, and we will encounter this below.

\subsubsection{Prescriptivity of Integrands and \emph{Purity} of Integrals}\label{subsubsec:prescriptivity_and_purity}
Besides simplifying the work involved in representing amplitudes, prescriptive integrand bases are valuable as they often have properties that simplify loop integration---which remains the hardest part of perturbation theory. This has been observed many times over recent years in the context of loop integrands which are entirely polylogarithmic (see e.g.~\cite{ArkaniHamed:2010gh,Drummond:2010cz,Drummond:2010mb,Arkani-Hamed:2014via}).

For the case of polylogarithmic Feynman integrals, an integrand basis is said to be \emph{pure} if it is satisfies a particular system of nilpotent differential equations (see e.g.~\cite{Henn:2013pwa,Henn:2014loa,Henn:2014qga}). Specifically, the space of master integrands must be organized by (some notion of `transcendental') \emph{weight}, and have the property that any derivatives of integrands with weight $w$ may be spanned entirely by those of lower weight (with coefficients algebraic coefficients). The notion of polylogarithmic purity is simple enough to understand: $\log(f(\g{z}))$ would be pure for any algebraic function $f(\g{z})$ because differentiating it with respect to $\g{z}$ would result in an algebraic function; in contrast, $g(\g{z})\log(f(\g{z}))$ would not be pure because its derivatives would involve both algebraic and transcendental pieces. This example generalizes naturally to higher-weight, multiple-polylogarithms. 

Thus, any Feynman integrand that integrates to a sum of polylogarithmic functions with \emph{constant} coefficients (independent of the kinematics) would be called pure. The coefficients of any multiple-polylogarithm may be accessed by contour integrals which encircle each of its simple poles; as such, a Feynman integrand which is pure should have the property of having `unit leading singularities' as defined in \cite{ArkaniHamed:2010gh,Arkani-Hamed:2014via}. More precisely, a Feynman integrand should be pure if \emph{all} its periods are kinematic-independent constants---on any choice of maximum-dimensional, compact contour of integration.

To understand how prescriptivity and purity are connected, consider for example the space of pentabox integrands defined by
\vspace{-10pt}\eq{\mathcal{I}(\r{N})\,\,\bigger{\Leftrightarrow}\!\!\pBox{\pBoxPlainEdges
\leg{(v1)}{-110}{$$}\leg{(v2)}{180}{$$}\legMassive{(v3)}{110}{$$}\leg{(v4)}{81}{$$}\legMassive{(v5)}{45}{$$}\legMassive{(v6)}{-45}{$$}
\coordinate (a1) at ($(v7)!.0!(v1)+(-72:0.25)$);\coordinate (a2) at ($(v1)!.5!(v2)+(-144:0.25)$);\coordinate (a3) at ($(v2)!.5!(v3)+(144:0.25)$);\coordinate (a4) at ($(v3)!.5!(v4)+(72:0.25)$);\coordinate (b1) at ($(v4)!.5!(v5)+(90:0.205)$);\coordinate (b2) at ($(v5)!.5!(v6)+(0:0.25)$);
\node[rdot] at (a1){};\node[rdot] at (a2){};\node[rdot] at (a3){};\node[rdot] at (a4){};\node[rdot] at (b1){};\node[rdot] at (b2){};
\coordinate (ella) at ($(v7)!.5!(v4)+(180:0.575)$);\coordinate (ellb) at ($(v7)!.5!(v4)+(0:\figScale*0.45)$);\node at ($(ella)$) {\text{{\normalsize${\ell_1}$}}};\node at ($(ellb)$) {\text{{\normalsize${\ell_2}$}}};
\node at ($(a1)+(-72:0.195)+(-90:0.05)+(0:0.0)$) {\text{{\normalsize$\r{a_1}$}}};
\node at ($(a2)+(-144:0.195)+(-90:0.1)+(0:0.2)$) {\text{{\normalsize$\r{a_2}$}}};
\node at ($(a3)+(144:0.195)+(90:0.05)+(0:0.2)$) {\text{{\normalsize$\r{a_3}$}}};
\node at ($(a4)+(72:0.195)$) {\text{{\normalsize$\r{a_4}$}}};
\node at ($(b1)+(80:0.195)+(0:0.1)+(80:0.05)$) {\text{{\normalsize$\r{b_1}$}}};
\node at ($(b2)+(0:0.25)$) {\text{{\normalsize$\r{b_2}$}}};
\draw[markedEdge] (v2)--(v3);
}\,.\label{alice_pentabox}\vspace{-10pt}}
This integrand is first relevant to planar scattering amplitudes for 9 particles. It can be obtained from the general case of (\ref{scalar_pentabox_integrand}) by identifying $\r{b_3}\!=\!\r{a_1}$, and taking the pairs of points $(\r{a_1},\r{a_2})$, $(\r{a_2},\r{a_3})$, and $(\r{a_4},\r{b_1})$ to be light-like separated. As before, we may describe an initial set of master integrands by their loop-dependent numerators chosen from $[\ell_1]$, which would be spanned by two `top-level' numerators and four contact-terms:
\vspace{-3pt}\eq{
\mathfrak{n}(\ell_1)\!\in\![\ell_1]=\mathrm{span}\Big\{\!\underbrace{\x{\ell_1}{\b{N_1^0}},\x{\ell_1}{\b{N_2^0}}}_{\text{`top-level'}},
\underbrace{\x{\ell_1}{\r{a_1}},\x{\ell_1}{\r{a_2}},\x{\ell_1}{\r{a_3}},\x{\ell_1}{\r{a_4}}}_{\text{`contact-terms'}}\!\Big\}\,.\vspace{-5pt}
\label{alice_numerators}\vspace{-3pt}}
Here, the precise form for the top-level numerators are not so important to us at the moment. However, it is interesting to note that the choice for these top-level integrands made by the authors of \cite{Bourjaily:2015jna} would \emph{not} be pure: upon loop integration, the `chiral' numerators defined in \cite{Bourjaily:2015jna} would result in an expression of the form:
\eq{\Big(\text{pure polylogarithms}\Big)+\b{f(\vec{p})}\!\times\!\Big(\text{other pure polylogarithms}\Big)\label{generic_structure_of_impure_alice_integral}}
where $\b{f(\vec{p})}$ is some \emph{algebraic} function of the external kinematics. The existence of such a prefactor prevents this expression from being pure or satisfying canonical sets of differential equations. The second sum of polylogarithms appearing (\ref{generic_structure_of_impure_alice_integral}) turns out to be nothing other than \emph{precisely} the result of integrating the (properly normalized) scalar double-box integral:
\vspace{-6pt}\eq{\int\limits_{\ell_{i}\in\mathbb{R}^{4}}\dBox{\dBoxPlainEdges\legMassive{(v1)}{-135}{$$}\legMassive{(v2)}{135}{$$}\leg{(v3)}{90}{$$}\legMassive{(v4)}{45}{$$}\legMassive{(v5)}{-45}{$$}
\coordinate(a1) at ($(v6)+(-90:0.3)$);\coordinate(a3) at ($(v1)!.5!(v2)+(180:0.3)$);\coordinate(a4) at ($(v2)!.5!(v3)+(90:0.2)$);\coordinate(b1) at ($(v3)!.5!(v4)+(90:0.2)$);\coordinate(b2) at ($(v4)!.5!(v5)+(0:0.3)$);\coordinate (ella) at ($(v6)!.5!(v3)+(180:\figScale*0.6)$);\coordinate (ellb) at ($(v6)!.5!(v3)+(0:\figScale*0.65)$);
\node[rdot] at (a1){};\node[rdot] at (a3){};\node[rdot] at (a4){};\node[rdot] at (b1){};\node[rdot] at (b2){};
\node at ($(a1)+(-72:0.195)+(-90:0.05)+(180:0.0)$) {\text{{\normalsize$\r{a_1}$}}};\node at ($(a3)+(-72:0.195)+(-90:0.03)+(0:0.03)+(90:0.38)$) {\text{{\normalsize$\r{a_3}$}}};\node at ($(b2)+(-72:0.195)+(-90:0.05)+(90:0.075)+(0:0.05)+(90:0.38)$) {\text{{\normalsize$\r{b_2}$}}};\node at ($(b1)+(80:0.195)+(0:0.1)+(80:0.05)$) {\text{{\normalsize$\r{b_1}$}}};\node at ($(a4)+(80:0.195)+(0:0.04)+(-90:0.045)+(80:0.05)$) {\text{{\normalsize$\r{a_4}$}}};\node at (ella) {\text{{\normalsize$\ell_1$}}};\node at (ellb) {\text{{\normalsize$\ell_2$}}};
}\propto\Big(\text{other pure polylogarithms}\Big)\,.
\vspace{-6pt}\label{general_form_of_alice}}
Thus, it is obvious that adding the appropriate amount of this contact-term would render the integral (\ref{generic_structure_of_impure_alice_integral}) pure---and this would simplify the resulting expression for the integral. Adding such a contact term would correspond to a particular but simple `rotation' in the space of initial master integrand numerators in (\ref{alice_numerators}). Without having the integrated expression on hand, it is natural to wonder how the initial integrand's \emph{impurity} could have been detected, and how the rotated basis could have been found?

The impurity of the initial Feynman integrand can easily be detected by the fact that some of its period integrals were kinematic-dependent. Let us see how this can be identified and cured by following the notion of prescriptivity.

For the six-dimensional space of master integrands defined in (\ref{alice_numerators}) for the pentabox (\ref{alice_pentabox}), consider the following (spanning set) of six, eight-dimensional compact contour integrals:
\eq{\begin{split}\hspace{65pt}\fwboxR{0pt}{\{\Omega_1,\ldots,\Omega_6\}\equivR\!\!\left\{\rule{0pt}{35pt}\right.\hspace{-10pt}}\pBox{\pBoxCoords\pBoxPlainEdges\coordinate (ella) at ($(v7)!.5!(v4)+(180:0.575)$);\coordinate (ellb) at ($(v7)!.5!(v4)+(0:\figScale*0.45)$);\coordinate (a1) at ($(v7)!.0!(v1)+(-72:0.25)$);\coordinate (a2) at ($(v1)!.5!(v2)+(-144:0.25)$);\coordinate (a3) at ($(v2)!.5!(v3)+(144:0.25)$);\coordinate (a4) at ($(v3)!.5!(v4)+(72:0.25)$);\coordinate (b1) at ($(v4)!.5!(v5)+(90:0.205)$);\coordinate (b2) at ($(v5)!.5!(v6)+(0:0.25)$);\leg{(v1)}{-110}{$$}\leg{(v2)}{-180}{$$}\legMassive{(v3)}{110}{$$}\leg{(v4)}{81}{$$}\legMassive{(v5)}{45}{$$}\legMassive{(v6)}{-45}{$$}\node[rdot] at (a1){};\node[rdot] at (a2){};\node[rdot] at (a3){};\node[rdot] at (a4){};\node[rdot] at (b1){};\node[rdot] at (b2){};\coordinate (ella) at ($(v7)!.5!(v4)+(180:0.575)$);\coordinate (ellb) at ($(v7)!.5!(v4)+(0:\figScale*0.45)$);\node at ($(ella)$) {\text{{\normalsize${\ell_1^*}$}}};\node at ($(ellb)$) {\text{{\normalsize${\ell_2^*}$}}};
\node at ($(a1)+(-72:0.195)+(-90:0.05)+(0:0.0)$) {\text{{\normalsize$\r{a_1}$}}};
\node at ($(a2)+(-144:0.195)+(-90:0.1)+(0:0.2)$) {\text{{\normalsize$\r{a_2}$}}};
\node at ($(a3)+(144:0.195)+(90:0.05)+(0:0.2)$) {\text{{\normalsize$\r{a_3}$}}};
\node at ($(a4)+(72:0.195)$) {\text{{\normalsize$\r{a_4}$}}};
\node at ($(b1)+(80:0.195)+(0:0.1)+(80:0.05)$) {\text{{\normalsize$\r{b_1}$}}};
\node at ($(b2)+(0:0.25)$) {\text{{\normalsize$\r{b_2}$}}};
\contourVerts{1}{2}{4}{4}{4}{4}{2}},\pBox{\pBoxCoords\pBoxPlainEdges\coordinate (ella) at ($(v7)!.5!(v4)+(180:0.575)$);\coordinate (ellb) at ($(v7)!.5!(v4)+(0:\figScale*0.45)$);\coordinate (a1) at ($(v7)!.0!(v1)+(-72:0.25)$);\coordinate (a2) at ($(v1)!.5!(v2)+(-144:0.25)$);\coordinate (a3) at ($(v2)!.5!(v3)+(144:0.25)$);\coordinate (a4) at ($(v3)!.5!(v4)+(72:0.25)$);\coordinate (b1) at ($(v4)!.5!(v5)+(90:0.205)$);\coordinate (b2) at ($(v5)!.5!(v6)+(0:0.25)$);\leg{(v1)}{-110}{$$}\leg{(v2)}{-180}{$$}\legMassive{(v3)}{110}{$$}\leg{(v4)}{81}{$$}\legMassive{(v5)}{45}{$$}\legMassive{(v6)}{-45}{$$}\node[rdot] at (a1){};\node[rdot] at (a2){};\node[rdot] at (a3){};\node[rdot] at (a4){};\node[rdot] at (b1){};\node[rdot] at (b2){};
\coordinate (ellb) at ($(v7)!.5!(v4)+(0:\figScale*0.45)$);\node at ($(ella)$) {\text{{\normalsize${\ell_1^*}$}}};\node at ($(ellb)$) {\text{{\normalsize${\ell_2^*}$}}};\node at ($(a1)+(-72:0.195)+(-90:0.05)+(0:0.0)$) {\text{{\normalsize$\r{a_1}$}}};
\node at ($(a2)+(-144:0.195)+(-90:0.1)+(0:0.2)$) {\text{{\normalsize$\r{a_2}$}}};
\node at ($(a3)+(144:0.195)+(90:0.05)+(0:0.2)$) {\text{{\normalsize$\r{a_3}$}}};
\node at ($(a4)+(72:0.195)$) {\text{{\normalsize$\r{a_4}$}}};
\node at ($(b1)+(80:0.195)+(0:0.1)+(80:0.05)$) {\text{{\normalsize$\r{b_1}$}}};
\node at ($(b2)+(0:0.25)$) {\text{{\normalsize$\r{b_2}$}}};\contourVerts{2}{1}{4}{4}{4}{4}{1}},
\dBox{\dBoxPlainEdges\leg{(v1)}{-135}{$$}\legMassive{(v2)}{135}{$$}\leg{(v3)}{90}{$$}\legMassive{(v4)}{45}{$$}\legMassive{(v5)}{-45}{$$}\leg{(v6)}{-90}{$$}
\coordinate(a1) at ($(v6)+(-90:0.3)$);\coordinate(a2) at ($(v6)!.5!(v1)+(-90:0.25)$);\coordinate(a3) at ($(v1)!.5!(v2)+(180:0.3)$);\coordinate(a4) at ($(v2)!.5!(v3)+(90:0.2)$);\coordinate(b1) at ($(v3)!.5!(v4)+(90:0.2)$);\coordinate(b2) at ($(v4)!.5!(v5)+(0:0.3)$);\coordinate(b3)at ($(v6)!.5!(v5)+(-90:0.25)$);\coordinate (ella) at ($(v6)!.5!(v3)+(180:\figScale*0.6)$);\coordinate (ellb) at ($(v6)!.5!(v3)+(0:\figScale*0.65)$);\node[rdot] at (a2){};\node[rdot] at (a3){};\node[rdot] at (a4){};\node[rdot] at (b1){};\node[rdot] at (b2){};\node[rdot]at(b3){};\node at ($(ella)$) {\text{{\normalsize${\ell_1^*}$}}};\node at ($(ellb)$) {\text{{\normalsize${\ell_2^*}$}}};
\node at ($(b3)+(-72:0.195)+(-90:0.05)+(180:0.0)$) {\text{{\normalsize$\r{a_1}$}}};
\node at ($(a2)+(-72:0.195)+(-90:0.05)+(180:0.0)$) {\text{{\normalsize$\r{a_2}$}}};\node at ($(a3)+(-72:0.195)+(-90:0.03)+(0:0.03)+(90:0.38)$) {\text{{\normalsize$\r{a_3}$}}};\node at ($(b2)+(-72:0.195)+(-90:0.05)+(90:0.075)+(0:0.05)+(90:0.38)$) {\text{{\normalsize$\r{b_2}$}}};\node at ($(b1)+(80:0.195)+(0:0.1)+(80:0.05)$) {\text{{\normalsize$\r{b_1}$}}};\node at ($(a4)+(80:0.195)+(0:0.04)+(-90:0.045)+(80:0.05)$) {\text{{\normalsize$\r{a_4}$}}};
\contourVerts{3}{4}{4}{4}{4}{4}{5}}\fwbox{0pt}{,}
\\
\hspace{65pt}\dBox{\dBoxPlainEdges\legMassive{(v1)}{-135}{$$}\legMassive{(v2)}{135}{$$}\leg{(v3)}{90}{$$}\legMassive{(v4)}{45}{$$}\legMassive{(v5)}{-45}{$$}
\coordinate(a1) at ($(v6)+(-90:0.3)$);\coordinate(a3) at ($(v1)!.5!(v2)+(180:0.3)$);\coordinate(a4) at ($(v2)!.5!(v3)+(90:0.2)$);\coordinate(b1) at ($(v3)!.5!(v4)+(90:0.2)$);\coordinate(b2) at ($(v4)!.5!(v5)+(0:0.3)$);\coordinate (ella) at ($(v6)!.5!(v3)+(180:\figScale*0.6)$);\coordinate (ellb) at ($(v6)!.5!(v3)+(0:\figScale*0.65)$);
\node[rdot] at (a1){};\node[rdot] at (a3){};\node[rdot] at (a4){};\node[rdot] at (b1){};\node[rdot] at (b2){};
\node at ($(a1)+(-72:0.195)+(-90:0.05)+(180:0.0)$) {\text{{\normalsize$\r{a_1}$}}};\node at ($(a3)+(-72:0.195)+(-90:0.03)+(0:0.03)+(90:0.38)$) {\text{{\normalsize$\r{a_3}$}}};\node at ($(b2)+(-72:0.195)+(-90:0.05)+(90:0.075)+(0:0.05)+(90:0.38)$) {\text{{\normalsize$\r{b_2}$}}};\node at ($(b1)+(80:0.195)+(0:0.1)+(80:0.05)$) {\text{{\normalsize$\r{b_1}$}}};\node at ($(a4)+(80:0.195)+(0:0.04)+(-90:0.045)+(80:0.05)$) {\text{{\normalsize$\r{a_4}$}}};\node at ($(ella)$) {\text{{\normalsize${\ell_1^*}$}}};\node at ($(ellb)$) {\text{{\normalsize${\ell_2^*}$}}};
\contourVerts{4}{4}{4}{4}{4}{3}{5}}
,
\dBox{\dBoxPlainEdges\leg{(v1)}{-135}{$$}\legMassive{(v2)}{135}{$$}\leg{(v3)}{90}{$$}\legMassive{(v4)}{45}{$$}\legMassive{(v5)}{-45}{$$}
\coordinate(a1) at ($(v6)+(-90:0.3)$);\coordinate(a3) at ($(v1)!.5!(v2)+(180:0.3)$);\coordinate(a4) at ($(v2)!.5!(v3)+(90:0.2)$);\coordinate(b1) at ($(v3)!.5!(v4)+(90:0.2)$);\coordinate(b2) at ($(v4)!.5!(v5)+(0:0.3)$);\coordinate (ella) at ($(v6)!.5!(v3)+(180:\figScale*0.6)$);\coordinate (ellb) at ($(v6)!.5!(v3)+(0:\figScale*0.65)$);
\node[rdot] at (a1){};\node[rdot] at (a3){};\node[rdot] at (a4){};\node[rdot] at (b1){};\node[rdot] at (b2){};
\node at ($(a1)+(-72:0.195)+(-90:0.05)+(180:0.0)$) {\text{{\normalsize$\r{a_1}$}}};\node at ($(a3)+(-72:0.195)+(-90:0.03)+(0:0.03)+(90:0.38)$) {\text{{\normalsize$\r{a_2}$}}};\node at ($(b2)+(-72:0.195)+(-90:0.05)+(90:0.075)+(0:0.05)+(90:0.38)$) {\text{{\normalsize$\r{b_2}$}}};\node at ($(b1)+(80:0.195)+(0:0.1)+(80:0.05)$) {\text{{\normalsize$\r{b_1}$}}};\node at ($(a4)+(80:0.195)+(0:0.04)+(-90:0.045)+(80:0.05)$) {\text{{\normalsize$\r{a_4}$}}};\node at ($(ella)$) {\text{{\normalsize${\ell_1^*}$}}};\node at ($(ellb)$) {\text{{\normalsize${\ell_2^*}$}}};
\contourVerts{3}{4}{4}{4}{4}{2}{5}}
,
\dBox{\draw[int](v6)--(v1);\draw[dashed](v1)--(v2);\draw[int](v2)--(v3);\draw[int](v3)--(v4);\draw[int](v4)--(v5);\draw[int](v5)--(v6);\draw[int](v6)--(v3);\leg{(v1)}{-135}{$$}\leg{(v2)}{135}{$$}\legMassive{(v3)}{90}{$$}\legMassive{(v4)}{45}{$$}\legMassive{(v5)}{-45}{$$}
\coordinate(a1) at ($(v6)+(-90:0.3)$);\coordinate(a3) at ($(v1)!.5!(v2)+(180:0.3)$);\coordinate(a4) at ($(v2)!.5!(v3)+(90:0.2)$);\coordinate(b1) at ($(v3)!.5!(v4)+(90:0.2)$);\coordinate(b2) at ($(v4)!.5!(v5)+(0:0.3)$);\coordinate (ella) at ($(v6)!.5!(v3)+(180:\figScale*0.6)$);\coordinate (ellb) at ($(v6)!.5!(v3)+(0:\figScale*0.65)$);
\node[rdot] at (a1){};\node[rdot] at (a3){};\node[rdot] at (a4){};\node[rdot] at (b1){};\node[rdot] at (b2){};
\node at ($(a1)+(-72:0.195)+(-90:0.05)+(180:0.0)$) {\text{{\normalsize$\r{a_1}$}}};\node at ($(a3)+(-72:0.195)+(-90:0.03)+(0:0.03)+(90:0.38)$) {\text{{\normalsize$\r{a_2}$}}};\node at ($(b2)+(-72:0.195)+(-90:0.05)+(90:0.075)+(0:0.05)+(90:0.38)$) {\text{{\normalsize$\r{b_2}$}}};\node at ($(b1)+(80:0.195)+(0:0.1)+(80:0.05)$) {\text{{\normalsize$\r{b_1}$}}};\node at ($(a4)+(80:0.195)+(0:0.04)+(-90:0.045)+(80:0.05)$) {\text{{\normalsize$\r{a_3}$}}};\node at ($(ella)$) {\text{{\normalsize${\ell_1^*}$}}};\node at ($(ellb)$) {\text{{\normalsize${\ell_2^*}$}}};
\contourVerts{0}{0}{4}{4}{4}{2}{5}}\fwboxL{0pt}{\hspace{-5pt}\left.\rule{0pt}{35pt}\right\}\fwboxL{0pt}{.}}
\end{split}\label{alice_contours}}
These pictures represent contours for integration as follows. For each contour, all its propagators should be cut (a contour encircling each of its poles); which particular solution to be taken is indicated by the color of each three-point vertex, with blue (white) indicating that the $\tilde{\lambda}$'s ($\lambda$'s) of the momenta at the vertex should be taken as proportional; a three-point vertex enclosed in a circle indicates a collinear pole (the intersection of the two solutions to the cut equations), and a dashed line indicates a momentum taken to be soft. Notice that the contours $\{\Omega_3,\Omega_5,\Omega_6\}$ \emph{directly} correspond to regions of loop momenta responsible for infrared divergences. 

Consider the initial choice of master integrands' numerators to be given by (\ref{alice_numerators}) with $\big|\b{N_i^0}\big)\equivR\big|\b{Q^{i}_{\r{\vec{a}}}}\big)$ where $\b{Q}^i_{\r{\vec{a}}}$ denotes one of the two solutions to the cut equations involving propagators $\x{\ell_1}{\r{a_i}}\!=\!0$, so that these numerators will each vanish on one of the two contours which involve cutting all four propagators $\x{\ell_1}{\r{a_i}}$. This are (not-yet-normalized) `chiral' numerators for the pentabox. 

If we compute the period matrix for these master integrands using the contours in (\ref{alice_contours}), it will take the form:
\eq{\fwboxL{0pt}{\raisebox{-9pt}{$\hspace{44pt}\left(\rule{0pt}{49pt}\right.$}}\begin{array}{@{}l@{}|@{$\;\;$}c@{}c@{}c@{}c@{}c@{}c@{}c@{}c@{}c@{}}
\fwbox{40pt}{\mathfrak{n}(\ell_1)\!\!\,\,}&\fwbox{38pt}{{\Omega_1}}&\fwbox{38pt}{\Omega_2}&\fwbox{38pt}{\Omega_3}&\fwbox{38pt}{\Omega_4}&\fwbox{38pt}{\Omega_5}&\fwbox{38pt}{\Omega_6}
\\\hline\\[-12pt]
\x{\ell_1}{\b{N_1^0}}&\b{f_1^0(\vec{p})}&0&\b{g_1^{\r{1}}(\vec{p})}&\b{g_1^{\r{2}}(\vec{p})}&\b{g_1^{\r{3}}(\vec{p})}&\b{g_1^{\r{4}}(\vec{p})}\\
\x{\ell_1}{\b{N_2^0}}&0&\b{f_2^0(\vec{p})}&\b{g_2^{\r{1}}(\vec{p})}&\b{g_2^{\r{2}}(\vec{p})}&\b{g_2^{\r{3}}(\vec{p})}&\b{g_2^{\r{4}}(\vec{p})}\\
\x{\ell_1}{\r{a_1}}&0&0&\b{h_{\r{1}}(\vec{p})}&0&0&0\\
\x{\ell_1}{\r{a_2}}&0&0&0&\b{h_{\r{2}}(\vec{p})}&0&0\\
\x{\ell_1}{\r{a_3}}&0&0&0&0&\b{h_{\r{3}}(\vec{p})}&0\\
\x{\ell_1}{\r{a_4}}&0&0&0&0&0&\b{h_{\r{4}}(\vec{p})}
\end{array}\fwboxL{0pt}{\raisebox{-9pt}{$\hspace{-6.5pt}\left.\rule{0pt}{49pt}\right)\!\!\!\equivL\mathbf{M}^0\,.$}}\label{alice_period_matrix}}
Notice that each of the double-box integrands only have support on the \emph{corresponding} double-box contours: they vanish (trivially) on the pentabox contours, as they each lack one of the requisite propagators, and they vanish on the \emph{other} double-box contours for the same reason. 

To construct a prescriptive basis from this initial basis, we need only diagonalize the period matrix (\ref{alice_period_matrix}). Considering its form, it is not hard to see that a prescriptive basis would correspond to the numerators chosen to be:
\eq{\begin{split}\x{\ell_1}{\b{N_i^0}}&\mapsto\x{\ell_1}{\b{N_i}}\equivR\frac{1}{\b{f_i^0(\vec{p})}}\Bigg[\x{\ell_1}{\b{N_i^0}}{\,-}\sum_{\r{j}=1}^4\frac{\b{g_i^{\r{j}}(\vec{p})}}{\b{h_{\r{j}}(\vec{p})}}\x{\ell_1}{\r{a_j}}\Bigg]\,;\\
\x{\ell_1}{\b{\r{a_j}}}&\mapsto\x{\ell_1}{\b{\r{a_j}}}/\b{h_{\r{j}}(\vec{p})}\,.
\end{split}\label{prescriptive_alice_numerators}
}
Notice that prescriptivity provides a \emph{precise} rule for adding contact-terms to some choice of initial top-level integrands. The new basis of master integrands with numerators given in (\ref{prescriptive_alice_numerators}) have several interesting features. For one thing, the regions of loop momenta responsible for infrared divergences are fully removed from the new top-level master integrands.\footnote{In fact, the initial choice of top-level integrands would have been infrared finite: by vanishing on one or the other solution to each three-particle cut, they will trivially vanish in the collinear (or soft) limits responsible for IR divergences. As such, the actual period matrix would have had $\b{g_i^{\r{j}}}(\vec{p})\!=\!0$ for all $\r{j}\!\neq\!2$. But this does not affect our argument, which is simpler in general.} For another, each of the masters associated with double-box contact-terms would be properly normalized to be pure polylogarithms. Finally, by vanishing on the cut $\Omega_3$ associated with the contact-term (\ref{general_form_of_alice}) the new top-level integrands involving numerators $\x{\ell_1}{\b{N_i}}$ will be pure.\\

Although the example above illustrates how a rotation of some initial choice of master integrands prescriptive often results in individually pure integrands, we should be clear that prescriptivity alone does not \emph{ensure} integrand purity. This is fairly obvious and trivial: a simple counter example would be a master integrand basis consisting of a single Feynman integrand; prescriptivity would only ensure the integrand's normalization on a single contour of integration, but this would be far from guaranteeing that its many other contour integrals are kinematically independent (for one thing, \emph{any} impurity associated with a contact-term would go undetected). 

More interesting exceptions arise when integrands involve both polylogarithmic contours enclosing simple poles and elliptic (or more rigid) substructures. As we will show in the following \mbox{section \ref{sec:stratifying_rigidity}}, although prescriptivity may always be used to define a particular choice of master integrands, it will not necessarily result in integrands which have definite rigidity. Indeed, in the case of pentabox integrands with box power-counting (those which can be rendered dual conformally invariant), there does not exist any choice of masters with definite rigidity and hence, no integrand basis which is pure. 

The unavoidable appearance of elliptic contributions to dual conformal pentabox integrals turns out to explain why the authors of \cite{Bourjaily:2015jna} overlooked the obviously superior choice of top-level integrand numerators for the pentabox (\ref{alice_pentabox}): they decided to treat all finite double-box integrands on equal footing, ignoring any distinction between those which were elliptic from those which were polylogarithmic (in the interest of uniformity and concision), and thereby overlooking an important and obvious improvement available for those such as (\ref{alice_pentabox}) which would have been polylogarithmic. 

To understand prescriptivity in the case of elliptic integrals, we first must better understand how they appear and the space of contour integrals which may be used to construct period matrices involving them.

\newpage
\section[Stratifications in Rigidity: Ellipticity, Polylogarithmicity, \& Purity]{Stratifications in Rigidity: Polylogarithmicity and Purity}\label{sec:stratifying_rigidity}

\subsection{Appearance of Non-Polylogarithmic Contributions at Two Loops}\label{subsec:source_of_rigidity}\vspace{0pt}
For the purposes of most of the following, the use of the language of momentum twistors and embedding space is---perhaps unfortunately---unavoidable. For those readers with limited (or no) familiarity with this material, we have supplied a short review of the necessary background in appendix~\ref{appendix:embedding_and_momentum_twistors}.

For planar theories at two loops involving massless propagators, there is essentially a unique source of non-vanishing rigidity in Feynman integrals: ellipticity associated with double-boxes of the form (\ref{scalar_double_box}), \cite{Caron-Huot:2012awx,Bourjaily:2017bsb}. The elliptic geometry can seen from the fact that the maximal cut (the contour integral which encloses the vanishing of each of its seven propagators)\footnote{To be clear, there are two solutions to the seven-cut equations; we define our contour as the \emph{odd} combination of these.} of double-box, denoted  $\Omega[\r{\vec{a}};\r{\vec{b}}\hspace{1pt}]$, is given by
\eq{\oint\limits_{\Omega[\r{\vec{a}};\r{\vec{b}}\hspace{1pt}]}\hspace{-5pt}\left[\rule{0pt}{35pt}\right.\hspace{-5pt}\dBox{\draw[int](v5)--(v1);\draw[int](v1)--(v2);\draw[int](v2)--(v3);\draw[int](v3)--(v4);\draw[int](v4)--(v5);\draw[int](v6)--(v3);\legMassive{(v1)}{-135}{$$}\legMassive{(v2)}{135}{$$}\legMassive{(v3)}{90}{$$}\legMassive{(v4)}{45}{$$}\legMassive{(v5)}{-45}{$$}\legMassive{(v6)}{-90}{$$}
\coordinate (ella) at ($(v6)!.5!(v1)+(90:\figScale*0.65)$);
\coordinate (ellb) at ($(v6)!.5!(v5)+(90:\figScale*0.65)$);
\coordinate (a1) at ($(v6)!.55!(v1)+(-90:0.25)$);
\coordinate (a2) at ($(v1)!.5!(v2)+(180:0.25)$);
\coordinate (a3) at ($(v3)!.55!(v2)+(90:0.25)$);
\coordinate (b1) at ($(v3)!.55!(v4)+(90:0.25)$);
\coordinate (b2) at ($(v4)!.5!(v5)+(0:0.25)$);
\coordinate (b3) at ($(v6)!.55!(v5)+(-90:0.25)$);
\node[rdot] at (a1){};
\node[rdot] at (a2){};
\node[rdot] at (a3){};
\node[rdot] at (b1){};
\node[rdot] at (b2){};
\node[rdot] at (b3){};
\node at ($(ella)$) {\text{\normalsize{$\ell_1$}}};
\node at ($(ellb)$) {\text{\normalsize{$\ell_2$}}};
\node at ($(a1)+(-90:0.25)$) {\text{{\normalsize$\r{a_1}$}}};
\node at ($(a2)+(180:0.275)+(90:0.0)$) {\text{{\normalsize$\r{a_2}$}}};
\node at ($(a3)+(90:0.25)$) {\text{{\normalsize$\r{a_3}$}}};
\node at ($(b1)+(0:0.05)+(90:0.25)$) {\text{{\normalsize$\r{b_1}$}}};
\node at ($(b2)+(0:0.275)+(90:0.05)$) {\text{{\normalsize$\r{b_2}$}}};
\node at ($(b3)+(-90:0.25)$) {\text{{\normalsize$\r{b_3}$}}};
}\hspace{-5pt}\left.\rule{0pt}{35pt}\right]\propto\frac{\dbar \g{\alpha}}{y(\g{\alpha})}\,,}
where $\g{\alpha}$ represents whatever integration variable remains from the original differential from $\dbar^4\!\ell_1\dbar^4\!\ell_2$ given in (\ref{double_box_in_dual_coordaintes}) and $y^2(\g{\alpha})\equivL Q(\g{\alpha})$ is a quartic polynomial. Such an integral is known to be associated with an elliptic curve. This curve becomes degenerate if any pair of the $\r{a_i}$'s or of the $\r{b_i}$'s become light-like separated, or if any $\r{a_i}$ is \emph{equal to} any $\r{b_j}$\footnote{In the former case, the integral will diverge in four dimensions; in the latter, it is finite---its integral over the Feynman contour is known and given by a weight-four polylogarithm \cite{MatthiasWilhelm}.}. Either degeneration results in a pair of roots of the quartic colliding, allowing us to define $y(\g{\alpha})\equivL(\g{\alpha}{-}q)\sqrt{q(\g{\alpha})}$ with $q(\g{\alpha})$ some quadratic; this allows for an additional \emph{polylogarithmic} contour integral to be taken (around $\g{\alpha}\!=\!q$) and the result of integration is indeed found to be polylogarithmic \cite{MatthiasWilhelm}. 

Thus, the following Feynman integrals would be elliptic multiple polylogarithms (have rigidity 1):
\eq{\fwbox{0pt}{\fwboxL{420pt}{\text{\textbf{elliptic:}}}\fwbox{0pt}{\hspace{-420pt}\left\{\rule{0pt}{30pt}\right.\hspace{-14pt}\dBox{\draw[int](v5)--(v1);\draw[int](v1)--(v2);\draw[int](v2)--(v3);\draw[int](v3)--(v4);\draw[int](v4)--(v5);\draw[int](v6)--(v3);\legMassive{(v1)}{-135}{$$}\legMassive{(v2)}{135}{$$}\legMassive{(v3)}{90}{$$}\legMassive{(v4)}{45}{$$}\legMassive{(v5)}{-45}{$$}\legMassive{(v6)}{-90}{$$}}\hspace{-7pt},\hspace{-7pt}\dBox{\draw[int](v5)--(v1);\draw[int](v1)--(v2);\draw[int](v2)--(v3);\draw[int](v3)--(v4);\draw[int](v4)--(v5);\draw[int](v6)--(v3);\legMassive{(v1)}{-135}{$$}\legMassive{(v2)}{135}{$$}\legMassive{(v3)}{90}{$$}\legMassive{(v4)}{45}{$$}\legMassive{(v5)}{-45}{$$}\leg{(v6)}{-90}{$$}}\hspace{-7pt},\hspace{-7pt}\dBox{\draw[int](v5)--(v1);\draw[int](v1)--(v2);\draw[int](v2)--(v3);\draw[int](v3)--(v4);\draw[int](v4)--(v5);\draw[int](v6)--(v3);\legMassive{(v1)}{-135}{$$}\legMassive{(v2)}{135}{$$}\leg{(v3)}{90}{$$}\legMassive{(v4)}{45}{$$}\legMassive{(v5)}{-45}{$$}\leg{(v6)}{-90}{$$}}
\hspace{-14pt}\left.\rule{0pt}{30pt}\right\}\fwboxL{0pt}{;}}}\label{elliptic_double_box_types}}
while any of the following would be strictly polylogarithmic (have rigidity 0):
\eq{\fwbox{0pt}{\fwboxL{420pt}{\text{\textbf{non-elliptic:}}}\fwbox{0pt}{\hspace{-420pt}\left\{\rule{0pt}{30pt}\right.\hspace{-14pt}\dBox{\draw[int](v5)--(v1);\draw[int](v1)--(v2);\draw[int](v2)--(v3);\draw[int](v3)--(v4);\draw[int](v4)--(v5);\draw[int](v6)--(v3);\leg{(v1)}{-135}{$$}\legMassive{(v2)}{135}{$$}\legMassive{(v3)}{90}{$$}\legMassive{(v4)}{45}{$$}\legMassive{(v5)}{-45}{$$}\legMassive{(v6)}{-90}{$$}}\hspace{-7pt},\hspace{-7pt}\dBox{\draw[int](v5)--(v1);\draw[int](v1)--(v2);\draw[int](v2)--(v3);\draw[int](v3)--(v4);\draw[int](v4)--(v5);\draw[int](v6)--(v3);\legMassive{(v1)}{-135}{$$}\legMassive{(v2)}{135}{$$}\legMassive{(v3)}{90}{$$}\legMassive{(v4)}{45}{$$}\legMassive{(v5)}{-45}{$$}}\hspace{-7pt},\hspace{-7pt}\dBox{\draw[int](v5)--(v1);\draw[int](v1)--(v2);\draw[int](v2)--(v3);\draw[int](v3)--(v4);\draw[int](v4)--(v5);\draw[int](v6)--(v3);\legMassive{(v1)}{-135}{$$}\legMassive{(v2)}{135}{$$}\legMassive{(v4)}{45}{$$}\legMassive{(v5)}{-45}{$$}}\hspace{-14pt}\left.\rule{0pt}{30pt}\right\}\fwboxL{0pt}{.}
}}\label{polylog_double_boxes}}

But these are not the only integrands we must consider, as is obvious in the embedding space formalism. In embedding space, all integrands are represented box power-counting in each loop momentum (achieved by adding appropriate powers of $\x{\ell_i}{\infX}\!\simeq\!1$ to the denominator as necessary); thus, any integrand with worse-than-box power-counting (with fewer than four propagators) will have additional `poles at infinity'---which are rendered manifest in embedding space. Thus, we can easily see that in addition to the double-boxes (\ref{elliptic_double_box_types}) shown above, generic `box-triangle' integrals would also be elliptic:
\eq{\bT{
\coordinate (ella) at ($(v5)!.5!(v1)+(90:\figScale*0.65)$);
\coordinate (ellb) at ($(v5)!.5!(v3)+(0:\figScale*0.325)$);
\coordinate (a1) at ($(v5)!.5!(v1)+(-90:0.25)$);
\coordinate (a2) at ($(v1)!.5!(v2)+(180:0.25)$);
\coordinate (a3) at ($(v2)!.5!(v3)+(90:0.25)$);
\coordinate (b1) at ($(v3)!.5!(v4)+(58:0.25)$);
\coordinate (b2) at ($(v4)!.5!(v5)+(-58:0.25)$);
\dimLines\draw[int] (ella)--(a1);
\draw[int] (ella)--(a2);
\draw[int] (ella)--(a3);
\draw[int] (ella)--(ellb);
\draw[int] (ellb)--(b1);
\draw[int] (ellb)--(b2);
\restoreDark
\draw[int](v5)--(v1);\draw[int](v1)--(v2);\draw[int](v2)--(v3);\draw[int](v3)--(v4);\draw[int](v4)--(v5);\draw[int](v5)--(v3);
\legMassive{(v1)}{-135}{$$}\legMassive{(v2)}{135}{$$}\legMassive{(v3)}{71.5}{$$}\legMassive{(v4)}{0}{$$}\legMassive{(v5)}{-71.5}{$$}
\node[rdot] at (a1){};
\node[rdot] at (a2){};
\node[rdot] at (a3){};
\node[rdot] at (b1){};
\node[rdot] at (b2){};
\node[ddot] at (ella){};
\node[ddot] at (ellb){};\node at ($(a1)+(-90:0.195)$) {\text{{\normalsize$\r{a_1}$}}};
\node at ($(a2)+(180:0.25)+(90:0.05)$) {\text{{\normalsize$\r{a_2}$}}};
\node at ($(a3)+(90:0.195)$) {\text{{\normalsize$\r{a_3}$}}};
\node at ($(b1)+(36:0.25)+(90:0.05)$) {\text{{\normalsize$\r{b_1}$}}};
\node at ($(b2)+(-36:0.25)+(-90:0.05)$) {\text{{\normalsize$\r{b_2}$}}};
}\hspace{-7pt}\bigger{\Rightarrow}\bT{
\coordinate (ella) at ($(v5)!.5!(v1)+(90:\figScale*0.65)$);
\coordinate (ellb) at ($(v5)!.5!(v3)+(0:\figScale*0.325)$);
\coordinate (a1) at ($(v5)!.5!(v1)+(-90:0.25)$);
\coordinate (a2) at ($(v1)!.5!(v2)+(180:0.25)$);
\coordinate (a3) at ($(v2)!.5!(v3)+(90:0.25)$);
\coordinate (b1) at ($(v3)!.5!(v4)+(58:0.25)$);
\coordinate (b2) at ($(v4)!.5!(v5)+(-58:0.25)$);
\dimLines\draw[int](v5)--(v1);\draw[int](v1)--(v2);\draw[int](v2)--(v3);\draw[int](v3)--(v4);\draw[int](v4)--(v5);\draw[int](v5)--(v3);\legMassive{(v1)}{-135}{$$}\legMassive{(v2)}{135}{$$}\legMassive{(v3)}{71.5}{$$}\legMassive{(v4)}{0}{$$}\legMassive{(v5)}{-71.5}{$$}
\restoreDark\draw[int] (ella)--(a1);
\draw[int] (ella)--(a2);
\draw[int] (ella)--(a3);
\draw[int] (ella)--(ellb);
\draw[int] (ellb)--(b1);
\draw[int] (ellb)--(b2);
\node[rdot] at (a1){};
\node[rdot] at (a2){};
\node[rdot] at (a3){};
\node[rdot] at (b1){};
\node[rdot] at (b2){};
\node[ddot] at (ella){};
\node[ddot] at (ellb){};\node at ($(a1)+(-90:0.195)$) {\text{{\normalsize$\r{a_1}$}}};
\node at ($(a2)+(180:0.25)+(90:0.05)$) {\text{{\normalsize$\r{a_2}$}}};
\node at ($(a3)+(90:0.195)$) {\text{{\normalsize$\r{a_3}$}}};
\node at ($(b1)+(36:0.25)+(90:0.05)$) {\text{{\normalsize$\r{b_1}$}}};
\node at ($(b2)+(-36:0.25)+(-90:0.05)$) {\text{{\normalsize$\r{b_2}$}}};
}\hspace{-7pt}\bigger{\simeq}\hspace{5pt}\dBox{\dimLines
\draw[int](v5)--(v1);\draw[int](v1)--(v2);\draw[int](v2)--(v3);\draw[int](v3)--(v4);\draw[int](v4)--(v5);\draw[int](v6)--(v3);\legMassive{(v1)}{-135}{$$}\legMassive{(v2)}{135}{$$}\legMassive{(v3)}{90}{$$}\legMassive{(v4)}{45}{$$}\legMassive{(v5)}{-45}{$$}\legMassive{(v6)}{-90}{$$}
\restoreDark
\coordinate (ella) at ($(v6)!.5!(v1)+(90:\figScale*0.65)$);
\coordinate (ellb) at ($(v6)!.5!(v5)+(90:\figScale*0.65)$);
\coordinate (a1) at ($(v6)!.55!(v1)+(-90:0.25)$);
\coordinate (a2) at ($(v1)!.5!(v2)+(180:0.25)$);
\coordinate (a3) at ($(v3)!.55!(v2)+(90:0.25)$);
\coordinate (b1) at ($(v3)!.55!(v4)+(90:0.25)$);
\coordinate (b2) at ($(v4)!.5!(v5)+(0:0.25)$);
\coordinate (b3) at ($(v6)!.55!(v5)+(-90:0.25)$);
\draw[int] (ella)--(a1);
\draw[int] (ella)--(a2);
\draw[int] (ella)--(a3);
\draw[int] (ella)--(ellb);
\draw[int] (ellb)--(b1);
\draw[int] (ellb)--(b2);
\draw[int] (ellb)--(b3);
\node[rdot] at (a1){};
\node[rdot] at (a2){};
\node[rdot] at (a3){};
\node[rdot] at (b1){};
\node[rdot] at (b2){};
\node[bldot] at (b3){};
\node[ddot] at (ella){};
\node[ddot] at (ellb){};
\node at ($(a1)+(-90:0.225)$) {\text{{\normalsize$\r{a_1}$}}};
\node at ($(a2)+(180:0.275)+(90:0.0)$) {\text{{\normalsize$\r{a_2}$}}};
\node at ($(a3)+(90:0.25)$) {\text{{\normalsize$\r{a_3}$}}};
\node at ($(b1)+(0:0.05)+(90:0.25)$) {\text{{\normalsize$\r{b_1}$}}};
\node at ($(b2)+(0:0.275)+(90:0.05)$) {\text{{\normalsize$\r{b_2}$}}};
\node at ($(b3)+(-90:0.15)+(0:0.2)$) {\text{{\normalsize$\,\,\,\,\b{\infX}$}}};
}\fwboxL{0pt}{\,\,\,,}
\label{box_triangle_ellipticity}}
while any `double-triangle' integral would be polylogarithmic, 
\eq{\dT{
\coordinate (ella) at ($(v4)!.5!(v2)+(180:0.325)$);
\coordinate (ellb) at ($(v4)!.5!(v2)+(0:0.325)$);
\coordinate (a1) at ($(v4)!.5!(v1)+(-135:0.25)$);
\coordinate (a2) at ($(v1)!.5!(v2)+(135:0.25)$);
\coordinate (b1) at ($(v2)!.5!(v3)+(45:0.25)$);
\coordinate (b2) at ($(v3)!.5!(v4)+(-45:0.25)$);
\dimLines
\draw[int] (ella)--(a1);
\draw[int] (ella)--(a2);
\draw[int] (ella)--(ellb);
\draw[int] (ellb)--(b1);
\draw[int] (ellb)--(b2);
\restoreDark
\draw[int](v4)--(v1);\draw[int](v1)--(v2);\draw[int](v2)--(v3);\draw[int](v3)--(v4);\draw[int](v4)--(v2);\legMassive{(v1)}{180}{$$}\legMassive{(v2)}{90}{$$}\legMassive{(v3)}{0}{$$}\legMassive{(v4)}{-90}{$$}
\node[ddot] at (ella) {};
\node[ddot] at (ellb) {};
\node[rdot] at (a1) {};\node[rdot] at (a2) {};\node[rdot] at (b1) {};\node[rdot] at (b2) {};
\node[anchor=north] at ($(a1)+(-135:0)+(-90:0.0)+(0:0)$) {\text{{\normalsize$\r{a_1}$}}};
\node[anchor=south] at ($(a2)+(135:0.)+(0:0.)$) {\text{{\normalsize$\r{a_2}$}}};
\node[anchor=south] at ($(b1)+(135:0.)+(0:0.)$) {\text{{\normalsize$\r{b_1}$}}};
\node[anchor=north] at ($(b2)+(-135:0)+(90:0.125)+(0:0)$) {\text{{\normalsize$\r{b_2}$}}};
}\hspace{-5pt}\bigger{\Rightarrow}\hspace{-5pt}\dT{
\coordinate (ella) at ($(v4)!.5!(v2)+(180:0.325)$);
\coordinate (ellb) at ($(v4)!.5!(v2)+(0:0.325)$);
\coordinate (a1) at ($(v4)!.5!(v1)+(-135:0.25)$);
\coordinate (a2) at ($(v1)!.5!(v2)+(135:0.25)$);
\coordinate (b1) at ($(v2)!.5!(v3)+(45:0.25)$);
\coordinate (b2) at ($(v3)!.5!(v4)+(-45:0.25)$);
\dimLines\draw[int](v4)--(v1);\draw[int](v1)--(v2);\draw[int](v2)--(v3);\draw[int](v3)--(v4);\draw[int](v4)--(v2);\legMassive{(v1)}{180}{$$}\legMassive{(v2)}{90}{$$}\legMassive{(v3)}{0}{$$}\legMassive{(v4)}{-90}{$$}
\restoreDark
\draw[int] (ella)--(a1);
\draw[int] (ella)--(a2);
\draw[int] (ella)--(ellb);
\draw[int] (ellb)--(b1);
\draw[int] (ellb)--(b2);
\node[ddot] at (ella) {};
\node[ddot] at (ellb) {};
\node[rdot] at (a1) {};\node[rdot] at (a2) {};\node[rdot] at (b1) {};\node[rdot] at (b2) {};
\node[anchor=north] at ($(a1)+(-135:0)+(-90:0.0)+(0:0)$) {\text{{\normalsize$\r{a_1}$}}};
\node[anchor=south] at ($(a2)+(135:0.)+(0:0.)$) {\text{{\normalsize$\r{a_2}$}}};
\node[anchor=south] at ($(b1)+(135:0.)+(0:0.)$) {\text{{\normalsize$\r{b_1}$}}};
\node[anchor=north] at ($(b2)+(-135:0)+(90:0.125)+(0:0)$) {\text{{\normalsize$\r{b_2}$}}};
}\hspace{-5pt}\bigger{\simeq}\hspace{6pt}\dBox{\dimLines
\draw[int](v5)--(v1);\draw[int](v1)--(v2);\draw[int](v2)--(v3);\draw[int](v3)--(v4);\draw[int](v4)--(v5);\draw[int](v6)--(v3);\legMassive{(v1)}{-135}{$$}\legMassive{(v2)}{135}{$$}\legMassive{(v3)}{90}{$$}\legMassive{(v4)}{45}{$$}\legMassive{(v5)}{-45}{$$}
\restoreDark
\coordinate (ella) at ($(v6)!.5!(v1)+(90:\figScale*0.65)$);
\coordinate (ellb) at ($(v6)!.5!(v5)+(90:\figScale*0.65)$);
\coordinate (a1) at ($(v6)!.0!(v1)+(-90:0.25)$);
\coordinate (a2) at ($(v1)!.5!(v2)+(180:0.25)$);
\coordinate (a3) at ($(v3)!.55!(v2)+(90:0.25)$);
\coordinate (b1) at ($(v3)!.55!(v4)+(90:0.25)$);
\coordinate (b2) at ($(v4)!.5!(v5)+(0:0.25)$);
\coordinate (b3) at ($(v6)!.55!(v5)+(-90:0.25)$);
\draw[int] (ella)--(a1);
\draw[int] (ella)--(a2);
\draw[int] (ella)--(a3);
\draw[int] (ella)--(ellb);
\draw[int] (ellb)--(b1);
\draw[int] (ellb)--(b2);
\draw[int] (ellb)--(a1);
\node[bldot] at (a1){};
\node[rdot] at (a2){};
\node[rdot] at (a3){};
\node[rdot] at (b1){};
\node[rdot] at (b2){};
\node[ddot] at (ella){};
\node[ddot] at (ellb){};
\node at ($(a2)+(180:0.275)+(90:0.0)$) {\text{{\normalsize$\r{a_1}$}}};
\node at ($(a3)+(90:0.25)$) {\text{{\normalsize$\r{a_2}$}}};
\node at ($(b1)+(0:0.05)+(90:0.25)$) {\text{{\normalsize$\r{b_1}$}}};
\node at ($(b2)+(0:0.275)+(90:0.05)$) {\text{{\normalsize$\r{b_2}$}}};
\node at ($(a1)+(-90:0.15)+(0:0.2)$) {\text{{\normalsize$\,\,\,\,\b{\infX}$}}};
}\fwboxL{0pt}{\,\,\,,}
}
being equivalent to the middle example of (\ref{polylog_double_boxes}). Indeed, these examples illustrate that the box-triangle and double-boxes are nothing but special cases of double-box integrals where one or more of the external dual-momentum coordinates involved has been replaced by the point at infinity $\infX$.

\subsection{Prescriptivity and Ellipticity: Contour Integrals and Periods}\label{subsec:prescriptivity_and_ellipticity}
The construction of integrand bases can be alternatively viewed as building bases of differential forms in which the amplitude integrand is expanded. In this language, the differential forms will be spanned by an underlying \emph{cohomology} group, which is heavily dependent on specific features of the amplitude, such as loop-level and multiplicity. In this language, prescriptive unitarity amounts to constructing a \emph{diagonal} homology basis dual to the basis of cohomology. This is very similar (but not identical) to the recent discussions of \emph{dual} Feynman integrals in recent research (see e.g.~\cite{Mastrolia:2018uzb,Cachazo:2018wvl,Frellesvig:2019kgj,Frellesvig:2019uqt,Frellesvig:2020qot,Casali:2019ihm,Caron-Huot:2021xqj,Caron-Huot:2021iev}).

In the absence of more complicated differential forms (such as elliptic ones), this problem is morally identical to the following rudimentary example in one dimension. Any meromorphic function $f(\g{z})$ on $\mathbb{P}^1$ with simple poles at locations $\{{z_{1}},\dots,z_{n},z_{\infty}\}$ (where `$z_{\infty}$' can represent any other location on $\mathbb{P}^1$ supporting a simple pole) may be expressed up to some \emph{holomorphic} function $\tilde{f}(\g{z})$ according to
\eq{f(\g{z})\equivL\tilde{f}(\g{z}){+}\sum_{i}\frac{a_{i}}{\g{z}{-}z_{i}}\quad\text{where}\quad a_i={-}{i}\!\!\!\!\!\!\oint\limits_{|\g{z}{-}z_i|=\epsilon}\!\!\!\!\!\dbar \g{z}\,\,\,f(\g{z})\,}
with the coefficients $a_i$ determined as simple period integrals---namely, \emph{residues}. In this elementary example, the cohomology group is $n$-dimensional and spanned by the differential forms
\eq{\ket{i}\equivR\frac{1}{\g{z}{-}z_{i}}\dbar \g{z}=\dbar\!\log(\g{z}{-}z_i)\,.}
(The holomorphic part $\tilde{f}(\g{z})$ of $f(\g{z})$ is cohomologically trivial, being a total derivative.) This space of differential forms is dual to the \emph{homology} group of the $(n{+}1)$-punctured Riemann sphere---the space on which $f(\g{z})$ is \emph{holomorphic}. This homology group is spanned by the one-(real-)dimensional algebraic varieties \mbox{$\Omega_j\equivR\big\{\g{z}\big|\,|z{-}z_j|\!=\!\epsilon\big\}$} $\equivL\bra{j}$ which encircle each of the punctures $z_i$ (including $z_{\infty}$). The connection between this cohomology (the differential forms) and homology (the contours) is defined via the \emph{period matrix} as seen in (\ref{general_period_matrix}) above:
\eq{\mathbf{M}_{i\,j}\equivR\braket{j|i}\equivR\oint\limits_{\Omega_j}\frac{\dbar \g{z}}{\g{z}{-}z_i}=i\,\delta_{i\,j}.\vspace{-3pt}}
(For some readers, this notation may be reminiscent of the dual Feynman diagram picture developed in \cite{Caron-Huot:2021xqj,Caron-Huot:2021iev}, where integrand coefficients are fixed by contracting with integrand duals furnished by the twisted cohomology. We emphasize that this is \emph{not} what we are doing here---although there are qualitative similarities which may be interesting to explore.)

We call a contour integral such as $\Omega_j$ above a `polylogarithmic' one, as it is dual to a differential form $\dbar\!\log(\g{z}{-}z_i)$. Feynman integrands which integrate to polylogarithmic integrals are spanned by wedge-products of such forms, and are dual to homologies which are constructed as intersections of such contours. Such integrands can always be expressed in a basis of `$\dbar\!\log$' differential forms---the coefficients of which can be matched by computing the period matrix and solving for coefficients as done for the matching of amplitudes in (\ref{general_solution_to_unitarity_problem_via_periods}).\\

The novelty of interest to us here is the case for Feynman integrands for which polylogarithmic differential forms (and contours) do not suffice. The simplest example of a non-polylogarithmic situation arises for the double-box integrand (\ref{scalar_double_box}) which takes the form 
\vspace{-5pt}\eq{\oint\limits_{\Omega[\r{\vec{a}};\r{\vec{b}}]}\left[\dBox{\draw[int](v5)--(v1);\draw[int](v1)--(v2);\draw[int](v2)--(v3);\draw[int](v3)--(v4);\draw[int](v4)--(v5);\draw[int](v6)--(v3);\legMassive{(v1)}{-135}{$$}\legMassive{(v2)}{135}{$$}\legMassive{(v3)}{90}{$$}\legMassive{(v4)}{45}{$$}\legMassive{(v5)}{-45}{$$}\legMassive{(v6)}{-90}{$$}
\coordinate (ella) at ($(v6)!.5!(v1)+(90:\figScale*0.65)$);
\coordinate (ellb) at ($(v6)!.5!(v5)+(90:\figScale*0.65)$);
\coordinate (a1) at ($(v6)!.55!(v1)+(-90:0.25)$);
\coordinate (a2) at ($(v1)!.5!(v2)+(180:0.25)$);
\coordinate (a3) at ($(v3)!.55!(v2)+(90:0.25)$);
\coordinate (b1) at ($(v3)!.55!(v4)+(90:0.25)$);
\coordinate (b2) at ($(v4)!.5!(v5)+(0:0.25)$);
\coordinate (b3) at ($(v6)!.55!(v5)+(-90:0.25)$);
\node[rdot] at (a1){};
\node[rdot] at (a2){};
\node[rdot] at (a3){};
\node[rdot] at (b1){};
\node[rdot] at (b2){};
\node[rdot] at (b3){};
\node at ($(ella)$) {\text{\normalsize{$\ell_1$}}};
\node at ($(ellb)$) {\text{\normalsize{$\ell_2$}}};
\node at ($(a1)+(-90:0.25)$) {\text{{\normalsize$\r{a_1}$}}};
\node at ($(a2)+(180:0.275)+(90:0.0)$) {\text{{\normalsize$\r{a_2}$}}};
\node at ($(a3)+(90:0.25)$) {\text{{\normalsize$\r{a_3}$}}};
\node at ($(b1)+(0:0.05)+(90:0.25)$) {\text{{\normalsize$\r{b_1}$}}};
\node at ($(b2)+(0:0.275)+(90:0.05)$) {\text{{\normalsize$\r{b_2}$}}};
\node at ($(b3)+(-90:0.25)$) {\text{{\normalsize$\r{b_3}$}}};}
\right]={-}i\,\frac{\dbar\g{\alpha}}{y(\g{\alpha})}\vspace{-5pt}}
where $\Omega[\r{\vec{a}};\r{\vec{b}}]$ denotes the odd combination of the two seven-dimensional contours which encircle each of the seven Feynman propagators, $\dbar\g{\alpha}$ represents whatever part of $\dbar^4\!\ell_1\dbar^4\!\ell_2$ remains after integration and $y^2(\g{\alpha})\equivL Q(\g{\alpha})$ is an irreducible quartic polynomial in $\g{\alpha}$. Importantly, $\dbar\g{\alpha}/y(\g{\alpha})$ is \emph{not} a $\dbar\!\log$ differential form; nevertheless, as always, meromorphic (or algebraic) differential forms are \emph{holomorphic} on the space of the complement of their divisors, and the space $\mathbb{P}^1\backslash\{y(\g{\alpha})\!=\!0\}$ describes the geometry of an elliptic curve, which we denote $\mathcal{E}$. 

Importantly, unlike a polylogarithmic contour encircling a simple pole, an elliptic curve (a Riemann surface of genus 1) admits a \emph{two-dimensional} space of contours of integration, a basis for which can be given by the non-contractable $a$- and $b$-cycles `around' the elliptic curve as viewed as a torus. (In terms of the quartic $Q(\g{\alpha})\!=\!y^2(\g{\alpha})$, these can be described as contours which enclose paths around branch-cuts connecting pairs of the roots of the quartic; but this level of detail is not necessary for us at present.) This homology has a dual cohomology which is \emph{also} \emph{two}-dimensional, and may be spanned by the differential forms 
\eq{\dbar\g{\alpha}\,\frac{1}{y(\g{\alpha})}\quad\text{and}\quad\dbar\g{\alpha}\,\frac{\g{\alpha}}{y(\g{\alpha})}\,.\label{basis_for_elliptic_cohomology}}
Of these, the latter represents a differential form with a simple pole at infinity, and would be cohomologous to $\dbar\g{\alpha}\,\,y(q)/(\g{\alpha}{-}q)y(\g{\alpha})$ for any point $q$ (not coincident with one of the roots of the quartic). Many of the Feynman integrands of interest to us here involve differential forms of the second type in (\ref{basis_for_elliptic_cohomology}), with additional simple poles arising from the existence of further, as-yet-uncut propagators in the original master integrand.

Suppose that we have the case of some Feynman integrand whose additional propagator becomes quadratic on some double-box seven-dimensional contour: 
\eq{\oint\limits_{\Omega[\r{\vec{a}};\r{\vec{b}}]}\!\!\!\mathcal{I}=\dbar\g{\alpha}\,\,\frac{n(\g{\alpha})}{(\g{\alpha}{-}q_+)(\g{\alpha}{-}q_+)y(\g{\alpha})}\,.}
The geometry on which we must discuss the homology of potential contours is now a doubly-punctured elliptic curve. This space would have a three-dimensional homology, with a dual cohomology spanned by the differential forms
\eq{\left\{\dbar\g{\alpha}\,\frac{1}{y(\g{\alpha})}\,,\dbar\g{\alpha}\,\frac{y(q_+)}{(\g{\alpha}{-}q_+)y(\g{\alpha})}\,,\dbar\g{\alpha}\,\frac{y(q_-)}{(\g{\alpha}{-}q_-)y(\g{\alpha})}\right\}\,.\label{eg_3d_basis_on_curve}}
The first is clearly elliptic, while the latter two would appear to have mixed rigidity---having support on both elliptic as well as polylogarithmic contours. We claim that one linear combination of these three differential forms is actually polylogarithmic, while two can be said to be elliptic (with definite rigidity). 

To see this, suppose that we were to construct the period matrix for these forms using the $a$-cycle, $b$-cycle, and either of the two polylogarithmic contours. Diagonalizing the forms against these contours would result in two forms which integrate to 1 on the $a$-cycle or $b$-cycle separately and zero on the polylogarithmic contour, while the dual of the polylogarithmic contour would integrate to zero on \emph{both} the $a$- and $b$-cycles of the elliptic curve. When the integral of a differential form vanishes on \emph{both} cycles of an elliptic curve, it must therefore be a total derivative \emph{on the curve} (cohomologically zero). Thus, we'd claim that it must be purely polylogarithmic---have definite rigidity zero. 

We'll see that the example discussed above is in fact exactly what arises in the case of Feynman integrands of interest.

\subsection{\emph{Definite} Rigidity for Master Integrands with Triangle Power-Counting}\label{subsec:diagonalizing_rigidity_box_triangle_basis}\vspace{-0pt}
In this section we describe our main result: that a master integrand basis exists for planar, two loop integrands involving massless propagators that can be fully stratified according to rigidity; in particular, this is achievable for the space of integrands with so-called triangle power-counting---and therefore also for any basis with worse ultraviolet behavior. 

The main technical step is the demonstration that the space of master integrands for box-triangles can be organized into masters of definite rigidity. Being the simplest case to consider, and one which recurs in all further examples, we have described this example in complete detail in \mbox{appendix \ref{appendix:box_triangle_masters}} and also in the ancillary files for this work.

\subsubsection{Diagonalizing Rigidity for Box-Triangle Integrands}\label{subsubsec:box_triangle}\vspace{-0pt}
The box-triangle is the arguably simplest planar integrand at two loops that supports elliptic leading singularities. With a loop-independent numerator, it is easy to see that the box-triangle integrates to an elliptic polylogarithm, being simply a case of the scalar double box, but involving a point at infinity (see eqn.~(\ref{box_triangle_ellipticity})):
\eq{\bT{
\coordinate (ella) at ($(v5)!.5!(v1)+(90:\figScale*0.65)$);
\coordinate (ellb) at ($(v5)!.5!(v3)+(0:\figScale*0.325)$);
\coordinate (a1) at ($(v5)!.5!(v1)+(-90:0.25)$);
\coordinate (a2) at ($(v1)!.5!(v2)+(180:0.25)$);
\coordinate (a3) at ($(v2)!.5!(v3)+(90:0.25)$);
\coordinate (b1) at ($(v3)!.5!(v4)+(58:0.25)$);
\coordinate (b2) at ($(v4)!.5!(v5)+(-58:0.25)$);
\dimLines\draw[int] (ella)--(a1);
\draw[int] (ella)--(a2);
\draw[int] (ella)--(a3);
\draw[int] (ella)--(ellb);
\draw[int] (ellb)--(b1);
\draw[int] (ellb)--(b2);
\restoreDark
\draw[int](v5)--(v1);\draw[int](v1)--(v2);\draw[int](v2)--(v3);\draw[int](v3)--(v4);\draw[int](v4)--(v5);\draw[int](v5)--(v3);
\legMassive{(v1)}{-135}{$$}\legMassive{(v2)}{135}{$$}\legMassive{(v3)}{71.5}{$$}\legMassive{(v4)}{0}{$$}\legMassive{(v5)}{-71.5}{$$}
\node[rdot] at (a1){};
\node[rdot] at (a2){};
\node[rdot] at (a3){};
\node[rdot] at (b1){};
\node[rdot] at (b2){};
\node[ddot] at (ella){};
\node[ddot] at (ellb){};\node at ($(a1)+(-90:0.195)$) {\text{{\normalsize$\r{a_1}$}}};
\node at ($(a2)+(180:0.25)+(90:0.05)$) {\text{{\normalsize$\r{a_2}$}}};
\node at ($(a3)+(90:0.195)$) {\text{{\normalsize$\r{a_3}$}}};
\node at ($(b1)+(36:0.25)+(90:0.05)$) {\text{{\normalsize$\r{b_1}$}}};
\node at ($(b2)+(-36:0.25)+(-90:0.05)$) {\text{{\normalsize$\r{b_2}$}}};
}\hspace{-7pt}\bigger{\Rightarrow}\bT{
\coordinate (ella) at ($(v5)!.5!(v1)+(90:\figScale*0.65)$);
\coordinate (ellb) at ($(v5)!.5!(v3)+(0:\figScale*0.325)$);
\coordinate (a1) at ($(v5)!.5!(v1)+(-90:0.25)$);
\coordinate (a2) at ($(v1)!.5!(v2)+(180:0.25)$);
\coordinate (a3) at ($(v2)!.5!(v3)+(90:0.25)$);
\coordinate (b1) at ($(v3)!.5!(v4)+(58:0.25)$);
\coordinate (b2) at ($(v4)!.5!(v5)+(-58:0.25)$);
\dimLines\draw[int](v5)--(v1);\draw[int](v1)--(v2);\draw[int](v2)--(v3);\draw[int](v3)--(v4);\draw[int](v4)--(v5);\draw[int](v5)--(v3);\legMassive{(v1)}{-135}{$$}\legMassive{(v2)}{135}{$$}\legMassive{(v3)}{71.5}{$$}\legMassive{(v4)}{0}{$$}\legMassive{(v5)}{-71.5}{$$}
\restoreDark\draw[int] (ella)--(a1);
\draw[int] (ella)--(a2);
\draw[int] (ella)--(a3);
\draw[int] (ella)--(ellb);
\draw[int] (ellb)--(b1);
\draw[int] (ellb)--(b2);
\node[rdot] at (a1){};
\node[rdot] at (a2){};
\node[rdot] at (a3){};
\node[rdot] at (b1){};
\node[rdot] at (b2){};
\node[ddot] at (ella){};
\node[ddot] at (ellb){};\node at ($(a1)+(-90:0.195)$) {\text{{\normalsize$\r{a_1}$}}};
\node at ($(a2)+(180:0.25)+(90:0.05)$) {\text{{\normalsize$\r{a_2}$}}};
\node at ($(a3)+(90:0.195)$) {\text{{\normalsize$\r{a_3}$}}};
\node at ($(b1)+(36:0.25)+(90:0.05)$) {\text{{\normalsize$\r{b_1}$}}};
\node at ($(b2)+(-36:0.25)+(-90:0.05)$) {\text{{\normalsize$\r{b_2}$}}};
}\hspace{-7pt}\bigger{\simeq}\hspace{5pt}\dBox{\dimLines
\draw[int](v5)--(v1);\draw[int](v1)--(v2);\draw[int](v2)--(v3);\draw[int](v3)--(v4);\draw[int](v4)--(v5);\draw[int](v6)--(v3);\legMassive{(v1)}{-135}{$$}\legMassive{(v2)}{135}{$$}\legMassive{(v3)}{90}{$$}\legMassive{(v4)}{45}{$$}\legMassive{(v5)}{-45}{$$}\legMassive{(v6)}{-90}{$$}
\restoreDark
\coordinate (ella) at ($(v6)!.5!(v1)+(90:\figScale*0.65)$);
\coordinate (ellb) at ($(v6)!.5!(v5)+(90:\figScale*0.65)$);
\coordinate (a1) at ($(v6)!.55!(v1)+(-90:0.25)$);
\coordinate (a2) at ($(v1)!.5!(v2)+(180:0.25)$);
\coordinate (a3) at ($(v3)!.55!(v2)+(90:0.25)$);
\coordinate (b1) at ($(v3)!.55!(v4)+(90:0.25)$);
\coordinate (b2) at ($(v4)!.5!(v5)+(0:0.25)$);
\coordinate (b3) at ($(v6)!.55!(v5)+(-90:0.25)$);
\draw[int] (ella)--(a1);
\draw[int] (ella)--(a2);
\draw[int] (ella)--(a3);
\draw[int] (ella)--(ellb);
\draw[int] (ellb)--(b1);
\draw[int] (ellb)--(b2);
\draw[int] (ellb)--(b3);
\node[rdot] at (a1){};
\node[rdot] at (a2){};
\node[rdot] at (a3){};
\node[rdot] at (b1){};
\node[rdot] at (b2){};
\node[bldot] at (b3){};
\node[ddot] at (ella){};
\node[ddot] at (ellb){};
\node at ($(a1)+(-90:0.225)$) {\text{{\normalsize$\r{a_1}$}}};
\node at ($(a2)+(180:0.275)+(90:0.0)$) {\text{{\normalsize$\r{a_2}$}}};
\node at ($(a3)+(90:0.25)$) {\text{{\normalsize$\r{a_3}$}}};
\node at ($(b1)+(0:0.05)+(90:0.25)$) {\text{{\normalsize$\r{b_1}$}}};
\node at ($(b2)+(0:0.275)+(90:0.05)$) {\text{{\normalsize$\r{b_2}$}}};
\node at ($(b3)+(-90:0.15)+(0:0.2)$) {\text{{\normalsize$\,\,\,\,\b{\infX}$}}};
}\fwboxL{0pt}{\,\,\,.}}
Providing this topology with triangle power-counting numerators, however, requires that we include an additional propagator $\x{\ell_1}{\infX}$ (supporting poles at infinite loop momentum) on the box ($\ell_1$) side as well, rendering the integrands effectively pentaboxes
\eq{\bT{
\coordinate (ella) at ($(v5)!.5!(v1)+(90:\figScale*0.65)$);
\coordinate (ellb) at ($(v5)!.5!(v3)+(0:\figScale*0.325)$);
\coordinate (a1) at ($(v5)!.5!(v1)+(-90:0.25)$);
\coordinate (a2) at ($(v1)!.5!(v2)+(180:0.25)$);
\coordinate (a3) at ($(v2)!.5!(v3)+(90:0.25)$);
\coordinate (b1) at ($(v3)!.5!(v4)+(58:0.25)$);
\coordinate (b2) at ($(v4)!.5!(v5)+(-58:0.25)$);
\dimLines
\restoreDark
\draw[int](v5)--(v1);\draw[int](v1)--(v2);\draw[int](v2)--(v3);\draw[int](v3)--(v4);\draw[int](v4)--(v5);\draw[int](v5)--(v3);
\legMassive{(v1)}{-135}{$$}\legMassive{(v2)}{135}{$$}\legMassive{(v3)}{71.5}{$$}\legMassive{(v4)}{0}{$$}\legMassive{(v5)}{-71.5}{$$}\draw[markedEdge] (v1)--(v2);
\node[rdot] at (a1){};
\node[rdot] at (a2){};
\node[rdot] at (a3){};
\node[rdot] at (b1){};
\node[rdot] at (b2){};
\node[] at (ella){$\ell_1$};
\node[] at (ellb){$\ell_2$};
\node at ($(a1)+(-90:0.195)$) {\text{{\normalsize$\r{a_1}$}}};
\node at ($(a2)+(180:0.25)+(90:0.05)$) {\text{{\normalsize$\r{a_2}$}}};
\node at ($(a3)+(90:0.195)$) {\text{{\normalsize$\r{a_3}$}}};
\node at ($(b1)+(36:0.25)+(90:0.05)$) {\text{{\normalsize$\r{b_1}$}}};
\node at ($(b2)+(-36:0.25)+(-90:0.05)$) {\text{{\normalsize$\r{b_2}$}}};
}\bigger{\Rightarrow}\pBox{\pBoxPlainEdges
\legMassive{(v1)}{-110}{$$}\legMassive{(v2)}{180}{$$}\legMassive{(v3)}{110}{$$}\legMassive{(v4)}{81}{$$}\legMassive{(v5)}{45}{$$}\legMassive{(v6)}{-45}{$$}\coordinate (ella) at ($(v7)!.5!(v4)+(180:0.575)$);\coordinate (ellb) at ($(v7)!.5!(v4)+(0:\figScale*0.45)$);\coordinate (a1) at ($(v7)!.0!(v1)+(-72:0.25)$);\coordinate (a2) at ($(v1)!.5!(v2)+(-144:0.25)$);\coordinate (a3) at ($(v2)!.5!(v3)+(144:0.25)$);\coordinate (a4) at ($(v3)!.5!(v4)+(72:0.25)$);\coordinate (b1) at ($(v4)!.5!(v5)+(90:0.205)$);\coordinate (b2) at ($(v5)!.5!(v6)+(0:0.25)$);
\node[bldot] at (a1){};\node[rdot] at (a2){};\node[rdot] at (a3){};\node[rdot] at (a4){};\node[rdot] at (b1){};\node[rdot] at (b2){};
\node at ($(ella)$) {\text{{\normalsize${\ell_1}$}}};\node at ($(ellb)$) {\text{{\normalsize${\ell_2}$}}};
\node at ($(a1)+(-72:0.195)+(-90:0.05)+(0:0.0)$) {\text{{\normalsize$\infX$}}};
\node at ($(a2)+(-144:0.195)+(-90:0.1)+(0:0.2)$) {\text{{\normalsize$\r{a_1}$}}};
\node at ($(a3)+(144:0.195)+(90:0.05)+(0:0.2)$) {\text{{\normalsize$\r{a_2}$}}};
\node at ($(a4)+(72:0.195)$) {\text{{\normalsize$\r{a_3}$}}};
\node at ($(b1)+(80:0.195)+(0:0.1)+(80:0.05)$) {\text{{\normalsize$\r{b_1}$}}};
\node at ($(b2)+(0:0.25)$) {\text{{\normalsize$\r{b_2}$}}};
\draw[markedEdge] (v2)--(v3);
}\bigger{\simeq}\pBox{
\coordinate (ella) at ($(v7)!.5!(v4)+(180:0.575)$);\coordinate (ellb) at ($(v7)!.5!(v4)+(0:\figScale*0.45)$);\coordinate (a1) at ($(v7)!.0!(v1)+(-72:0.25)$);\coordinate (a2) at ($(v1)!.5!(v2)+(-144:0.25)$);\coordinate (a3) at ($(v2)!.5!(v3)+(144:0.25)$);\coordinate (a4) at ($(v3)!.5!(v4)+(72:0.25)$);\coordinate (b1) at ($(v4)!.5!(v5)+(90:0.205)$);\coordinate (b2) at ($(v5)!.5!(v6)+(0:0.25)$);
\dimLines\pBoxPlainEdges
\legMassive{(v1)}{-110}{$$}\legMassive{(v2)}{180}{$$}\legMassive{(v3)}{110}{$$}\legMassive{(v4)}{81}{$$}\legMassive{(v5)}{45}{$$}\legMassive{(v6)}{-45}{$$}\restoreDark
\draw[int](ella)--(a1);\draw[int](ella)--(a2);\draw[int](ella)--(a3);\draw[int](ella)--(a4);\draw[int](ella)--(ellb);\draw[int](ellb)--(b1);\draw[int](ellb)--(a1);\draw[int](ellb)--(b2);
\node[ddot]at (ella){};\node[ddot]at (ellb){};\node[bldot] at (a1){};\node[rdot] at (a2){};\node[rdot] at (a3){};\node[rdot] at (a4){};\node[rdot] at (b1){};\node[rdot] at (b2){};
\node at ($(a1)+(-72:0.195)+(-90:0.05)+(0:0.0)$) {\text{{\normalsize$\infX$}}};
\node at ($(a2)+(-144:0.195)+(-90:0.1)+(0:0.2)$) {\text{{\normalsize$\r{a_1}$}}};
\node at ($(a3)+(144:0.195)+(90:0.05)+(0:0.2)$) {\text{{\normalsize$\r{a_2}$}}};
\node at ($(a4)+(72:0.195)$) {\text{{\normalsize$\r{a_3}$}}};
\node at ($(b1)+(80:0.195)+(0:0.1)+(80:0.05)$) {\text{{\normalsize$\r{b_1}$}}};
\node at ($(b2)+(0:0.25)$) {\text{{\normalsize$\r{b_2}$}}};
}\fwboxL{0pt}{\,\,\,\,,}}
corresponding to the six-dimensional space of master integrands of the form
\eq{\mathcal{I}_i^0\equivR\dbar^4\!\ell_1\dbar^4\!\ell_2\frac{\x{\ell_1}{\b{N_i^0}}}{\x{\ell_1}{\infX}\x{\ell_1}{\r{a_1}}\x{\ell_1}{\r{a_2}}\x{\ell_1}{\r{a_3}}\x{\ell_1}{\ell_2}\x{\ell_2}{\r{b_1}}\x{\ell_2}{\r{b_2}}\x{\ell_2}{\b{X}}\,,
}}
with $\x{\ell_1}{\b{N_i^0}}\!\in\![\ell_1]$ chosen from some initial basis. For an arbitrary element in this space of masters with some numerator $\x{\ell_1}{\b{N}}$, it is easy to see that the integral will support both elliptic and polylogarithmic contours---therefore representing an integrand with indefinite rigidity. The polylogarithmic contours are identifiable as maximal cuts of the pentabox and the leading singularities of the double-triangle contact terms, while the (mixed, polylogarithmic and) elliptic contours correspond to the $a$- and $b$-cycles of the elliptic heptacut $\Omega[\r{\vec{a}};(\r{b_1},\r{b_2},\infX)]$.

By choosing an initial basis of for the master integrands' numerators of the form
\eq{\begin{split}\mathfrak{n}(\ell_1)\!\in\![\ell_1]&=\mathrm{span}\Big\{\!\underbrace{\x{\ell_1}{\b{X}},\x{\ell_1}{\b{N_2^0}},\x{\ell_1}{\b{N_3^0}}}_{\text{`top-level'}},\underbrace{\x{\ell_1}{\r{a_1}},\x{\ell_1}{\r{a_2}},\x{\ell_1}{\r{a_3}}}_{\text{`contact-terms'}}\!\Big\}\\
&\equivL\,\mathrm{span}\big\{\x{\ell_1}{\b{N_i^0}}\big\}\bigger{\oplus}\,\,\mathrm{span}\big\{\x{\ell_1}{\r{N_i^0}}\big\}\,,\end{split}}
it is easy to see that the contact terms will represent integrands of definite, polylogarithmic rigidity, and the top-level integrand $\mathcal{I}_{\b{1}}^0$ with numerator $\x{\ell_1}{\b{X}}$---corresponding to the scalar box-triangle integrand---will have definite rigidity 1 corresponding to an elliptic polylogarithm. However, the remaining two, top-level master integrands will still have indefinite rigidity, as these integrands take the form 
\eq{\oint\limits_{\fwbox{0pt}{\Omega[\r{\vec{a}};(\r{b_1},\r{b_2},\b{X})]}} \mathcal{I}_{\b{i}}^0 =  {-}i\,\dbar\g{\alpha}\,\,\frac{\x{\ell_1(\g{\alpha})}{\b{N_i^0}}}{\x{\ell_1(\g{\alpha})}{\infX}y(\g{\alpha})}}
on the (odd combination of the) elliptic-containing heptacut $\Omega[\r{\vec{a}};(\r{b_1},\r{b_2},\b{X})]$. Because of the as-yet-uncut propagator $\x{\ell_1(\g{\alpha})}{\infX}$ supports additional poles, it is easy to see that these three differential forms can be expanded exactly into the basis described in (\ref{eg_3d_basis_on_curve})

Consider the basis for homology consisting of the contours $\{\Omega_a,\Omega_b,\Omega_{\text{poly}}\}$ where $\Omega_{\text{poly}}$ corresponds to some choice of polylogarithmic contour encircling the final pole $\x{\ell_1}{\infX}\!=\!0$; for this set of integration contours, we may compute the period matrix for the basis according to
\eq{
\mathbf{M}_{\b{i}\,a}\equivR\oint\limits_{\substack{\fwbox{0pt}{\Omega[\r{\vec{a}};(\r{b_1},\r{b_2},\b{X})]}\\\text{a-cycle}}}\!\!\!\mathcal{I}_{\b{i}}^0\,,\qquad
\mathbf{M}_{\b{i}\,b}\equivR\oint\limits_{\substack{\fwbox{0pt}{\Omega[\r{\vec{a}};(\r{b_1},\r{b_2},\b{X})]}\\\text{b-cycle}}}\!\!\!\mathcal{I}_{\b{i}}^0\,,\qquad
\mathbf{M}_{\b{i}\,\text{poly}}\equivR\oint\limits_{\substack{\fwbox{0pt}{\Omega[\r{\vec{a}};(\r{b_1},\r{b_2},\b{X})]}\\\x{\ell_1}{\b{X}}=0}}\!\!\!\mathcal{I}_{\b{i}}^0\,.
}
We have constructed this period matrix explicitly, and checked that it is of full rank. For details, please consult \mbox{appendix \ref{appendix:box_triangle_numerators_detail}}. By diagonalizing this period matrix according to
\eq{\big|\b{N_i^0}\big)\mapsto\big|\b{N_i}\big)\equivR\mathbf{M}^{-1}_{\b{i}\,\b{j}}\big|\b{N_i^0}\big)\,,}
we find prescriptive master integrands. Notice that this new, prescriptive integrand basis has the property that $\mathcal{I}_{\b{1,2}}$ have support exclusively on the $a$- or $b$-cycles of the elliptic curve, respectively, and vanish on the polylogarithmic contour of integration---leading us to conclude that these integrands have \emph{definite} rigidity of 1. On the other hand, $\mathcal{I}_{\b{3}}$ has been constructed to vanish on \emph{both} the $a$- and $b$-cycle of the elliptic curve but integrates to 1 on the choice of polylogarithmic contour; as such $\mathcal{I}_{\b{3}}$ must be \emph{cohomologically trivial} on the elliptic curve---a total derivative---and we may conclude that $\mathcal{I}_{\b{3}}$ has \emph{definite} rigidity 0.

One interesting aspect of this construction is that although it is still true that the diagonalized master integrands numerators are elements of $[\ell_1]$, they require coefficients that are not merely algebraic functions of the external kinematics---but rather the ratios of complete elliptic integrals (and polylogarithmic periods) as appearing in the inverse of the period matrix $\mathbf{M}^{-1}$.\\

Although we have shown that there exists a master integrand basis for the box-triangle involving two elliptic polylogarithms of definite rigidity and four purely polylogarithmic integrals (one top-level, and three contact-terms), we have not yet been able to integrate $\mathcal{I}_{\b{3}}$ or determine its symbol alphabet. Being an intensely interesting open problem, we have provided complete details on the initial basis of master integrands and the diagonalized, prescriptive ones as linear combinations involving $\mathbf{M}^{-1}$ both in \mbox{appendix \ref{appendix:box_triangle_masters}} and in the ancillary files for this work.

\subsubsection{Diagonalizing Rigidity for Double-Box Integrands}\label{subsubsec:double_boxes}\vspace{-0pt}
In the space of master integrands with triangle power-counting, we also must consider the case of double-box integrands. By endowing these integrand topologies with triangle power-counting numerators $\mathfrak{n}(\ell_1,\ell_2)\!\in\![\ell_1][\ell_2]$, these master integrands take the form of double-pentagons in embedding space according to
\eq{\dBox{
\coordinate (ella) at ($(v6)!.5!(v1)+(90:\figScale*0.65)$);\coordinate (ellb) at ($(v6)!.5!(v5)+(90:\figScale*0.65)$);\coordinate (a1) at ($(v6)!.55!(v1)+(-90:0.25)$);\coordinate (a2) at ($(v1)!.5!(v2)+(180:0.25)$);\coordinate (a3) at ($(v3)!.55!(v2)+(90:0.25)$);\coordinate (b1) at ($(v3)!.55!(v4)+(90:0.25)$);\coordinate (b2) at ($(v4)!.5!(v5)+(0:0.25)$);\coordinate (b3) at ($(v6)!.55!(v5)+(-90:0.25)$);\draw[int](v5)--(v1);\draw[int](v1)--(v2);\draw[int](v2)--(v3);\draw[int](v3)--(v4);\draw[int](v4)--(v5);\draw[int](v6)--(v3);\legMassive{(v1)}{-135}{$$}\legMassive{(v2)}{135}{$$}\legMassive{(v3)}{90}{$$}\legMassive{(v4)}{45}{$$}\legMassive{(v5)}{-45}{$$}\legMassive{(v6)}{-90}{$$}\draw[markedEdge](v1)--(v2);\draw[markedEdge](v4)--(v5);
\node[rdot] at (a1){};
\node[rdot] at (a2){};
\node[rdot] at (a3){};
\node[rdot] at (b1){};
\node[rdot] at (b2){};
\node[rdot] at (b3){};
\node at ($(ella)$) {\text{\normalsize{$\ell_1$}}};
\node at ($(ellb)$) {\text{\normalsize{$\ell_2$}}};
\node at ($(a1)+(-90:0.25)$) {\text{{\normalsize$\r{a_1}$}}};
\node at ($(a2)+(180:0.275)+(90:0.0)$) {\text{{\normalsize$\r{a_2}$}}};
\node at ($(a3)+(90:0.25)$) {\text{{\normalsize$\r{a_3}$}}};
\node at ($(b1)+(0:0.05)+(90:0.25)$) {\text{{\normalsize$\r{b_1}$}}};
\node at ($(b2)+(0:0.275)+(90:0.05)$) {\text{{\normalsize$\r{b_2}$}}};
\node at ($(b3)+(-90:0.25)$) {\text{{\normalsize$\r{b_3}$}}};}\bigger{\Rightarrow}\;\;\;
\pBox{
\def\ephScale{1.2*\figScale}\coordinate(v8)at($(0,-\ephScale*0.5)$);\coordinate(v1)at($(v8)+(198:\ephScale)$);\coordinate(v2)at($(v1)+(198-72:\ephScale)$);\coordinate(v3)at($(v2)+(198-144:\ephScale)$);\coordinate(v4)at($(v3)+(-18:\ephScale)$);\coordinate(v5)at($(v4)+(18:\ephScale)$);\coordinate(v6)at($(v5)+(18-72:\ephScale)$);\coordinate(v7)at($(v6)+(18-144:\ephScale)$);\coordinate(ella)at($(v8)!.5!(v4)+(180:\ephScale*0.7)$);\coordinate(ellb)at($(v8)!.5!(v4)+(0:\ephScale*0.7)$);\coordinate(x)at($(v8)+(-90:0.405)$);\coordinate(a1)at($(v1)!.5!(v2)+(-144:0.275)$);\coordinate(a2)at($(v2)!.5!(v3)+(-144-72:0.275)$);\coordinate(a3)at($(v3)!.5!(v4)+(72:0.275)$);\coordinate(b1)at($(v4)!.5!(v5)+(108:0.275)$);\coordinate(b2)at($(v5)!.5!(v6)+(108-72:0.275)$);\coordinate(b3)at($(v6)!.5!(v7)+(108-144:0.275)$);
\draw[int](v8)--(v1);\draw[int](v1)--(v2);\draw[int](v2)--(v3);\draw[int](v3)--(v4);\draw[int](v4)--(v8);\draw[int](v4)--(v5);\draw[int](v5)--(v6);\draw[int](v6)--(v7);\draw[int](v7)--(v8);\draw[markedEdge](v2)--(v3);\draw[markedEdge](v5)--(v6);\legMassive{(v1)}{180+72}{};\legMassive{(v2)}{180}{};\legMassive{(v3)}{180-72}{};\legMassive{(v4)}{90}{};\legMassive{(v5)}{72}{};\legMassive{(v6)}{0}{};\legMassive{(v7)}{-72}{};
\node[bldot] at (x){};\node[rdot] at (a1){};\node[rdot] at (a2){};\node[rdot] at (a3){};\node[rdot] at (b1){};\node[rdot] at (b2){};\node[rdot] at (b3){};
\node[anchor=north east,inner sep=0pt] at ($(a1)+(-144:0)$) {\text{{\normalsize$\r{a_1}$}}};\node[anchor=south east,inner sep=1pt] at ($(a2)+(-144+72:0)$) {\text{{\normalsize$\r{a_2}$}}};\node[anchor=south,inner sep=2pt] at ($(a3)+(72:0)$) {\text{{\normalsize$\r{a_3}$}}};\node[anchor=south,inner sep=2pt] at ($(b1)+(0.2,-0.05)$) {\text{{\normalsize$\r{b_1}$}}};\node[anchor=south west,inner sep=1pt] at ($(b2)+(-0,-0.05)$) {\text{{\normalsize$\r{b_2}$}}};\node[anchor=north west,inner sep=0pt] at ($(b3)+(0.1,0.1)$) {\text{{\normalsize$\r{b_3}$}}};\node[anchor=north,inner sep=2pt]at(x) {\text{{\normalsize$\b{X}$}}};\node[] at (ella) {\text{{\normalsize$\ell_1$}}};\node[] at (ellb) {\text{{\normalsize$\ell_2$}}};
}\;\;\;\bigger{\simeq}\;\;\;\pBox{
\def\ephScale{1.2*\figScale}\coordinate(v8)at($(0,-\ephScale*0.5)$);\coordinate(v1)at($(v8)+(198:\ephScale)$);\coordinate(v2)at($(v1)+(198-72:\ephScale)$);\coordinate(v3)at($(v2)+(198-144:\ephScale)$);\coordinate(v4)at($(v3)+(-18:\ephScale)$);\coordinate(v5)at($(v4)+(18:\ephScale)$);\coordinate(v6)at($(v5)+(18-72:\ephScale)$);\coordinate(v7)at($(v6)+(18-144:\ephScale)$);\coordinate(ella)at($(v8)!.5!(v4)+(180:\ephScale*0.7)$);\coordinate(ellb)at($(v8)!.5!(v4)+(0:\ephScale*0.7)$);\coordinate(x)at($(v8)+(-90:0.405)$);\coordinate(a1)at($(v1)!.5!(v2)+(-144:0.275)$);\coordinate(a2)at($(v2)!.5!(v3)+(-144-72:0.275)$);\coordinate(a3)at($(v3)!.5!(v4)+(72:0.275)$);\coordinate(b1)at($(v4)!.5!(v5)+(108:0.275)$);\coordinate(b2)at($(v5)!.5!(v6)+(108-72:0.275)$);\coordinate(b3)at($(v6)!.5!(v7)+(108-144:0.275)$);
\dimLines\draw[int](v8)--(v1);\draw[int](v1)--(v2);\draw[int](v2)--(v3);\draw[int](v3)--(v4);\draw[int](v4)--(v8);\draw[int](v4)--(v5);\draw[int](v5)--(v6);\draw[int](v6)--(v7);\draw[int](v7)--(v8);\legMassive{(v1)}{180+72}{};\legMassive{(v2)}{180}{};\legMassive{(v3)}{180-72}{};\legMassive{(v4)}{90}{};\legMassive{(v5)}{72}{};\legMassive{(v6)}{0}{};\legMassive{(v7)}{-72}{};
\restoreDark\draw[int](ella)--(x);\draw[int](ella)--(a1);\draw[int](ella)--(a2);\draw[int](ella)--(a3);\draw[int](ella)--(ellb);\draw[int](ellb)--(b1);\draw[int](ellb)--(b2);\draw[int](ellb)--(b3);\draw[int](ellb)--(x);
\node[bldot] at (x){};\node[rdot] at (a1){};\node[rdot] at (a2){};\node[rdot] at (a3){};\node[rdot] at (b1){};\node[rdot] at (b2){};\node[rdot] at (b3){};\node[ddot] at (ella){};\node[ddot] at (ellb){};
\node[anchor=north east,inner sep=0pt] at ($(a1)+(-144:0)$) {\text{{\normalsize$\r{a_1}$}}};\node[anchor=south east,inner sep=1pt] at ($(a2)+(-144+72:0)$) {\text{{\normalsize$\r{a_2}$}}};\node[anchor=south,inner sep=2pt] at ($(a3)+(72:0)$) {\text{{\normalsize$\r{a_3}$}}};\node[anchor=south,inner sep=2pt] at ($(b1)+(0.2,-0.05)$) {\text{{\normalsize$\r{b_1}$}}};\node[anchor=south west,inner sep=1pt] at ($(b2)+(-0,-0.05)$) {\text{{\normalsize$\r{b_2}$}}};\node[anchor=north west,inner sep=0pt] at ($(b3)+(0.1,0.1)$) {\text{{\normalsize$\r{b_3}$}}};\node[anchor=north,inner sep=2pt]at(x) {\text{{\normalsize$\b{X}$}}};}\fwboxL{0pt}{\,\,\,.}
}
Specifically, we consider the space of integrands of the form
\eq{\mathcal{I}_{\b{i\,j}}\equivR\dbar^4\!\ell_1\dbar^4\!\ell_2\frac{\x{\ell_1}{\b{N_i}}\x{\ell_2}{\b{N_j}}}{\x{\ell_1}{\infX}\x{\ell_1}{\r{a_1}}\x{\ell_1}{\r{a_2}}\x{\ell_1}{\r{a_3}}\x{\ell_1}{\ell_2}\x{\ell_2}{\r{b_1}}\x{\ell_2}{\r{b_2}}\x{\ell_2}{\r{b_3}}\x{\ell_2}{\b{X}}}\,.\label{double_box_with_triangle_power_counting}}
Notice that the `scalar' double-box would correspond to the integrand with particular numerator $\x{\ell_1}{\infX}\x{\ell_2}{\infX}$---which could be chosen as a master integrand basis element, or expressed as a linear combination of some other set of master integrands.

Interestingly, while the scalar double-box integrand itself would have definite rigidity 1, a \emph{generic} loop-dependent numerator $\x{\ell_1}{\b{N_i}}\x{\ell_2}{\b{N_j}}$ would actually lead to an integral which depended upon \emph{seven different} elliptic curves(!): the one involving the external propagators $\Omega[\r{\vec{a}};\r{\vec{b}}]$, but also those associated with the six box-triangle contact terms: $\Omega[\r{\vec{a}};(\r{b_i},\r{b_j},\b{X})]$ and $\Omega[(\r{a_i},\r{a_j},\b{X});\r{\vec{b}}]$---elliptic curves involving the point at infinity $\infX$. 

Of course, we know that the {\color{totalCount}36}-dimensional space of master integrands can be mostly spanned by contact terms---specifically, breaking down according to
\eq{\begin{split}\underset{d=4}{\mathrm{rank}}\big([\ell_1][\ell_2]\big)={\color{totalCount}36}=&\hspace{16pt}{\color{topCount}8}\big(\text{top-level integrands}\big){+}{\color{topCount}3}\!\times\!\!\big(6\text{ box-triangles}\big)\\&{+}{\color{topCount}1}\!\times\!\!\big(9\text{ double-triangles}\big){+}{\color{topCount}1}\!\times\!\!\big(1\text{ kissing-triangle}\big)\,.\end{split}}
In representing the space of master integrands in this way and using our work above in \mbox{section \ref{subsubsec:box_triangle}}, we see that each box-triangle contact-term subspace may be represented by a single polylogarithmic integrand and two integrands of definite elliptic rigidity; moreover, prescriptivity of the total space will ensure that each of the {\color{topCount}8} top-level master integrands will vanish on the $a$- and $b$-cycles of each elliptic curve associated with the contact terms. Therefore, the 28-dimensional space of contact-term master integrands will include $6\!\times\!2{=}12$ integrals with definite, elliptic rigidity (associated with each box-triangle contact-term) and $6\!\times\!1{+}9{+}{=}16$ pure and strictly polylogarithmic master integrands. The {\color{topCount}8} top-level master integrands can be ensured to vanish on all the contours used for the prescriptivity of the contact-terms. 

Viewing the double-box with triangle power-counting as a double-pentagon in embedding space suggests that the {\color{topCount}8} top-level numerators may be chosen from
\eq{\Big\{\x{\ell_1}{\infX}\x{\ell_2}{\infX},\x{\ell_1}{\infX}\x{\ell_2}{\b{Q_{\r{b_1},\r{b_2},\r{b_3},\b{X}}^i}},\x{\ell_1}{\b{Q_{\r{a_1},\r{a_2},\r{a_3},\b{X}}^i}}\x{\ell_2}{\infX},\x{\ell_1}{\b{Q_{\r{a_1},\r{a_2},\r{a_3},\b{X}}^i}}\x{\ell_2}{\b{Q_{\r{b_1},\r{b_2},\r{b_3},\b{X}}^j}}\Big\}\,,\nonumber}
where the last set of set of $2\!\times\!2$ numerators spans a space that includes the numerator $\x{\ell_1}{\ell_2}$---
\eq{\x{\ell_1}{\ell_2}\propto\sum_{\b{i},\b{j}=1}^2\x{\ell_1}{\b{Q_{\r{a_1},\r{a_2},\r{a_3},\b{X}}^i}}\x{\ell_2}{\b{Q_{\r{b_1},\r{b_2},\r{b_3},\b{X}}^j}}}
---which represents the kissing-triangle contact-term, and so should be excluded from this space of top-level numerators.

Although the master integrand with numerator $\x{\ell_1}{\b{X}}\x{\ell_2}{\b{X}}$ corresponds to a scalar double-box which has definite elliptic rigidity, all seven of the other top-level master integrands would generically have support on both elliptic and polylogarithmic contours. 

The solution is obvious: we may construct a basis of masters to be prescriptive with respect to the $a$- and $b$-cycles of the elliptic curve, and any choice of 6 polylogarithmic contours; by diagonalizing the basis with respect to these contours we will ensure that exactly two master integrands are elliptic, with the six integrands dual to the polylogarithmic cycles will be purely polylogarithmic. We have checked that such a choice of contours exists that gives rise to a full-rank period matrix.\\

Thus, the {\color{totalCount}36}-dimensional space of master integrands associated with the double-box given triangle power-counting can be organized into 14 pure elliptic polylogarithmic integrands (2 for each elliptic curve) and 22 pure polylogarithmic integrands. 

\newpage
\subsubsection{Diagonalizing Rigidity for Pentabox Integrands}\label{subsubsec:pentaboxes}\vspace{-0pt}
The last case of two-loop master integrands which seem to give rise to integrands with indefinite rigidity are the pentaboxes, which when viewed in embedding space, 
\eq{\pBox{\pBoxPlainEdges
\legMassive{(v1)}{-110}{$$}\legMassive{(v2)}{180}{$$}\legMassive{(v3)}{110}{$$}\legMassive{(v4)}{80}{$$}\legMassive{(v5)}{45}{$$}\legMassive{(v6)}{-45}{$$}\legMassive{(v7)}{-80}{$$}\draw[markedEdge] (v2)--(v3);\draw[markedEdge] (v1)--(v2);\draw[markedEdge] (v5)--(v6);\coordinate (ella) at ($(v7)!.5!(v4)+(180:0.575)$);
\coordinate (ellb) at ($(v7)!.5!(v4)+(0:\figScale*0.45)$);
\coordinate (a1) at ($(v7)!.5!(v1)+(-72:0.25)$);
\coordinate (a2) at ($(v1)!.5!(v2)+(-144:0.25)$);
\coordinate (a3) at ($(v2)!.5!(v3)+(144:0.25)$);
\coordinate (a4) at ($(v3)!.5!(v4)+(72:0.25)$);
\coordinate (b1) at ($(v4)!.5!(v5)+(90:0.205)$);
\coordinate (b2) at ($(v5)!.5!(v6)+(0:0.25)$);
\coordinate (b3) at ($(v6)!.5!(v7)-(90:0.205)$);
\node[rdot] at (a1){};
\node[rdot] at (a2){};
\node[rdot] at (a3){};
\node[rdot] at (a4){};
\node[rdot] at (b1){};
\node[rdot] at (b2){};
\node[rdot] at (b3){};\node at ($(ella)$) {\text{{\normalsize${\ell_1}$}}};
\node at ($(ellb)$) {\text{{\normalsize${\ell_2}$}}};
\node at ($(a1)+(-72:0.195)+(-90:0.05)+(0:0.0)$) {\text{{\normalsize$\r{a_1}$}}};
\node at ($(a2)+(-144:0.195)+(-90:0.1)+(0:0.2)$) {\text{{\normalsize$\r{a_2}$}}};
\node at ($(a3)+(144:0.195)+(90:0.05)+(0:0.2)$) {\text{{\normalsize$\r{a_3}$}}};
\node at ($(a4)+(72:0.195)$) {\text{{\normalsize$\r{a_4}$}}};
\node at ($(b1)+(80:0.195)+(0:0.1)+(80:0.05)$) {\text{{\normalsize$\r{b_1}$}}};
\node at ($(b2)+(0:0.25)$) {\text{{\normalsize$\r{b_2}$}}};
\node at ($(b3)+(-80:0.195)+(0:0.2)+(90:0.05)$) {\text{{\normalsize$\r{b_3}$}}};
}
\bigger{\Rightarrow}\;\;\;\;\;
\pBox{\def\ephScale{1.0*\figScale}\coordinate(v9)at(0,-\ephScale*0.5);\coordinate(v1)at($(v9)+(-90-60:\ephScale)$);\coordinate(v2)at($(v1)+(-90-2*60:\ephScale)$);\coordinate(v3)at($(v2)+(-90-3*60:\ephScale)$);\coordinate(v4)at($(v3)+(-90-4*60:\ephScale)$);\coordinate(v5)at($(v4)+(-90-5*60:\ephScale)$);\coordinate(v6)at($(v5)+(18:\ephScale)$);\coordinate(v7)at($(v6)+(18-72:\ephScale)$);\coordinate(v8)at($(v7)+(18-144:\ephScale)$);
\coordinate(x)at($(v9)+(-84:0.405)$);\coordinate(a1)at($(v1)!.5!(v2)+(-120+0*60:0.205)$);\coordinate(a2)at($(v2)!.5!(v3)+(-120-1*60:0.205)$);\coordinate(a3)at($(v3)!.5!(v4)+(-120-2*60:0.205)$);\coordinate(a4)at($(v4)!.4!(v5)+(-120-3*60:0.205)$);\coordinate(b1)at($(v5)!.6!(v6)+(108:0.205)$);\coordinate(b2)at($(v6)!.5!(v7)+(108-72:0.205)$);\coordinate(b3)at($(v7)!.5!(v8)+(108-2*72:0.205)$);\coordinate(ella)at($(v9)!.5!(v5)+(180:\ephScale*0.85)$);\coordinate(ellb)at($(v9)!.5!(v5)+(0:\ephScale*0.7)$);
\draw[int](v9)--(v1);\draw[int](v1)--(v2);\draw[int](v2)--(v3);\draw[int](v3)--(v4);\draw[int](v4)--(v5);\draw[int](v5)--(v9);\draw[int](v5)--(v6);\draw[int](v6)--(v7);\draw[int](v7)--(v8);\draw[int](v8)--(v9);\draw[markedEdge](v1)--(v2);\draw[markedEdge](v3)--(v4);\draw[markedEdge](v6)--(v7);
\legMassive{(v7)}{0}{};\legMassive{(v6)}{72}{};\legMassive{(v8)}{-72}{};\legMassive{(v1)}{-90}{};\legMassive{(v2)}{-90-60}{};\legMassive{(v3)}{-90-2*60}{};\legMassive{(v4)}{-90-3*60}{};
\legMassive{(v5)}{84}{};
\node[bldot]at(x){};\node[rdot]at(a1){};\node[rdot]at(a2){};\node[rdot]at(a3){};\node[rdot]at(a4){};\node[rdot]at(b1){};\node[rdot]at(b2){};\node[rdot]at(b3){};
\node[anchor=north east,inner sep=0pt] at ($(a1)+(-120:0)$) {\text{{\normalsize$\r{a_1}$}}};\node[anchor=east,inner sep=2pt] at ($(a2)+(-120:0)$) {\text{{\normalsize$\r{a_2}$}}};\node[anchor=south east,inner sep=1pt] at ($(a3)+(120:0)$) {\text{{\normalsize$\r{a_3}$}}};\node[anchor=south,inner sep=1pt] at ($(a4)+(120:0)$) {\text{{\normalsize$\r{a_4}$}}};\node[anchor=south,inner sep=2pt] at ($(b1)+(0.2,-0.05)$) {\text{{\normalsize$\r{b_1}$}}};\node[anchor=south west,inner sep=1pt] at ($(b2)+(-0,-0.05)$) {\text{{\normalsize$\r{b_2}$}}};\node[anchor=north west,inner sep=0pt] at ($(b3)+(0.1,0.1)$) {\text{{\normalsize$\r{b_3}$}}};
\node[] at (ella) {\text{{\normalsize$\ell_1$}}};\node[] at (ellb) {\text{{\normalsize$\ell_2$}}};\node[anchor=north,inner sep=2pt]at(x) {\text{{\normalsize$\b{X}$}}};
}\;\;\;\bigger{\simeq}\;\;\;\;\;
\pBox{\def\ephScale{1.0*\figScale}\coordinate(v9)at(0,-\ephScale*0.5);\coordinate(v1)at($(v9)+(-90-60:\ephScale)$);\coordinate(v2)at($(v1)+(-90-2*60:\ephScale)$);\coordinate(v3)at($(v2)+(-90-3*60:\ephScale)$);\coordinate(v4)at($(v3)+(-90-4*60:\ephScale)$);\coordinate(v5)at($(v4)+(-90-5*60:\ephScale)$);\coordinate(v6)at($(v5)+(18:\ephScale)$);\coordinate(v7)at($(v6)+(18-72:\ephScale)$);\coordinate(v8)at($(v7)+(18-144:\ephScale)$);
\coordinate(x)at($(v9)+(-84:0.405)$);\coordinate(a1)at($(v1)!.5!(v2)+(-120+0*60:0.205)$);\coordinate(a2)at($(v2)!.5!(v3)+(-120-1*60:0.205)$);\coordinate(a3)at($(v3)!.5!(v4)+(-120-2*60:0.205)$);\coordinate(a4)at($(v4)!.4!(v5)+(-120-3*60:0.205)$);\coordinate(b1)at($(v5)!.6!(v6)+(108:0.205)$);\coordinate(b2)at($(v6)!.5!(v7)+(108-72:0.205)$);\coordinate(b3)at($(v7)!.5!(v8)+(108-2*72:0.205)$);\coordinate(ella)at($(v9)!.5!(v5)+(180:\ephScale*0.85)$);\coordinate(ellb)at($(v9)!.5!(v5)+(0:\ephScale*0.7)$);
\dimLines\draw[int](v9)--(v1);\draw[int](v1)--(v2);\draw[int](v2)--(v3);\draw[int](v3)--(v4);\draw[int](v4)--(v5);\draw[int](v5)--(v9);\draw[int](v5)--(v6);\draw[int](v6)--(v7);\draw[int](v7)--(v8);\draw[int](v8)--(v9);\legMassive{(v7)}{0}{};\legMassive{(v6)}{72}{};\legMassive{(v8)}{-72}{};\legMassive{(v1)}{-90}{};\legMassive{(v2)}{-90-60}{};\legMassive{(v3)}{-90-2*60}{};\legMassive{(v4)}{-90-3*60}{};
\legMassive{(v5)}{84}{};\restoreDark\draw[int](ella)--(ellb);\draw[int](ella)--(x);\draw[int](ellb)--(x);\draw[int](ella)--(a1);\draw[int](ella)--(a2);\draw[int](ella)--(a3);\draw[int](ella)--(a4);\draw[int](ellb)--(b1);\draw[int](ellb)--(b2);\draw[int](ellb)--(b3);
\node[ddot]at(ella){};\node[ddot]at(ellb){};\node[bldot]at(x){};\node[rdot]at(a1){};\node[rdot]at(a2){};\node[rdot]at(a3){};\node[rdot]at(a4){};\node[rdot]at(b1){};\node[rdot]at(b2){};\node[rdot]at(b3){};
\node[anchor=north east,inner sep=0pt] at ($(a1)+(-120:0)$) {\text{{\normalsize$\r{a_1}$}}};\node[anchor=east,inner sep=2pt] at ($(a2)+(-120:0)$) {\text{{\normalsize$\r{a_2}$}}};\node[anchor=south east,inner sep=1pt] at ($(a3)+(120:0)$) {\text{{\normalsize$\r{a_3}$}}};\node[anchor=south,inner sep=1pt] at ($(a4)+(120:0)$) {\text{{\normalsize$\r{a_4}$}}};\node[anchor=south,inner sep=2pt] at ($(b1)+(0.2,-0.05)$) {\text{{\normalsize$\r{b_1}$}}};\node[anchor=south west,inner sep=1pt] at ($(b2)+(-0,-0.05)$) {\text{{\normalsize$\r{b_2}$}}};\node[anchor=north west,inner sep=0pt] at ($(b3)+(0.1,0.1)$) {\text{{\normalsize$\r{b_3}$}}};\node[anchor=north,inner sep=2pt]at(x) {\text{{\normalsize$\b{X}$}}};
}\fwboxL{0pt}{\,\,,}
}
corresponds to the set of master integrands of the form:
\eq{\fwbox{0pt}{\hspace{-19pt}\mathcal{I}_{(\b{i},\b{j}),\b{k}}\equivR\dbar^4\!\ell_1\dbar^4\!\ell_2\frac{\x{\ell_1}{\b{N_i}}\x{\ell_1}{\b{N_j}}\x{\ell_2}{\b{N_k}}}{\x{\ell_1}{\infX}\x{\ell_1}{\r{a_1}}\x{\ell_1}{\r{a_2}}\x{\ell_1}{\r{a_3}}\x{\ell_1}{\r{a_4}}\x{\ell_1}{\ell_2}\x{\ell_2}{\r{b_1}}\x{\ell_2}{\r{b_2}}\x{\ell_2}{\r{b_3}}\x{\ell_2}{\b{X}}}\!.}\label{pentabox_with_triangle_power_counting}}
This space has $\big(\mathrm{rank}\big([\ell_1]^2[\ell_2]\big){=}\big){\color{totalCount}120}$-dimensional. For a generic numerator, this integrand will integrate to an impure function of indefinite rigidity and depend on elliptic polylogarithms of \textbf{22(!)} distinct elliptic curves. Notice that this number of elliptic curves is substantially larger than the case of the pentabox with box power-counting discussed in the introduction (\ref{scalar_pentabox_integrand}) which only involved the four elliptic curves associated with the double-box contact terms; for triangle power-counting, the integrand now supports and additional $6{+}12$ elliptic curves corresponding to box-triangle contact-terms.

Naturally, of course, the {\color{totalCount}120}-dimensional space of master integrands can be represented in terms of a 116-dimensional subspace spanned by contact-terms and {\color{topCount}4} top-level master integrands. For each of the 22 elliptic curves (all associated with contact terms), we choose both $a$- and $b$-cycles as defining contours of the prescriptive basis; for this choice, any non-vanishing support an initial choice of top-level master integrands may have had on these elliptic contours would be removed directly by the prescriptivity requirement; to illustrate this subtraction once more, consider the case where an initial choice had support on the $a$-cycle of the elliptic curve $\Omega[(\r{a_1},\r{a_2},\r{a_3});(\r{b_1},\r{b_2},\b{X})]$; then we would manifestly remove this support by rotating the basis (as required by prescriptivity) according to the subtraction
\eq{\mathcal{I}_{(\b{i},\b{j}),\b{k}}^0\mapsto\mathcal{I}_{(\b{i},\b{j}),\b{k}}\equivR\mathcal{I}_{(\b{i},\b{j}),\b{k}}^0{-}\x{\ell_1}{\r{a_4}}\x{\ell_1}{\b{X}}\x{\ell_2}{\r{b_3}}\oint\limits_{\substack{\fwbox{0pt}{\Omega[(\r{a_1},\r{a_2},\r{a_3});(\r{b_1},\r{b_2},\b{X})]}\\\text{a-cycle}}}\!\!\!\mathcal{I}_{(\b{i},\b{j}),\b{k}}^0\,.}

Thus, using prescriptivity, we can uniquely ensure that each top-level master integrand vanishes on all the contours used to define the contact terms; because all elliptic contours are associated with contact-terms, we can ensure that each of the {\color{topCount}4} top-level master integrands vanish on each of the 22 elliptic curves present in the general case. The {\color{topCount}4} top-level master integrands can be made dual to the four solutions to the maximal cut equations of the pentabox, which would then be purely polylogarithmic.

In summary, the {\color{totalCount}120}-dimensional space of master integrands for a pentabox with triangle power-counting can be fully stratified into 44 pure elliptic integrals of definite rigidity (depending on a single elliptic curve each), and 76 pure polylogarithmic integrals.

\subsection{General, \emph{Indefinite} Rigidity of Dual-Conformal Master Integrands}\label{subsec:mixed_rigidity}
Let us now revisit the general obstruction to obtaining integrands with definite rigidity in the case of a basis with box power-counting. (Any such basis can be rendered term-wise dual-conformally invariant (DCI) by appropriate normalization, and so we often speak of this space of integrands as `DCI'.) 

The principal obstruction should be fairly clear. Consider the case of pentabox integrands with box power-counting, corresponding to 
\eq{\fwbox{0pt}{\hspace{-40pt}\pBox{\pBoxPlainEdges
\legMassive{(v1)}{-110}{$$}\legMassive{(v2)}{180}{$$}\legMassive{(v3)}{110}{$$}\legMassive{(v4)}{80}{$$}\legMassive{(v5)}{45}{$$}\legMassive{(v6)}{-45}{$$}\legMassive{(v7)}{-80}{$$}\draw[markedEdge] (v2)--(v3);\coordinate (ella) at ($(v7)!.5!(v4)+(180:0.575)$);
\coordinate (ellb) at ($(v7)!.5!(v4)+(0:\figScale*0.45)$);
\coordinate (a1) at ($(v7)!.5!(v1)+(-72:0.25)$);
\coordinate (a2) at ($(v1)!.5!(v2)+(-144:0.25)$);
\coordinate (a3) at ($(v2)!.5!(v3)+(144:0.25)$);
\coordinate (a4) at ($(v3)!.5!(v4)+(72:0.25)$);
\coordinate (b1) at ($(v4)!.5!(v5)+(90:0.205)$);
\coordinate (b2) at ($(v5)!.5!(v6)+(0:0.15)$);
\coordinate (b3) at ($(v6)!.5!(v7)-(90:0.205)$);
\node[rdot] at (a1){};
\node[rdot] at (a2){};
\node[rdot] at (a3){};
\node[rdot] at (a4){};
\node[rdot] at (b1){};
\node[rdot] at (b2){};
\node[rdot] at (b3){};\node at ($(ella)$) {\text{{\normalsize${\ell_1}$}}};
\node at ($(ellb)$) {\text{{\normalsize${\ell_2}$}}};
\node at ($(a1)+(-72:0.195)+(-90:0.05)+(0:0.0)$) {\text{{\normalsize$\r{a_1}$}}};
\node at ($(a2)+(-144:0.195)+(-90:0.1)+(0:0.2)$) {\text{{\normalsize$\r{a_2}$}}};
\node at ($(a3)+(144:0.195)+(90:0.05)+(0:0.2)$) {\text{{\normalsize$\r{a_3}$}}};
\node at ($(a4)+(72:0.195)$) {\text{{\normalsize$\r{a_4}$}}};
\node at ($(b1)+(80:0.195)+(0:0.1)+(80:0.05)$) {\text{{\normalsize$\r{b_1}$}}};
\node at ($(b2)+(0:0.25)$) {\text{{\normalsize$\r{b_2}$}}};
\node at ($(b3)+(-80:0.195)+(0:0.2)+(90:0.05)$) {\text{{\normalsize$\r{b_3}$}}};
}\bigger{\!\!\!\Rightarrow}\mathcal{I}_{\b{i}}\equivR\!\dbar^4\!\ell_1\dbar^4\!\ell_2\frac{\x{\ell_1}{\b{N_i^0}}}{\x{\ell_1}{\r{a_1}}\x{\ell_1}{\r{a_2}}\x{\ell_1}{\r{a_3}}\x{\ell_1}{\r{a_4}}\x{\ell_1}{\ell_2}\x{\ell_2}{\r{b_1}}\x{\ell_2}{\r{b_2}}\x{\ell_2}{\r{b_3}}}\!\!.}}
These represent a {\color{totalCount}6}-dimensional space of master integrands. For general external masses, there are four elliptic contours supported by such integrands: $\Omega[(\r{a_i},\r{a_j},\r{a_k});\r{\vec{b}}]$ for any choice of subsets $\r{a_{i,j,k}}$ of the relevant four external points; and there are two independent polylogarithmic contours associated with the maximal cuts of the pentabox. Even choosing a maximal number (all {\color{topCount}6}) of elliptic contours to define a prescriptive basis, the top-level integrands would have support on at least 2 different elliptic curves \emph{and} also the polylogarithmic contours associated with the leading singularities. 

Arguably the best one can do would be to use one of the two elliptic cycles for each of the four elliptic curves to define the contact-terms; in this case, at least 4 of the {\color{totalCount}6} master integrands would be pure, elliptic polylogarithms with definite rigidity; the 2 remaining (top-level) master integrands, however, would have indefinite rigidity and depend on multiple elliptic curves. 

\paragraph{\emph{An Interesting Exception to the General Rule}}~\\
\indent It is interesting to note that the obstruction described above for bases with box power-counting can be avoided for special configurations of external masses. Perhaps the simplest example where this obstruction can be avoided is in the case of the following (first relevant to the scattering of 10 massless particles):
\vspace{-4pt}\eq{\fwbox{0pt}{\hspace{-40pt}\pBox{\pBoxPlainEdges
\leg{(v1)}{-110}{$$}\leg{(v2)}{180}{$$}\legMassive{(v3)}{110}{$$}
\leg{(v4)}{81}{$$}
\legMassive{(v5)}{45}{$$}
\legMassive{(v6)}{-45}{$$}\leg{(v7)}{-81}{$$}
\coordinate (ella) at ($(v7)!.5!(v4)+(180:0.575)$);\coordinate (ellb) at ($(v7)!.5!(v4)+(0:\figScale*0.45)$);\coordinate (a1) at ($(v7)!.5!(v1)+(-72:0.25)$);\coordinate (a2) at ($(v1)!.5!(v2)+(-144:0.25)$);\coordinate (a3) at ($(v2)!.5!(v3)+(144:0.25)$);\coordinate (a4) at ($(v3)!.5!(v4)+(72:0.25)$);\coordinate (b1) at ($(v4)!.5!(v5)+(90:0.205)$);\coordinate (b2) at ($(v5)!.5!(v6)+(0:0.15)$);\coordinate (b3) at ($(v6)!.5!(v7)+(-90:0.205)$);
\node[rdot] at (a1){};\node[rdot] at (a2){};\node[rdot] at (a3){};\node[rdot] at (a4){};\node[rdot] at (b1){};\node[rdot] at (b2){};\node[rdot] at (b3){};
\node at ($(ella)$) {\text{{\normalsize${\ell_1}$}}};
\node at ($(ellb)$) {\text{{\normalsize${\ell_2}$}}};
\node at ($(a1)+(-72:0.195)+(-90:0.05)+(0:0.0)$) {\text{{\normalsize$\r{a_1}$}}};
\node at ($(a2)+(-144:0.195)+(-90:0.1)+(0:0.2)$) {\text{{\normalsize$\r{a_2}$}}};
\node at ($(a3)+(144:0.195)+(90:0.05)+(0:0.2)$) {\text{{\normalsize$\r{a_3}$}}};
\node at ($(a4)+(72:0.195)$) {\text{{\normalsize$\r{a_4}$}}};
\node at ($(b1)+(80:0.195)+(0:0.1)+(80:0.05)$) {\text{{\normalsize$\r{b_1}$}}};
\node at ($(b2)+(0:0.25)$) {\text{{\normalsize$\r{b_2}$}}};
\node at ($(b3)+(-80:0.195)+(0:0.2)+(90:0.05)$) {\text{{\normalsize$\r{b_3}$}}};
\draw[markedEdge](v1)--(v2);
}\bigger{\!\!\!\Rightarrow}\mathcal{I}_{\b{i}}^0\equivR\!\dbar^4\!\ell_1\dbar^4\!\ell_2\frac{\x{\ell_1}{\b{N_i^0}}}{\x{\ell_1}{\r{a_1}}\x{\ell_1}{\r{a_2}}\x{\ell_1}{\r{a_3}}\x{\ell_1}{\r{a_4}}\x{\ell_1}{\ell_2}\x{\ell_2}{\r{b_1}}\x{\ell_2}{\r{b_2}}\x{\ell_2}{\r{b_3}}}\!\!.}\vspace{-4pt}}
In this case, we restricted the pairs $(\r{a_1},\r{a_2}), (\r{a_2},\r{a_3}),(\r{a_4},\r{b_1}),$ and $(\r{b_3},\r{a_1})$ to be light-like separated (corresponding to incoming, massless momenta indicated by thin-stroke lines in the figure above). In this case, there is exactly one elliptic curve accessible via the contour $\Omega[(\r{a_1},\r{a_3},\r{a_4}),\r{\vec{b}}]$---with all other contact terms being polylogarithmic (and infrared divergent in four dimensions).

Let us start with an initial basis of master integrands with numerators
\eq{\Big\{\big|\b{N_1^0}\big),\ldots,\big|\b{N_6^0}\big)\Big\}\equivR\Big\{\big|\b{Q_{\r{a_1},\r{a_2},\r{a_3},\r{a_4}}^1}\big),\big|\b{Q_{\r{a_1},\r{a_2},\r{a_3},\r{a_4}}^2}\big),\big|\r{a_2}\big),\big|\r{a_1}\big),\big|\r{a_3}\big),\big|\r{a_4}\big)\Big\}\,,}
and use the following contours to define a prescriptive basis of master integrands:
\eq{\begin{split}\hspace{65pt}\fwboxR{0pt}{\{\Omega_1,\ldots,\Omega_6\}\equivR\!\!\left\{\rule{0pt}{35pt}\right.\hspace{-10pt}}
\pBox{\pBoxCoords\pBoxPlainEdges\coordinate (ella) at ($(v7)!.5!(v4)+(180:0.575)$);\coordinate (ellb) at ($(v7)!.5!(v4)+(0:\figScale*0.45)$);\coordinate (a1) at ($(v7)!.5!(v1)+(-72:0.25)$);\coordinate (a2) at ($(v1)!.5!(v2)+(-144:0.25)$);\coordinate (a3) at ($(v2)!.5!(v3)+(144:0.25)$);\coordinate (a4) at ($(v3)!.5!(v4)+(72:0.25)$);\coordinate (b1) at ($(v4)!.5!(v5)+(90:0.205)$);\coordinate (b2) at ($(v5)!.5!(v6)+(0:0.15)$);\coordinate (b3) at ($(v6)!.5!(v7)+(-90:0.205)$);
\leg{(v1)}{-110}{$$}\leg{(v2)}{180}{$$}\legMassive{(v3)}{110}{$$}
\leg{(v4)}{81}{$$}\legMassive{(v5)}{45}{$$}\legMassive{(v6)}{-45}{$$}\leg{(v7)}{-81}{$$}
\node[rdot] at (a1){};\node[rdot] at (a2){};\node[rdot] at (a3){};\node[rdot] at (a4){};\node[rdot] at (b1){};\node[rdot] at (b2){};\node[rdot] at (b3){};
\node at ($(ella)$) {\text{{\normalsize${\ell_1^*}$}}};
\node at ($(ellb)$) {\text{{\normalsize${\ell_2}^*$}}};
\node at ($(a1)+(-72:0.195)+(-90:0.05)+(0:0.0)$) {\text{{\normalsize$\r{a_1}$}}};
\node at ($(a2)+(-144:0.195)+(-90:0.1)+(0:0.2)$) {\text{{\normalsize$\r{a_2}$}}};
\node at ($(a3)+(144:0.195)+(90:0.05)+(0:0.2)$) {\text{{\normalsize$\r{a_3}$}}};
\node at ($(a4)+(72:0.195)$) {\text{{\normalsize$\r{a_4}$}}};
\node at ($(b1)+(80:0.195)+(0:0.1)+(80:0.05)$) {\text{{\normalsize$\r{b_1}$}}};
\node at ($(b2)+(0:0.25)$) {\text{{\normalsize$\r{b_2}$}}};
\node at ($(b3)+(-80:0.195)+(0:0.2)+(90:0.05)$) {\text{{\normalsize$\r{b_3}$}}};
\contourVerts{1}{2}{4}{4}{4}{4}{4}}
,
\begin{array}{@{}c@{}}~\\[-6pt]\dBox{\dBoxPlainEdges\legMassive{(v1)}{-135}{$$}\legMassive{(v2)}{135}{$$}\leg{(v3)}{90}{$$}\legMassive{(v4)}{45}{$$}\legMassive{(v5)}{-45}{$$}\leg{(v6)}{-90}{$$}
\coordinate (a1) at ($(v6)!.55!(v1)+(-90:0.25)$);
\coordinate (a2) at ($(v1)!.5!(v2)+(180:0.25)$);
\coordinate (a3) at ($(v3)!.55!(v2)+(90:0.25)$);
\coordinate (b1) at ($(v3)!.55!(v4)+(90:0.25)$);
\coordinate (b2) at ($(v4)!.5!(v5)+(0:0.25)$);
\coordinate (b3) at ($(v6)!.55!(v5)+(-90:0.25)$);
\node[rdot] at (a1){};\node[rdot] at (a2){};\node[rdot] at (a3){};\node[rdot] at (b1){};\node[rdot] at (b2){};\node[rdot] at (b3){};
\node at ($(a1)+(-90:0.25)$) {\text{{\normalsize$\r{a_1}$}}};
\node at ($(a2)+(180:0.275)+(90:0.0)$) {\text{{\normalsize$\r{a_3}$}}};
\node at ($(a3)+(90:0.25)$) {\text{{\normalsize$\r{a_4}$}}};
\node at ($(b1)+(0:0.05)+(90:0.25)$) {\text{{\normalsize$\r{b_1}$}}};
\node at ($(b2)+(0:0.275)+(90:0.05)$) {\text{{\normalsize$\r{b_2}$}}};
\node at ($(b3)+(-90:0.20)$) {\text{{\normalsize$\r{b_3}$}}};
\node at ($(ella)$) {\text{{\normalsize${\ell_1^*}$}}};\node at ($(ellb)$) {\text{{\normalsize${\ell_2^*}$}}};
\contourVerts{4}{4}{4}{4}{4}{4}{5}}\\[-6pt]\text{{\footnotesize(a-cycle)}}\end{array}
,
\begin{array}{@{}c@{}}~\\[-6pt]\dBox{\dBoxPlainEdges\legMassive{(v1)}{-135}{$$}\legMassive{(v2)}{135}{$$}\leg{(v3)}{90}{$$}\legMassive{(v4)}{45}{$$}\legMassive{(v5)}{-45}{$$}\leg{(v6)}{-90}{$$}
\coordinate (a1) at ($(v6)!.55!(v1)+(-90:0.25)$);
\coordinate (a2) at ($(v1)!.5!(v2)+(180:0.25)$);
\coordinate (a3) at ($(v3)!.55!(v2)+(90:0.25)$);
\coordinate (b1) at ($(v3)!.55!(v4)+(90:0.25)$);
\coordinate (b2) at ($(v4)!.5!(v5)+(0:0.25)$);
\coordinate (b3) at ($(v6)!.55!(v5)+(-90:0.25)$);
\node[rdot] at (a1){};\node[rdot] at (a2){};\node[rdot] at (a3){};\node[rdot] at (b1){};\node[rdot] at (b2){};\node[rdot] at (b3){};
\node at ($(a1)+(-90:0.25)$) {\text{{\normalsize$\r{a_1}$}}};
\node at ($(a2)+(180:0.275)+(90:0.0)$) {\text{{\normalsize$\r{a_3}$}}};
\node at ($(a3)+(90:0.25)$) {\text{{\normalsize$\r{a_4}$}}};
\node at ($(b1)+(0:0.05)+(90:0.25)$) {\text{{\normalsize$\r{b_1}$}}};
\node at ($(b2)+(0:0.275)+(90:0.05)$) {\text{{\normalsize$\r{b_2}$}}};
\node at ($(b3)+(-90:0.20)$) {\text{{\normalsize$\r{b_3}$}}};
\node at ($(ella)$) {\text{{\normalsize${\ell_1^*}$}}};\node at ($(ellb)$) {\text{{\normalsize${\ell_2^*}$}}};
\contourVerts{4}{4}{4}{4}{4}{4}{5}}\\[-6pt]\text{{\footnotesize(b-cycle)}}\end{array}
\fwboxL{0pt}{,}
\\
\hspace{65pt}\dBox{\dBoxPlainEdges\leg{(v1)}{-135}{$$}\legMassive{(v2)}{135}{$$}\leg{(v3)}{90}{$$}\legMassive{(v4)}{45}{$$}\legMassive{(v5)}{-45}{$$}\legMassive{(v6)}{-90}{$$}
\coordinate(a1) at ($(v6)+(-90:0.3)$);\coordinate(a2) at ($(v6)!.5!(v1)+(-90:0.25)$);\coordinate(a3) at ($(v1)!.5!(v2)+(180:0.3)$);\coordinate(a4) at ($(v2)!.5!(v3)+(90:0.2)$);\coordinate(b1) at ($(v3)!.5!(v4)+(90:0.2)$);\coordinate(b2) at ($(v4)!.5!(v5)+(0:0.3)$);\coordinate (b3) at ($(v6)!.55!(v5)+(-90:0.25)$);\coordinate (ella) at ($(v6)!.5!(v3)+(180:\figScale*0.6)$);\coordinate (ellb) at ($(v6)!.5!(v3)+(0:\figScale*0.65)$);\node[rdot] at (a2){};\node[rdot] at (a3){};\node[rdot] at (a4){};\node[rdot] at (b1){};\node[rdot] at (b2){};\node[rdot]at(b3){};\node at ($(ella)$) {\text{{\normalsize${\ell_1^*}$}}};\node at ($(ellb)$) {\text{{\normalsize${\ell_2^*}$}}};
\node at ($(b3)+(-72:0)+(-90:0.2)+(180:0.0)$) {\text{{\normalsize$\r{b_3}$}}};
\node at ($(a2)+(-72:0.195)+(-90:0.05)+(180:0.0)$) {\text{{\normalsize$\r{a_2}$}}};\node at ($(a3)+(-72:0.195)+(-90:0.03)+(0:0.03)+(90:0.38)$) {\text{{\normalsize$\r{a_3}$}}};\node at ($(b2)+(-72:0.195)+(-90:0.05)+(90:0.075)+(0:0.05)+(90:0.38)$) {\text{{\normalsize$\r{b_2}$}}};\node at ($(b1)+(80:0.195)+(0:0.1)+(80:0.05)$) {\text{{\normalsize$\r{b_1}$}}};\node at ($(a4)+(80:0.195)+(0:0.04)+(-90:0.045)+(80:0.05)$) {\text{{\normalsize$\r{a_4}$}}};
\contourVerts{3}{4}{4}{4}{4}{4}{5}}
,
\dBox{\dBoxPlainEdges\leg{(v1)}{-135}{$$}\legMassive{(v2)}{135}{$$}\leg{(v3)}{90}{$$}\legMassive{(v4)}{45}{$$}\legMassive{(v5)}{-45}{$$}\leg{(v6)}{-90}{$$}
\coordinate(a1) at ($(v6)+(-90:0.3)$);\coordinate(a2) at ($(v6)!.5!(v1)+(-90:0.25)$);\coordinate(a3) at ($(v1)!.5!(v2)+(180:0.3)$);\coordinate(a4) at ($(v2)!.5!(v3)+(90:0.2)$);\coordinate(b1) at ($(v3)!.5!(v4)+(90:0.2)$);\coordinate(b2) at ($(v4)!.5!(v5)+(0:0.3)$);\coordinate (b3) at ($(v6)!.55!(v5)+(-90:0.25)$);\coordinate (ella) at ($(v6)!.5!(v3)+(180:\figScale*0.6)$);\coordinate (ellb) at ($(v6)!.5!(v3)+(0:\figScale*0.65)$);\node[rdot] at (a2){};\node[rdot] at (a3){};\node[rdot] at (a4){};\node[rdot] at (b1){};\node[rdot] at (b2){};\node[rdot]at(b3){};\node at ($(ella)$) {\text{{\normalsize${\ell_1^*}$}}};\node at ($(ellb)$) {\text{{\normalsize${\ell_2^*}$}}};
\node at ($(b3)+(-72:0)+(-90:0.2)+(180:0.0)$) {\text{{\normalsize$\r{b_3}$}}};
\node at ($(a2)+(-72:0.195)+(-90:0.05)+(180:0.0)$) {\text{{\normalsize$\r{a_1}$}}};\node at ($(a3)+(-72:0.195)+(-90:0.03)+(0:0.03)+(90:0.38)$) {\text{{\normalsize$\r{a_2}$}}};\node at ($(b2)+(-72:0.195)+(-90:0.05)+(90:0.075)+(0:0.05)+(90:0.38)$) {\text{{\normalsize$\r{b_2}$}}};\node at ($(b1)+(80:0.195)+(0:0.1)+(80:0.05)$) {\text{{\normalsize$\r{b_1}$}}};\node at ($(a4)+(80:0.195)+(0:0.04)+(-90:0.045)+(80:0.05)$) {\text{{\normalsize$\r{a_4}$}}};
\contourVerts{3}{4}{4}{4}{4}{4}{5}}
,
\dBox{\dBoxPlainEdges\leg{(v1)}{-135}{$$}\leg{(v2)}{135}{$$}\legMassive{(v3)}{90}{$$}\legMassive{(v4)}{45}{$$}\legMassive{(v5)}{-45}{$$}\leg{(v6)}{-90}{$$}
\coordinate(a1) at ($(v6)+(-90:0.3)$);\coordinate(a2) at ($(v6)!.5!(v1)+(-90:0.25)$);\coordinate(a3) at ($(v1)!.5!(v2)+(180:0.3)$);\coordinate(a4) at ($(v2)!.5!(v3)+(90:0.2)$);\coordinate(b1) at ($(v3)!.5!(v4)+(90:0.2)$);\coordinate(b2) at ($(v4)!.5!(v5)+(0:0.3)$);\coordinate (b3) at ($(v6)!.55!(v5)+(-90:0.25)$);\coordinate (ella) at ($(v6)!.5!(v3)+(180:\figScale*0.6)$);\coordinate (ellb) at ($(v6)!.5!(v3)+(0:\figScale*0.65)$);\node[rdot] at (a2){};\node[rdot] at (a3){};\node[rdot] at (a4){};\node[rdot] at (b1){};\node[rdot] at (b2){};\node[rdot]at(b3){};\node at ($(ella)$) {\text{{\normalsize${\ell_1^*}$}}};\node at ($(ellb)$) {\text{{\normalsize${\ell_2^*}$}}};
\node at ($(b3)+(-72:0)+(-90:0.2)+(180:0.0)$) {\text{{\normalsize$\r{b_3}$}}};
\node at ($(a2)+(-72:0.195)+(-90:0.05)+(180:0.0)$) {\text{{\normalsize$\r{a_1}$}}};\node at ($(a3)+(-72:0.195)+(-90:0.03)+(0:0.03)+(90:0.38)$) {\text{{\normalsize$\r{a_2}$}}};\node at ($(b2)+(-72:0.195)+(-90:0.05)+(90:0.075)+(0:0.05)+(90:0.38)$) {\text{{\normalsize$\r{b_2}$}}};\node at ($(b1)+(80:0.195)+(0:0.1)+(80:0.05)$) {\text{{\normalsize$\r{b_1}$}}};\node at ($(a4)+(80:0.195)+(0:0.04)+(-90:0.045)+(80:0.05)$) {\text{{\normalsize$\r{a_3}$}}};
\contourVerts{0}{0}{4}{4}{4}{4}{5}}\fwboxL{0pt}{\hspace{-5pt}\left.\rule{0pt}{35pt}\right\}\fwboxL{0pt}{.}}
\end{split}\label{dci_counterexample_contours}}
With these contours, our initial master integrands would have a period matrix of the form 
\eq{\fwboxL{0pt}{\raisebox{-9pt}{$\hspace{43pt}\left(\rule{0pt}{49pt}\right.$}}\begin{array}{@{}l@{}|@{$\;\;$}c@{}c@{}c@{}c@{}c@{}c@{}c@{}c@{}c@{}}
\fwbox{40pt}{\mathfrak{n}(\ell_1)\!\!\,\,}&\fwbox{38pt}{{\Omega_1}}&\fwbox{38pt}{\Omega_2}&\fwbox{38pt}{\Omega_3}&\fwbox{38pt}{\Omega_4}&\fwbox{38pt}{\Omega_5}&\fwbox{38pt}{\Omega_6}
\\\hline\\[-12pt]
\x{\ell_1}{\b{N_1^0}}&\b{f_1^{\r{1}}(\vec{p})}&\b{f_1^{\r{2}}(\vec{p})}&\b{f_1^{\r{3}}(\vec{p})}&\b{g_1^{\r{4}}(\vec{p})}&\b{g_1^{\r{5}}(\vec{p})}&\b{g_1^{\r{6}}(\vec{p})}\\
\x{\ell_1}{\b{N_2^0}}&0&\b{f_2^{\r{2}}(\vec{p})}&\b{f_2^{\r{3}}(\vec{p})}&\b{g_2^{\r{4}}(\vec{p})}&\b{g_2^{\r{5}}(\vec{p})}&\b{g_2^{\r{6}}(\vec{p})}\\
\x{\ell_1}{\r{a_2}}&0&\b{f_3^{\r{2}}(\vec{p})}&\b{f_3^{\r{3}}(\vec{p})}&0&0&0\\
\x{\ell_1}{\r{a_1}}&0&0&0&\b{h_{\r{4}}(\vec{p})}&0&0\\
\x{\ell_1}{\r{a_3}}&0&0&0&0&\b{h_{\r{5}}(\vec{p})}&0\\
\x{\ell_1}{\r{a_4}}&0&0&0&0&0&\b{h_{\r{6}}(\vec{p})}
\end{array}\fwboxL{0pt}{\raisebox{-9pt}{$\hspace{-6.5pt}\left.\rule{0pt}{49pt}\right)\!\!\!\equivL\mathbf{M}^0$}}\label{dci_counterexample_period_matrix}}
which is easy to confirm is full-rank. Diagonalizing the space of master integrands with respect to this period matrix would result in a space of 2 pure, elliptic polylogarithms of definite rigidity and 4 pure polylogarithmic master integrands. 

As such a `counterexample' to the general rule would seem only marginally useful---and fails for the asymptotic majority of pentabox integrands with box power-counting---it is not clear how valuable such a case is for the actual representations of amplitudes; however, it would be interesting to study this example in further detail as perhaps a particularly nice example of finding purely polylogarithmic integrands with non-vanishing support on elliptic contours (which are in fact total derivatives).

\newpage
\vspace{4pt}
\section{Open Problems and Future Directions}\label{sec:conclusions}\vspace{0pt}

In this paper, we have seen that problems associated with impurities, indefinite-rigidity, and dependence on multiple elliptic curves that arise in the case of sufficiently generic two-loop integrands can all be remedied via prescriptivity---provided the basis of master integrands is sufficiently large. In particular, provided there are sufficient degrees of freedom to diagonalize support on \emph{both} homological cycles on curves, then prescriptivity can ensure that every integrand (other than those two dual to the elliptic cycles) can be made to vanish (cohomologically) on that elliptic curve. And we have shown that triangle power-counting provides a sufficiently large basis to achieve this separation in the case of planar integrands involving massless particles at two-loops. 

It remains an interesting and open problem to determine the analytic form of integrands dual to polylogarithmic cycles when elliptic curves are present. In particular, it would be extremely interesting to determine the polylogarithmic master, top-level integrand for the case of the box-triangle discussed in \mbox{section \ref{subsec:diagonalizing_rigidity_box_triangle_basis}} (with complete details described in \mbox{appendix \ref{appendix:box_triangle_masters}}). Even simply the \emph{symbology} would be extremely interesting---which may be accessible through an investigation of the differential equations satisfied by these master integrands, which should be intimately related to their discontinuities \cite{Bourjaily:2020wvq}. Similarly, it would be interesting to know the symbology of the elliptic basis elements \cite{Broedel:2018iwv,Wilhelm:2022wow}. However, we must leave such questions to future work.

One may naturally wonder whether the conclusions here may extend to other cases where indefinite or higher rigidity has been observed---in particular, for integrands involving massive particles, those beyond the planar limit, or for those at higher loop-orders. We briefly discuss each of these open directions for future work in turn below.

\subsection{Beyond Massless Propagators at Two Loops}\label{subsec:multiple_curves}
The appearance of Feynman integrands that depend on \emph{multiple} elliptic curves has been the subject of much research in recent years. The examples we described here---such as a massless double-box depending on 7 elliptic curves (\ref{double_box_with_triangle_power_counting}) or the a pentabox depending on 22 (\ref{pentabox_with_triangle_power_counting}) elliptic curves---arise only for amplitudes involving a relatively large number of external states; more phenomenologically relevant examples arise for much lower multiplicity when massive propagators are considered (see e.g.~\cite{Remiddi:2016gno, Adams:2018bsn, Adams:2018kez, Adams:2018yqc, Muller:2022gec}). (It is well known that all one-loop amplitudes (involving arbitrary masses in any number of dimensions) are polylogarithmic (see e.g.~\cite{Bourjaily:2019exo}).) 

In cases involving massive propagators, rigidity may arise through a massive sunrise contact-term---which would be invisible within any basis with triangle power-counting. However, like the examples discussed in this work, every known instance where multiple elliptic curves appear to arise has the property that each elliptic curve may be identified with a \emph{specific} and distinct contact-term of the integrand being investigated. Thus, provided the space of master integrands is sufficiently large to diagonalize with respect to both homological cycles of each curve, we strongly suspect that bases with definite rigidity can be constructed for  these cases---provided the power-counting is sufficiently increased. But we must leave such conjectures to future research.

\subsection{Beyond Planarity at Two Loops: Higher Rigidity}\label{subsec:non_planar_two_loops}
As was shown in \cite{Bourjaily:2018yfy}, two-loop involving massless particles are generically \emph{more rigid} than elliptic polylogarithms beyond the planar limit. The simplest example of a Feynman integrand which involves two-dimensional Calabi-Yau (K3) geometry (and therefore having rigidity 2) arises in the case of the `tardigrade': 
\vspace{-4pt}\eq{\hspace{-10pt}\begin{tikzpicture}[scale=1.0*\figScale,baseline=-3.05]\useasboundingbox ($(-1.75,-1.5)$) rectangle ($(1.75,1.5)$);\draw[int,line width=0.1,red,draw=\boundingDraw] ($(-1.75,-1.5)$) rectangle ($(1.75,1.5)$);
\coordinate(i1)at(-1.0,0.65);
\coordinate(i2)at(-1.0,-0.65);
\coordinate(o1)at(1.0,1);
\coordinate(o2)at(1.0,0);
\coordinate(o3)at(1.0,-1);
\draw[under]($(i1)!.4!(o1)$)--($(i1)!.6!(o1)$);\draw[int](i1)--(o1);\draw[under]($(i2)!.4!(o1)$)--($(i2)!.6!(o1)$);\draw[int](i2)--(o1);
\draw[under]($(i1)!.3!(o2)$)--($(i1)!.6!(o2)$);\draw[int](i1)--(o2);\draw[under]($(i2)!.3!(o2)$)--($(i2)!.6!(o2)$);\draw[int](i2)--(o2);
\draw[under]($(i1)!.3!(o3)$)--($(i1)!.6!(o3)$);\draw[int](i1)--(o3);\draw[under]($(i2)!.3!(o3)$)--($(i2)!.6!(o3)$);\draw[int](i2)--(o3);
\draw[directedEdge](i2)--(o3);\draw[directedEdge](i1)--(o1);
\node[anchor=south]at($(i1)!.5!(o1)$){$\ell_1$};\node[anchor=north]at($(i2)!.5!(o3)$){$\ell_2$};
\leg{(i1)}{180-10}{$p_1\,$}\leg{(i2)}{180+10}{$p_2\,$}\legMassive{(o1)}{20}{$\,\,k_1$}\legMassive{(o2)}{0}{$\,\,k_3$}\legMassive{(o3)}{-20}{$\,\,k_2$}
\end{tikzpicture}\bigger{\,\,\,\Rightarrow\,}\frac{\dbar^4\!\ell_1\dbar^4\!\ell_2}{\ell_1^2\,(\ell_1{+}k_1)^2\ell_2^2\,(\ell_2{+}k_2)^2(\ell_1{-}\ell_2{+}p_1{+}k_1)^2(\ell_2{-}\ell_1{+}k_2{+}p_1)^2}\,.\vspace{-4pt}\label{tardigrade_example}}
Here, we have made it clear that the integrand's rigidity does not require $p_{1,2}$ to be massless; as such, this example is first relevant for the phenomenologically relevant case of five-particle scattering (at least three of which are massive). (Moreover, such integrands appear term-wise in the representation of MHV amplitudes in maximally supersymmetric Yang-Mills theory as given in \cite{Bourjaily:2018omh,Bourjaily:2019iqr,Bourjaily:2019gqu,Bourjaily:2021vyj,Bourjaily:2021iyq}.)

This integrand's maximal cut surface describes the geometry of a Calabi-Yau 2-fold---a `K3' surface. If the K3 were generically smooth, its (middle-dimensional) homology would span a 22-dimensional space, requiring at least 22 degrees of freedom to diagonalize a basis with respect to each K3 cycle of integration. This seems unlikely to be possible even for triangle power-counting (for which the number of {\color{topCount}top-level} numerators spans a {\color{topCount}6}-dimensional space \cite{Bourjaily:2020qca}); but perhaps possible if furnished with numerators defined with bubble power-counting (for which this integrand would have {\color{topCount}32 top-level} masters). 

Another possibility, however, would be that perhaps the K3 surface associated with a tardigrade (\ref{tardigrade_example}) is \emph{far from smooth}; in general, singularities of the surface reduce the rank of its period matrix (with the `shrunk cycles' of the geometry not adding to the rank). Thus, the question of whether or not a given space of numerators may suffice to diagonalize with respect to such examples of rigidity depends strongly on how singular such geometries are. 

We do not know of any generic way of computing the rank of the full period matrix for a particular K3 surface (even numerically). But it would be very interesting to know its rank both for generic external kinematics, and in particular kinematic limits.

\subsection{Higher Rigidity for Planar Integrands at Higher Loops}\label{subsec:higher_loops}
As discussed in \mbox{section \ref{subsec:source_of_rigidity}}, the only source of rigidity for planar integrands at two loops is the double-box integrand, with rigidity 1. Beyond two loops, there are multiple sources of ellipticity and also higher rigidity. Although we strongly suspect that definite rigidity for elliptic integrands can be achieved in a basis with triangle power-counting, our ability to stratify integrands with higher rigidity is much less certain. 

Consider for example the case of the three-loop `traintrack' integrand: 
\eq{\fwbox{0pt}{\hspace{-20pt}\begin{tikzpicture}[scale=1.0*\figScale,baseline=-3.05]\useasboundingbox ($(-2,-1.5)$) rectangle ($(2,1.5)$);\draw[int,line width=0.1,red,draw=\boundingDraw] ($(-2,-1.5)$) rectangle ($(2,1.5)$);\def\ephScale{1.2*\figScale}
\coordinate(ellb)at(0,0);
\coordinate(ella)at(-1,0);
\coordinate(ellc)at(1,0);
\coordinate(v1)at(-0.5,-0.5);\coordinate(v2)at(-1.5,-0.5);\coordinate(v3)at(-1.5,0.5);\coordinate(v4)at(-0.5,0.5);\coordinate(v5)at(0.5,0.5);\coordinate(v6)at(1.5,0.5);\coordinate(v7)at(1.5,-0.5);\coordinate(v8)at(0.5,-0.5);
\coordinate(a1)at(-1,-0.65);\coordinate(b2)at(-0,-0.65);\coordinate(c3)at(1,-0.65);
\coordinate(a2)at(-1.65,0);\coordinate(c2)at(1.65,0);
\coordinate(a3)at(-1,0.65);\coordinate(b1)at(0,0.65);\coordinate(c1)at(1,0.65);
\node[rdot]at(a1){};\node[rdot]at(a2){};\node[rdot]at(a3){};\node[rdot]at(b1){};\node[rdot]at(b2){};\node[rdot]at(c1){};\node[rdot]at(c2){};\node[rdot]at(c3){};
\draw[int](v1)--(v2)--(v3)--(v4)--(v5)--(v6)--(v7)--(v8)--(v1);
\draw[int](v1)--(v4);
\draw[int](v8)--(v5);\legMassive{(v1)}{-90}{}\legMassive{(v8)}{-90}{}\legMassive{(v2)}{-135}{}\legMassive{(v3)}{135}{}\legMassive{(v4)}{90}{}\legMassive{(v5)}{90}{}\legMassive{(v6)}{45}{}\legMassive{(v7)}{-45}{}
\node[]at(ella){$\ell_1$};\node[]at(ellb){$\ell_2$};\node[]at(ellc){$\ell_3$};
\node[anchor=north,inner sep=3pt]at(a1){$\r{a_1}$};\node[anchor=north,inner sep=0pt]at(b2){$\r{b_2}$};\node[anchor=north,inner sep=3pt]at(c3){$\r{c_3}$};
\node[anchor=south,inner sep=1pt]at(a3){$\,\,\r{a_3}$};\node[anchor=south,inner sep=0pt]at(b1){$\,\,\r{b_1}$};\node[anchor=south,inner sep=0pt]at(c1){$\,\,\r{c_1}$};
\node[anchor=east,inner sep=2pt]at(a2){$\r{a_2}$};\node[anchor=west,inner sep=2pt]at(c2){$\r{c_2}$};
\end{tikzpicture}\bigger{\,\Rightarrow}\frac{\dbar^{4}\ell_{1} \dbar^{4}\ell_{2} \dbar^{4}\ell_{3}}{\x{\ell_1}{\r{a_1}} \x{\ell_1}{\r{a_2}} \x{\ell_1}{\r{a_3}} \x{\ell_1}{\ell_2} \x{\ell_2}{\r{b_1}} \x{\ell_2}{\r{b_2}} \x{\ell_2}{\ell_3}\x{\ell_3}{\r{c_1}} \x{\ell_3}{\r{c_2}} \x{\ell_3}{\r{c_3}}}\!.}\nonumber}
The maximal cut of this integrand---a contour encircling the vanishing of each of its ten propagators---is known to have the geometry of a K3 surface \cite{Bourjaily:2018ycu}. 

As for the non-planar tardigrade, it is unclear whether triangle power-counting provides a space of integrands sufficiently large to saturate the generically 22 cycles of the K3 period matrix. Luckily, however, for triangle power-counting, the three-loop traintrack would involve {\color{topCount}30 top-level} master integrands \cite{Bourjaily:2020qca}. We suspect that this space allows us to fully diagonalize this basis with respect to the 22-dimensional (middle-dimensional) homology of the K3---ensuring that all integrands including the traintrack as a contact term would, once prescriptively ensured to vanish on each cycle, would be cohomologically trivial on the K3, thereby ensuring the definiteness of their rigidity. However, this would only be answered by explicitly checking that the triangle power-counting integrands can saturate the (possibly less than maximal) rank of the K3 surface. This is a question we must leave to future work.

A second source of higher rigidity at three loops arises in the case of the wheel integrand:
\eq{\hspace{-10pt}\begin{tikzpicture}[scale=1.0*\figScale,baseline=-3.05]\useasboundingbox ($(-2.,-1.5)$) rectangle ($(2.,1.5)$);\draw[int,line width=0.1,red,draw=\boundingDraw] ($(-2.,-1.5)$) rectangle ($(2.,1.5)$);
\coordinate (v0) at (0,0);
\coordinate (v1) at ($(v0)+(150:0.6)$);
\coordinate (v2) at ($(v0)+(30:0.6)$);
\coordinate (v3) at ($(v0)+(-90:0.6)$);
\coordinate (v12) at ($(v0)+(90:1.2)$);
\coordinate (v23) at ($(v0)+(-30:1.2)$);
\coordinate (v31) at ($(v0)+(-150:1.2)$);
\coordinate (ella)at($(v0)+(90:0.5)$);
\coordinate (ellb)at($(v0)+(-30:0.5)$);
\coordinate (ellc)at($(v0)+(-150:0.5)$);
\coordinate (a1)at($(v0)+(120:1.0)$);
\coordinate (a2)at($(v0)+(60:1.0)$);
\coordinate (b1)at($(v0)+(0:1.0)$);
\coordinate (b2)at($(v0)+(-60:1.0)$);
\coordinate (c1)at($(v0)+(-120:1.0)$);
\coordinate (c2)at($(v0)+(-180:1.0)$);
\node[rdot]at(a1){};\node[rdot]at(a2){};\node[rdot]at(b1){};\node[rdot]at(b2){};\node[rdot]at(c1){};\node[rdot]at(c2){};
\node[]at(ella){$\ell_1$};\node[]at(ellb){$\ell_2$};\node[]at(ellc){$\ell_3$};
\node[anchor=south east,inner sep=0pt]at(a1){$\r{a_1}$};\node[anchor=south west,inner sep=0pt]at(a2){$\r{a_2}$};
\node[anchor=west,inner sep=1.5pt]at(b1){$\r{b_1}$};\node[anchor=north,inner sep=0pt]at(b2){$\r{b_2}$};
\node[anchor=north,inner sep=1.5pt]at(c1){$\r{c_1}$};\node[anchor=east,inner sep=1pt]at(c2){$\r{c_2}$};
\draw[int] (v0)--(v1);\draw[int] (v0)--(v2);\draw[int] (v0)--(v3);
\draw[int] (v2)--(v12);\draw[int] (v1)--(v12);\draw[int] (v2)--(v23);\draw[int] (v3)--(v23);\draw[int] (v3)--(v31);\draw[int] (v31)--(v1);
\legMassive{(v12)}{90}{}\legMassive{(v23)}{-30}{}\legMassive{(v31)}{-150}{}\legMassive{(v1)}{150}{}\legMassive{(v2)}{30}{}\legMassive{(v3)}{-90}{}
\end{tikzpicture}\hspace{-15pt}\bigger{\Rightarrow}\frac{\dbar^{4}\!\ell_{1} \dbar^{4}\!\ell_{2} \dbar^{4}\!\ell_{3}}{\x{\ell_1}{\r{a_1}} \x{\ell_1}{\r{a_2}} \x{\ell_1}{\ell_2} \x{\ell_2}{\r{b_1}} \x{\ell_2}{\r{b_2}} \x{\ell_2}{\ell_3}\x{\ell_3}{\r{c_1}} \x{\ell_3}{\r{c_2}} \x{\ell_3}{\ell_1}}\,.\label{wheel_integrand_example}}
An analysis of the Feynman-parametric representation of this integrand in \cite{Bourjaily:2019hmc} suggested that this integrand may have rigidity 3 (depending on a Calabi-Yau three-fold); but direct inspection of the maximal cut surface makes it clear that this 3-dimensional geometry supports an additional polylogarithmic contour in four dimensions---corresponding to the collinear limit of the three internal propagators of the graph (\ref{wheel_integrand_example}). Thus, we suspect that this integrand has rigidity (at most) two---and should be described by a K3 surface like that of the traintrack discussed above. 

When this integrand is endowed with numerators according to triangle power-counting, there are {\color{topCount}52 top-level} degrees of freedom. This seems large enough to saturate the (possibly reduced) rank of the K3 period matrix, but as before, this conclusion must wait until we have the technology to compute these periods explicitly (even numerically).\\

As should be clear from the discussions above, whether or not master integrand bases may be stratified by their rigidity is a detailed question of both the space of numerators considered (as dictated by power-counting or some other rule) and \emph{also} the degree of singularity of the (Calabi-Yau) manifolds involved. It is therefore critical to develop some understanding of the periods supported by these surfaces of higher rigidity. This could involve simply (robust) numeric algorithms for computing the compact contour integrals of a K3 surface on the (real) two-dimensional contours which span its middle-homology; or this may be addressed by better understanding of the differential equations satisfied by these integrands. We leave both of these promising directions for concrete study to future work.

\vspace{\fill}\vspace{-4pt}
\section*{Acknowledgements}%
\vspace{-4pt}
\noindent The authors gratefully acknowledge fruitful conversations with Nima Arkani-Hamed, Claude Duhr, Johannes Henn, Brenda Penante, Cristian Vergu, and Matthias Wilhelm. This project has been supported by an ERC Starting Grant \mbox{(No.\ 757978)}, a grant from the Villum Fonden \mbox{(No.\ 15369)}, by a grant from the US Department of Energy \mbox{(No.\ DE-SC00019066)}.

\newpage
\appendix

\section[Relation between Embedding Space and Momentum-Twistor Space]{Relation between Embedding Space and Momentum-Twistors}\label{appendix:embedding_and_momentum_twistors}\vspace{-0pt}
\subsection{Review of the Embedding Space Formalism}\label{appendix_subsec:embedding_space}\vspace{0pt}
In this work, we will make frequent use of the equivalence between Feynman integrands expressed in dual-momentum coordinates and their representations in embedding space. Although we will not review this formalism in detail here, it is worth emphasizing a few details and some of the notational advantages of viewing these integrands in embedding space. Readers interested in a more comprehensive summary of this formalism may consult the appendix of \cite{Bourjaily:2019exo}. 

In the embedding-space formalism, the loop integrand's differential form 
\eq{d^d\ell_i\mapsto\frac{d^{d+2}|\ell_i)}{GL(1)}\delta\big(\hspace{-1pt}\x{\ell_i}{\ell_i}\hspace{-1pt}\big)\,,\label{embedding_space_image}}
where the new integration variable `$|\ell_i)$'$\in\mathbb{P}^{d+1}$ is represented in $(d{+}2)$ homogeneous coordinates and the factor of `$GL(1)$' in the denominator is an instruction to deprojectivize the $GL(1)$-redundancy of their homogeneous coordinates. Now expressed as an integral over projective space, the loop integrand (not including the differential form \ref{embedding_space_image}) \emph{must} have scaling weight $({-}d)$---which is achieved by adding the requisite factors of `$\x{\ell_i}{\infX}$' involving the point `at infinity' $|\infX)\!\in\!\mathbb{P}^{d+1}$. We apologize to the reader for the abuse in notation for our using `$\x{\r{a}}{\b{b}}$' to denote both `$(x_{\r{a}}{-}x_{\b{b}})^2$' in ordinary, dual loop-momentum space and also the inner product in embedding space---the metric for which is given in \cite{Bourjaily:2019exo}. In most cases, our expressions are well defined in either interpretation of their meaning (justifying our slight abuse of notation). 

For us, the key advantage of using the embedding space formalism is simply that it manifests poles/contours involving `infinite' loop momentum. For example a triangle integrand in four dimensions would be represented by
\eq{\tikzBoxDeux{
\coordinate (v0) at (0,0);\coordinate (v1) at ($(v0)+(0:0.8)$);\coordinate(v2) at ($(v0)+(-120:0.8)$);\coordinate(v3) at ($(v0)+(120:0.8)$);
\coordinate (a1) at ($(v0)+(180:0.75)$);\coordinate(a2) at ($(v0)+(60:0.75)$);\coordinate(a3) at ($(v0)+(-60:0.75)$);\dimLines\draw[int] (v0)--(a1);\draw[int](v0)--(a2);\draw[int](v0)--(a3);\restoreDark\draw[int] (v1)--(v2)--(v3)--(v1);
\legMassive{(v1)}{0}{$$};\legMassive{(v2)}{-120}{$$};\legMassive{(v3)}{120}{$$};\node[rdot] at (a1) {$$};\node[rdot] at (a2) {$$};\node[rdot] at (a3) {};\node[ddot] at (v0) {};\node[anchor=east,inner sep=1pt] at (a1) {$\r{a_1}$};\node[anchor=south west,inner sep=1pt] at (a2) {$\r{a_2}$};\node[anchor=north west,inner sep=1pt] at (a3) {$\r{a_3}$};
}\bigger{\Rightarrow}
\tikzBoxDeux{
\coordinate (v0) at (0,0);\coordinate (v1) at ($(v0)+(-135:0.6928)$);\coordinate(v2) at ($(v0)+(135:0.6928)$);\coordinate(v3) at ($(v0)+(45:0.6928)$);\coordinate(v4) at ($(v0)+(-45:0.6928)$);
\coordinate (a1) at ($(v0)+(180:0.65)$);\coordinate(a2) at ($(v0)+(90:0.65)$);\coordinate(a3) at ($(v0)+(0:0.65)$);\coordinate(a4) at ($(v0)+(-90:0.65)$);\dimLines\draw[int] (v0)--(a1);\draw[int](v0)--(a2);\draw[int](v0)--(a3);\draw[int](v0)--(a3);\draw[int](v0)--(a4);\restoreDark\draw[int] (v1)--(v2)--(v3)--(v4)--(v1);
\legMassive{(v1)}{-135}{$$};\legMassive{(v2)}{135}{$$};\legMassive{(v3)}{45}{$$};\legMassive{(v4)}{-45}{$$};\node[rdot] at (a1) {$$};\node[rdot] at (a2) {$$};\node[rdot] at (a3) {};\node[bldot] at (a4) {};\node[ddot] at (v0) {};\node[anchor=east,inner sep=1pt] at (a1) {$\r{a_1}$};\node[anchor=south,inner sep=1pt] at (a2) {$\r{a_2}$};\node[anchor=west,inner sep=1pt] at (a3) {$\r{a_3}$};\node[anchor=north,inner sep=2pt] at (a4) {$\infX$};
}\bigger{\Leftrightarrow}\hspace{5pt}\frac{1}{\x{\ell}{\r{a_1}}\x{\ell}{\r{a_2}}\x{\ell}{\r{a_3}}\x{\ell}{\infX}}\label{eg_triangle_in_embedding_space}}
where the extra factor of $\x{\ell}{\infX}$ in the denominator makes manifest the fact that the triangle integral does indeed support contours associated with co-dimension four residues which `cut' the three manifest propagators $\x{\ell}{\r{a_i}}$ and then also sending $\ell\!\to\!\infty$. Including $\x{\ell}{\infX}$ as a `propagator' in the expression merely makes this fact more transparent. 

Thus, in the embedding space formalism, all four-dimensional integrands will be represented with effective box power-counting upon inclusion of some power of additional propagators $\x{\ell}{\infX}$. Those \emph{initially} with box power-counting will be independent of $\infX$---leaving open the question of them being dual-conformally invariant; those with triangle power-counting would include additional poles at infinity, and those with bubble or worse power-counting would necessarily have \emph{double} (or higher-degree) poles at infinity. These higher-degree poles (at infinity or otherwise) signal that an integrand will integrate to something with contributions of less than maximal (`transcendental') weight. This fact is of course familiar at one loop, where scalar bubble integrands integrate to functions with transcendental weight $1$ while boxes and triangles both integrate to functions of weight 2. 

Beyond one-loop order, the same analysis applies. However, it is important to note that the point `$\infX$' at infinity is the same for all loop momenta. This is necessitated by the translational invariance of the original loop integrand which would allow, for example, for the change of variables from $(\ell_1,\ell_2)\!\mapsto\!(\ell_1',\ell_2')\equivR(\ell_1,(\ell_1{+}\ell_2))$. Such mixing of loop integration variables necessitates that the point at infinity $\infX$ which defines the map to embedding space must be the same for all loop momenta.

Before discussing what is arguably the most useful manifestation of embedding space for four-dimensional theories, it is worth noting that while in dual-momentum coordinates the space-time points $\r{a_i}$ are strictly, cyclically ordered (clockwise around the graph) by the map connecting these expressions to their momentum-space manifestations, when drawing dual-graphs such as (\ref{eg_triangle_in_embedding_space}), it is important to observe that the point $\infX$ is \emph{not} part of this cyclic ordering: it is time-like separated from all other points. 

\subsection{Momentum-Twistors as Manifestation of Embedding Space}\label{appendix_subsec:momentum_twistors_review}

There is a close technical connection between embedding space and momentum-twistor space, which will be valuable to review briefly here. The map from dual-momentum coordinates to momentum twistor variables defined by Hodges in \cite{Hodges:2009hk} associates with each point $x_{\r{a}}\!\in\!\mathbb{R}^4$ in dual-momentum space the \emph{line} in momentum-twistor space `$(\hspace{-1.4pt}\r{a}\hspace{-1.4pt})$'$\equivR$`$(\r{A}\,\r{a})$'$\equivR\mathrm{span}\hspace{-1pt}\{z_{\r{A}},z_{\r{a}}\}$. Points in twistor space (`momentum twistors') $z_{\r{a}}\!\in\!\mathbb{P}^{3}$ are usually expressed in homogeneous coordinates as four-vectors. Therefore, lines in momentum-twistor space can be viewed as the \emph{span} of a pair of vectors in four dimensions. We can represent them concretely as $2\!\times\!4$ matrix defined up to an $GL(2)$ equivalence class which acts on this matrix by left multiplication (they are elements of the `\emph{Grassmannian}' $Gr(2,4)$). The precise map from dual-momentum space to momentum-twistor space makes use of a canonical line `at infinity', typically `$(\b{I_{\infty}})$'.

The norm-squared separation of points in dual-momentum space, $\x{\r{a}}{\b{b}}\!=\!(x_{\r{a}}{-}x_{\b{b}})^2$ is proportional to $\text{det}\big\{z_{\r{A}},z_{\r{a}},z_{\b{B}},z_{\b{b}}\big\}\equivL\ab{\r{A}\r{a}\,\b{B}\b{b}}\equivL\ab{(\hspace{-1.4pt}\r{a}\hspace{-1.4pt})\,(\hspace{-1.4pt}\b{b}\hspace{-1.4pt})}$; as such, two points in dual-momentum space are light-like separated iff their corresponding lines in twistor space intersect (if they span a combined space of rank $\leq3$). 

Momentum twistors are most useful for the representation of massless particles, where each massless momentum $p_{\r{a}}$ is associated with spinors according to $p_{\r{a}}^\mu\!\mapsto\!p_{\r{a}}^{\alpha\,\dot{\alpha}}\equivR p_{\r{a}}^{\mu}\sigma_{\mu}^{\alpha\,\dot{\alpha}}\equivL\lambda_{\r{a}}^{\alpha}\tilde{\lambda}_{\r{a}}^{\dot{\alpha}}$ where the spinors $(\lambda_{\r{a}},\tilde{\lambda}_{\r{a}})$ are defined up to a $GL(1)$ redundancy from scaling $\lambda_{\r{a}}$ with a compensatory, inverse scaling of $\tilde{\lambda}_{\r{a}}$---the action of the little group. 

In dual-momentum coordinates, momenta are defined via $p_{\r{a}}\equivR(x_{\r{a+1}}{-}x_{\r{a}})$ (with cyclic labeling understood), and points in dual-momentum coordinates by consecutively-labeled pairs of twistors \mbox{$x_{\r{a+1}}\!\Leftrightarrow\!(\r{a{+}1}\,\r{a})$} and \mbox{$x_{\r{a}}\!\Leftrightarrow\!(\r{a}\,\r{a{-}1})$}, thus trivializing the on-shell condition that $p_{\r{a}}^2\!=\!(x_{\r{a+1}}{-}x_{\r{a}})^2\!\propto\!\ab{\r{a{+}1}\,\r{a}\,\r{a}\,\r{a{-}1}}\!=\!0$. The precise map between dual-momentum coordinates and momentum twistor space makes use of the fact that $\lambda_{\r{a}}\!=\!z_{\r{a}}\tncap(\b{I_{\infty}})^{\perp}$ where $(\b{I_{\infty}})^{\perp}$ denotes the two-dimensional span \emph{complementary to} $(\b{I_{\infty}})$; thus, $\ab{\r{a}\,\r{b}\,(\b{I_{\infty}})}\equivL\ab{\r{a}\,\r{b}}\equivR\mathrm{det}\{\lambda_{\r{a}},\lambda_{\r{b}}\}$ and $(x_{\r{a}}{-}x_{\b{b}})^2\!=\!\ab{\r{A}\r{a}\,\b{B}\b{b}}/(\ab{\r{A}\r{a}}\ab{\b{B}\b{b}})$.

The space of `lines' in momentum-twistor space is five-dimensional projective space $\mathbb{P}^5$---spanned homogeneously, \emph{e.g.}\ by the $\binom{4}{2}$ pairs of twistors forming a basis for homogeneous coordinates on $\mathbb{P}^3$. The images of dual-momentum coordinates, being represented by the spans of \emph{pairs} of momentum-twistors satisfy an additional condition of `simplicity': that $\ab{(\hspace{-1.4pt}\r{a}\hspace{-1.4pt})\,(\hspace{-1.4pt}\r{a}\hspace{-1.4pt})}\!=\!0$. The space of \emph{simple} lines in momentum-twistor space is therefore four dimensional. 

Mapping the loop integration variable $x_{\ell_i}$ to its image as a line in momentum-twistor space, therefore, manifests the embedding space formalism---with the restriction $\delta\big(\x{\ell_i}{\ell_i}\big)$ in (\ref{embedding_space_image}) being simply the statement of bi-twistor simplicity. Moreover, the line at infinity in momentum-twistor space $(\b{I_{\infty}})$ can be viewed as equivalent to what we have denoted $\infX$ throughout this work. 

Thus, in addition to the possibility of interpreting our expressions either in \mbox{(dual-)}momentum space or embedding space, it may sometimes be useful to view loop integrands as being defined in momentum-twistor space. This leads to yet another abuse of notation throughout our work (limited for the most part to a few concrete examples discussed in the following section): we occasionally choose interpret a loop-dependent numerator `$\x{\ell_i}{\b{N}}$' not merely as involving a point $x_{\b{N}}$ in dual-momentum space or by an element $|\b{N})$ in embedding space, but by a \emph{line} in momentum-twistor space spanned by a pair of momentum-twistors $(z_{\b{N_1}}\,z_{\b{N_2}})$. As the maps between these interpretations are all very explicit, we hope the reader will forgive the occasional, explicit formula given in one interpretation or the other.  

To be clear, although momentum-twistors do prove very useful, the connections between embedding space and momentum-twistors can be largely ignored by most readers. And we have made sure that the majority of this work can be understood without needing such interpretations. That being said, it turns out that momentum-twistors provide a much better formalism to give explicit expressions for solutions to cut-equations due largely to the fact that many more solutions are rational as opposed to algebraic (see \emph{e.g.}~\cite{Bourjaily:2018aeq} for a thorough discussion of this fact). This is why momentum twistors make an appearance in a few concrete examples discussed below and used also in the concrete expressions provided in the ancillary files to this work.

\subsection{Momentum-Twistor Notation and Geometry}\label{appendix_subsec:momentum_twistor_notation}
In momentum-twistor space, the cut-condition for an inverse propagator to vanish $\x{\ell}{\r{a}}\!\propto\!\ab{\ell^1\,\ell^2\,\r{A}\,\r{a}}\equivL\ab{(\hspace{-1.25pt}\r{\ell}\hspace{-1.25pt})\,(\hspace{-1.4pt}\r{a}\hspace{-1.4pt})}$ becomes the geometric condition that the pair of lines `$(\hspace{-1.0pt}\ell\hspace{-1.0pt})$' and `$(\hspace{-1.4pt}\r{a}\hspace{-1.4pt})$' intersect. The the solution to such equations, it is often useful to discuss quantities such as the `plane' spanned by three momentum-twistors as in
\eq{\big(\r{a}\,\r{b}\,\r{c}\big)\equivR\mathrm{span}\{z_{\r{a}},z_{\r{b}},z_{\r{c}}\}\,.}
Sometimes, we use notation such as $\big(\!(\hspace{-1.4pt}\r{a}\hspace{-1.4pt})\r{b}\big)\equivR\big(\r{A}\,\r{a}\,\r{b}\big)$ to discus planes involving a \emph{line} $(\hspace{-1.4pt}\r{a}\hspace{-1.4pt})$ directly associated with a pair of momentum twistors encoding a point in dual-momentum coordinates $x_{\r{a}}$. We do appreciate how difficult this makes some of our formulae to read. To be clear: a single argument in parenthesis will \emph{always} represent a bi-twistor $(\hspace{-1.4pt}\r{a}\hspace{-1.4pt})\equivR(z_{\r{A}}\,z_{\r{a}})$.

One particularly useful geometric construction is the point of intersection between a line and a plane, denoted
\eq{\big(\r{a}\,\r{b}\big)\tcap\big(\b{c}\,\b{d}\,\b{e}\big)\equivR z_{\r{a}}\ab{\r{b}\,\b{c}\,\b{d}\,\b{e}}{+}z_{\r{b}}\ab{\b{c}\,\b{d}\,\b{e}\,\r{a}}={-}\big[z_{\b{c}}\ab{\b{d}\,\b{e}\,\r{a}\,\r{b}}{+}z_{\b{d}}\ab{\b{e}\,\r{a}\,\r{b}\,\b{c}}{+}z_{\b{e}}\ab{\r{a}\,\r{b}\,\b{c}\,\b{d}}\big]\label{line_with_plane_definition}}
which follows directly from Cramer's rule (the general five-term identity satisfied by any four-vectors). Similarly useful is the \emph{line} spanned by the intersection of two planes:
\eq{\begin{split}\big(\r{a}\,\r{b}\,\r{c}\big)\tcap\big(\b{d}\,\b{e}\,\b{f}\big)&\equivR\,\,\,\,\, (\r{a}\,\r{b})\ab{\r{c}\,\b{d}\,\b{e}\,\b{f}}{+}(\r{b}\,\r{c})\ab{\r{a}\,\b{d}\,\b{e}\,\b{f}}{+}(\r{c}\,\r{a})\ab{\r{b}\,\b{d}\,\b{e}\,\b{f}}\\
&=\!{-}\!\big[(\b{d}\,\b{e})\ab{\b{f}\,\r{a}\,\r{b}\,\r{c}}{+}(\b{e}\,\b{f})\ab{\b{d}\,\r{a}\,\r{b}\,\r{c}}{+}(\b{f}\,\b{d})\ab{\b{e}\,\r{a}\,\r{b}\,\r{c}}\big]\,.
\end{split}\label{two_planes_intersection_definition}}

\section{Details of the Box-Triangle Master Integrand Basis}\label{appendix:box_triangle_masters}
In this appendix, we provide complete details for the space of master integrands for the box-triangle, its period matrices, and the master integrands diagonalized with respect to rigidity. 

Let us begin with a concrete description of the space of master integrands for triangle power-counting. As described in \mbox{section \ref{subsubsec:box_triangle}}, these integrands take the form of pentaboxes in the embedding formalism (which manifests their support on poles at infinite loop momentum):
\eq{\bT{
\coordinate (ella) at ($(v5)!.5!(v1)+(90:\figScale*0.65)$);
\coordinate (ellb) at ($(v5)!.5!(v3)+(0:\figScale*0.325)$);
\coordinate (a1) at ($(v5)!.5!(v1)+(-90:0.25)$);
\coordinate (a2) at ($(v1)!.5!(v2)+(180:0.25)$);
\coordinate (a3) at ($(v2)!.5!(v3)+(90:0.25)$);
\coordinate (b1) at ($(v3)!.5!(v4)+(58:0.25)$);
\coordinate (b2) at ($(v4)!.5!(v5)+(-58:0.25)$);
\dimLines
\restoreDark
\draw[int](v5)--(v1);\draw[int](v1)--(v2);\draw[int](v2)--(v3);\draw[int](v3)--(v4);\draw[int](v4)--(v5);\draw[int](v5)--(v3);
\legMassive{(v1)}{-135}{$$}\legMassive{(v2)}{135}{$$}\legMassive{(v3)}{71.5}{$$}\legMassive{(v4)}{0}{$$}\legMassive{(v5)}{-71.5}{$$}\draw[markedEdge] (v1)--(v2);
\node[rdot] at (a1){};
\node[rdot] at (a2){};
\node[rdot] at (a3){};
\node[rdot] at (b1){};
\node[rdot] at (b2){};
\node[] at (ella){$\ell_1$};
\node[] at (ellb){$\ell_2$};
\node at ($(a1)+(-90:0.195)$) {\text{{\normalsize$\r{a_1}$}}};
\node at ($(a2)+(180:0.25)+(90:0.05)$) {\text{{\normalsize$\r{a_2}$}}};
\node at ($(a3)+(90:0.195)$) {\text{{\normalsize$\r{a_3}$}}};
\node at ($(b1)+(36:0.25)+(90:0.05)$) {\text{{\normalsize$\r{b_1}$}}};
\node at ($(b2)+(-36:0.25)+(-90:0.05)$) {\text{{\normalsize$\r{b_2}$}}};
}\bigger{\Rightarrow}\pBox{\pBoxPlainEdges
\legMassive{(v1)}{-110}{$$}\legMassive{(v2)}{180}{$$}\legMassive{(v3)}{110}{$$}\legMassive{(v4)}{81}{$$}\legMassive{(v5)}{45}{$$}\legMassive{(v6)}{-45}{$$}
\coordinate (ella) at ($(v7)!.5!(v4)+(180:0.575)$);\coordinate (ellb) at ($(v7)!.5!(v4)+(0:\figScale*0.45)$);\coordinate (a1) at ($(v7)!.0!(v1)+(-72:0.25)$);\coordinate (a2) at ($(v1)!.5!(v2)+(-144:0.25)$);\coordinate (a3) at ($(v2)!.5!(v3)+(144:0.25)$);\coordinate (a4) at ($(v3)!.5!(v4)+(72:0.25)$);\coordinate (b1) at ($(v4)!.5!(v5)+(90:0.205)$);\coordinate (b2) at ($(v5)!.5!(v6)+(0:0.25)$);
\node[bldot] at (a1){};\node[rdot] at (a2){};\node[rdot] at (a3){};\node[rdot] at (a4){};\node[rdot] at (b1){};\node[rdot] at (b2){};
\node at ($(ella)$) {\text{{\normalsize${\ell_1}$}}};\node at ($(ellb)$) {\text{{\normalsize${\ell_2}$}}};
\node at ($(a1)+(-72:0.195)+(-90:0.05)+(0:0.0)$) {\text{{\normalsize$\infX$}}};
\node at ($(a2)+(-144:0.195)+(-90:0.1)+(0:0.2)$) {\text{{\normalsize$\r{a_1}$}}};
\node at ($(a3)+(144:0.195)+(90:0.05)+(0:0.2)$) {\text{{\normalsize$\r{a_2}$}}};
\node at ($(a4)+(72:0.195)$) {\text{{\normalsize$\r{a_3}$}}};
\node at ($(b1)+(80:0.195)+(0:0.1)+(80:0.05)$) {\text{{\normalsize$\r{b_1}$}}};
\node at ($(b2)+(0:0.25)$) {\text{{\normalsize$\r{b_2}$}}};
\draw[markedEdge] (v2)--(v3);
}\bigger{\simeq}\pBox{
\coordinate (ella) at ($(v7)!.5!(v4)+(180:0.575)$);\coordinate (ellb) at ($(v7)!.5!(v4)+(0:\figScale*0.45)$);\coordinate (a1) at ($(v7)!.0!(v1)+(-72:0.25)$);\coordinate (a2) at ($(v1)!.5!(v2)+(-144:0.25)$);\coordinate (a3) at ($(v2)!.5!(v3)+(144:0.25)$);\coordinate (a4) at ($(v3)!.5!(v4)+(72:0.25)$);\coordinate (b1) at ($(v4)!.5!(v5)+(90:0.205)$);\coordinate (b2) at ($(v5)!.5!(v6)+(0:0.25)$);
\dimLines\pBoxPlainEdges
\legMassive{(v1)}{-110}{$$}\legMassive{(v2)}{180}{$$}\legMassive{(v3)}{110}{$$}\legMassive{(v4)}{81}{$$}\legMassive{(v5)}{45}{$$}\legMassive{(v6)}{-45}{$$}\restoreDark
\draw[int](ella)--(a1);\draw[int](ella)--(a2);\draw[int](ella)--(a3);\draw[int](ella)--(a4);\draw[int](ella)--(ellb);\draw[int](ellb)--(b1);\draw[int](ellb)--(a1);\draw[int](ellb)--(b2);
\node[ddot]at (ella){};\node[ddot]at (ellb){};\node[bldot] at (a1){};\node[rdot] at (a2){};\node[rdot] at (a3){};\node[rdot] at (a4){};\node[rdot] at (b1){};\node[rdot] at (b2){};
\node at ($(a1)+(-72:0.195)+(-90:0.05)+(0:0.0)$) {\text{{\normalsize$\infX$}}};
\node at ($(a2)+(-144:0.195)+(-90:0.1)+(0:0.2)$) {\text{{\normalsize$\r{a_1}$}}};
\node at ($(a3)+(144:0.195)+(90:0.05)+(0:0.2)$) {\text{{\normalsize$\r{a_2}$}}};
\node at ($(a4)+(72:0.195)$) {\text{{\normalsize$\r{a_3}$}}};
\node at ($(b1)+(80:0.195)+(0:0.1)+(80:0.05)$) {\text{{\normalsize$\r{b_1}$}}};
\node at ($(b2)+(0:0.25)$) {\text{{\normalsize$\r{b_2}$}}};
}\label{box_triangle_as_pentabox}}
Therefore, the space of master integrands in which we are interested take the form 
\eq{\mathcal{I}_i^0\equivR\dbar^4\!\ell_1\dbar^4\!\ell_2\frac{\x{\ell_1}{\b{N_i^0}}}{\x{\ell_1}{\infX}\x{\ell_1}{\r{a_1}}\x{\ell_1}{\r{a_2}}\x{\ell_1}{\r{a_3}}\x{\ell_1}{\ell_2}\x{\ell_2}{\r{b_1}}\x{\ell_2}{\r{b_2}}\x{\ell_2}{\b{X}}\,}\label{initial_master_integrands_for_box_triangle}}
representing a six-dimensional space of integrands. For our initial choice of the master integrands' numerators, we will divide the space according to
\eq{\begin{split}\mathfrak{n}(\ell_1)\!\in\![\ell_1]&=\mathrm{span}\Big\{\!\underbrace{\x{\ell_1}{\b{N_1^0}},\x{\ell_1}{\b{N_2^0}},\x{\ell_1}{\b{N_3^0}}}_{\text{`top-level'}},\underbrace{\x{\ell_1}{\r{a_1}},\x{\ell_1}{\r{a_2}},\x{\ell_1}{\r{a_3}}}_{\text{`contact-terms'}}\!\Big\}\\
&\equivL\,\mathrm{span}\big\{\x{\ell_1}{\b{N_i^0}}\big\}\bigger{\oplus}\,\,\mathrm{span}\big\{\x{\ell_1}{\r{N_i^0}}\big\}\,.\end{split}}
Although we will mostly be interested in the top-level degrees of freedom, it is worth remarking that the contact terms correspond to scalar double-triangle integrals of the form of polylogarithmic double-boxes in embedding space:
\eq{\dT{
\coordinate (ella) at ($(v4)!.5!(v2)+(180:0.325)$);
\coordinate (ellb) at ($(v4)!.5!(v2)+(0:0.325)$);
\coordinate (a1) at ($(v4)!.5!(v1)+(-135:0.25)$);
\coordinate (a2) at ($(v1)!.5!(v2)+(135:0.25)$);
\coordinate (b1) at ($(v2)!.5!(v3)+(45:0.25)$);
\coordinate (b2) at ($(v3)!.5!(v4)+(-45:0.25)$);
\dimLines
\draw[int] (ella)--(a1);
\draw[int] (ella)--(a2);
\draw[int] (ella)--(ellb);
\draw[int] (ellb)--(b1);
\draw[int] (ellb)--(b2);
\restoreDark
\draw[int](v4)--(v1);\draw[int](v1)--(v2);\draw[int](v2)--(v3);\draw[int](v3)--(v4);\draw[int](v4)--(v2);\legMassive{(v1)}{180}{$$}\legMassive{(v2)}{90}{$$}\legMassive{(v3)}{0}{$$}\legMassive{(v4)}{-90}{$$}
\node[ddot] at (ella) {};
\node[ddot] at (ellb) {};
\node[rdot] at (a1) {};\node[rdot] at (a2) {};\node[rdot] at (b1) {};\node[rdot] at (b2) {};
\node[anchor=north] at ($(a1)+(-135:0)+(-90:0.0)+(0:0)$) {\text{{\normalsize$\r{a_j}$}}};
\node[anchor=south] at ($(a2)+(135:0.)+(0:0.)$) {\text{{\normalsize$\r{a_k}$}}};
\node[anchor=south] at ($(b1)+(135:0.)+(0:0.)$) {\text{{\normalsize$\r{b_1}$}}};
\node[anchor=north] at ($(b2)+(-135:0)+(90:0.125)+(0:0)$) {\text{{\normalsize$\r{b_2}$}}};
}\hspace{-5pt}\bigger{\Rightarrow}\hspace{-5pt}\dT{
\coordinate (ella) at ($(v4)!.5!(v2)+(180:0.325)$);
\coordinate (ellb) at ($(v4)!.5!(v2)+(0:0.325)$);
\coordinate (a1) at ($(v4)!.5!(v1)+(-135:0.25)$);
\coordinate (a2) at ($(v1)!.5!(v2)+(135:0.25)$);
\coordinate (b1) at ($(v2)!.5!(v3)+(45:0.25)$);
\coordinate (b2) at ($(v3)!.5!(v4)+(-45:0.25)$);
\dimLines\draw[int](v4)--(v1);\draw[int](v1)--(v2);\draw[int](v2)--(v3);\draw[int](v3)--(v4);\draw[int](v4)--(v2);\legMassive{(v1)}{180}{$$}\legMassive{(v2)}{90}{$$}\legMassive{(v3)}{0}{$$}\legMassive{(v4)}{-90}{$$}
\restoreDark
\draw[int] (ella)--(a1);
\draw[int] (ella)--(a2);
\draw[int] (ella)--(ellb);
\draw[int] (ellb)--(b1);
\draw[int] (ellb)--(b2);
\node[ddot] at (ella) {};
\node[ddot] at (ellb) {};
\node[rdot] at (a1) {};\node[rdot] at (a2) {};\node[rdot] at (b1) {};\node[rdot] at (b2) {};
\node[anchor=north] at ($(a1)+(-135:0)+(-90:0.0)+(0:0)$) {\text{{\normalsize$\r{a_j}$}}};
\node[anchor=south] at ($(a2)+(135:0.)+(0:0.)$) {\text{{\normalsize$\r{a_k}$}}};
\node[anchor=south] at ($(b1)+(135:0.)+(0:0.)$) {\text{{\normalsize$\r{b_1}$}}};
\node[anchor=north] at ($(b2)+(-135:0)+(90:0.125)+(0:0)$) {\text{{\normalsize$\r{b_2}$}}};
}\hspace{-5pt}\bigger{\simeq}\hspace{6pt}\dBox{\dimLines
\draw[int](v5)--(v1);\draw[int](v1)--(v2);\draw[int](v2)--(v3);\draw[int](v3)--(v4);\draw[int](v4)--(v5);\draw[int](v6)--(v3);\legMassive{(v1)}{-135}{$$}\legMassive{(v2)}{135}{$$}\legMassive{(v3)}{90}{$$}\legMassive{(v4)}{45}{$$}\legMassive{(v5)}{-45}{$$}
\restoreDark
\coordinate (ella) at ($(v6)!.5!(v1)+(90:\figScale*0.65)$);
\coordinate (ellb) at ($(v6)!.5!(v5)+(90:\figScale*0.65)$);
\coordinate (a1) at ($(v6)!.0!(v1)+(-90:0.25)$);
\coordinate (a2) at ($(v1)!.5!(v2)+(180:0.25)$);
\coordinate (a3) at ($(v3)!.55!(v2)+(90:0.25)$);
\coordinate (b1) at ($(v3)!.55!(v4)+(90:0.25)$);
\coordinate (b2) at ($(v4)!.5!(v5)+(0:0.25)$);
\coordinate (b3) at ($(v6)!.55!(v5)+(-90:0.25)$);
\draw[int] (ella)--(a1);
\draw[int] (ella)--(a2);
\draw[int] (ella)--(a3);
\draw[int] (ella)--(ellb);
\draw[int] (ellb)--(b1);
\draw[int] (ellb)--(b2);
\draw[int] (ellb)--(a1);
\node[bldot] at (a1){};
\node[rdot] at (a2){};
\node[rdot] at (a3){};
\node[rdot] at (b1){};
\node[rdot] at (b2){};
\node[ddot] at (ella){};
\node[ddot] at (ellb){};
\node at ($(a2)+(180:0.275)+(90:0.0)$) {\text{{\normalsize$\r{a_j}$}}};
\node at ($(a3)+(90:0.25)$) {\text{{\normalsize$\r{a_k}$}}};
\node at ($(b1)+(0:0.05)+(90:0.25)$) {\text{{\normalsize$\r{b_1}$}}};
\node at ($(b2)+(0:0.275)+(90:0.05)$) {\text{{\normalsize$\r{b_2}$}}};
\node at ($(a1)+(-90:0.15)+(0:0.2)$) {\text{{\normalsize$\,\,\,\,\b{\infX}$}}};
}\fwboxL{0pt}{\,\,\,.}
}
The appropriate choice of contour on which to normalize such integrands corresponds to the seven-cut encircling all seven propagators, $\Omega[(\b{X},\r{a_j},\r{a_k});(\r{b_1},\r{b_2},\b{X})]$, followed by the parity-even sum (the difference) of simple poles-contours defined by the collinearity of the momenta flowing through the three propagators involved in the three-point vertex, $\x{\ell_1}{\b{X}}, \x{\ell_2}{\b{X}},\x{\ell_1}{\ell_2}$. On this contour, 
\vspace{5pt}\eq{\fwbox{0pt}{\hspace{-30pt}\oint\limits_{\substack{\Omega[(\b{X},\r{a_j},\r{a_k});(\r{b_1},\r{b_2},\b{X})]\hspace{-40pt}\\\text{collinear-pole}\hspace{-40pt}}}
\hspace{-15pt}\dbar^4\!\ell_1\dbar^4\!\ell_2\frac{\x{\ell_1}{\r{a_i}}}{\x{\ell_1}{\infX}\x{\ell_1}{\r{a_1}}\x{\ell_1}{\r{a_2}}\x{\ell_1}{\r{a_3}}\x{\ell_1}{\ell_2}\x{\ell_2}{\r{b_1}}\x{\ell_2}{\r{b_2}}\x{\ell_2}{\b{X}}}=\underset{[(\r{a_j}\r{a_k});\hspace{-0pt}(\r{b_1}\r{b_2})]}{\Delta^{-1}}}}
where $\{\r{a_j},\r{a_k}\}\equivR\{\r{a_1},\r{a_2},\r{a_3}\}\backslash\{\r{a_i}\}$ and where 
\vspace{-10pt}\eq{\begin{split}\hspace{-5pt}\underset{\fwbox{0pt}{[(\r{a_1}\r{a_2});\hspace{-1pt}(\r{b_1}\r{b_2})]}}{\Delta^{2}}\equivR\hspace{6pt}&\fwboxL{330pt}{\Big(\!\ab{\big(\b{X}(\hspace{-1.4pt}\r{a_1}\hspace{-1.4pt})\!\big)\tcap\big(\!(\hspace{-1.4pt}\r{a_2}\hspace{-1.4pt})\!\big)\big(\!(\hspace{-1.4pt}\r{b_1}\hspace{-1.4pt})\!\big)\tcap\big(\!(\hspace{-1.4pt}\r{b_2}\hspace{-1.4pt})\,\b{x}\big)(\hspace{-1.4pt}\b{X}\hspace{-1.4pt})}{+}
\ab{\big(\b{x}(\hspace{-1.4pt}\r{a_1}\hspace{-1.4pt})\!\big)\tcap\big(\!(\hspace{-1.4pt}\r{a_2}\hspace{-1.4pt})\!\big)\big(\!(\hspace{-1.4pt}\r{b_1}\hspace{-1.4pt})\!\big)\tcap\big(\!(\hspace{-1.4pt}\r{b_2}\hspace{-1.4pt})\,\b{X}\big)(\hspace{-1.4pt}\b{X}\hspace{-1.4pt})}\!\Big)^2}\\
&\fwboxL{330pt}{{-}4\,\ab{\big(\b{X}(\hspace{-1.4pt}\r{a_1}\hspace{-1.4pt})\!\big)\tcap\big(\!(\hspace{-1.4pt}\r{a_2}\hspace{-1.4pt})\!\big)\big(\!(\hspace{-1.4pt}\r{b_1}\hspace{-1.4pt})\!\big)\tcap\big(\!(\hspace{-1.4pt}\r{b_2}\hspace{-1.4pt})\,\b{X}\big)(\hspace{-1.4pt}\b{X}\hspace{-1.4pt})}\ab{\big(\b{x}(\hspace{-1.4pt}\r{a_1}\hspace{-1.4pt})\!\big)\tcap\big(\!(\hspace{-1.4pt}\r{a_2}\hspace{-1.4pt})\!\big)\big(\!(\hspace{-1.4pt}\r{b_1}\hspace{-1.4pt})\!\big)\tcap\big(\!(\hspace{-1.4pt}\r{b_2}\hspace{-1.4pt})\,\b{x}\big)(\hspace{-1.4pt}\b{X}\hspace{-1.4pt})}}
\end{split}\label{alice_normalization}}
with the point at infinity represented by the pair of twistors $(\hspace{-1.4pt}\b{X}\hspace{-1.4pt})\equivR(\b{X}\,\b{x})$. Appropriately normalized to be prescriptive on these contours, the contact-term master integrands with numerators given by
\eq{\biketB{\r{N_i}}\equivR\biketC{\r{a_i}}\,\,\,\underset{\fwbox{0pt}{[(\r{a_j}\r{a_k});\hspace{-1pt}(\r{\vec{b}})]}}{\Delta}\label{normalized_alice_contact_terms}}
will be dual-conformally invariant and integrate to pure, weight-four polylogarithms---the expressions for which are known \cite{MatthiasWilhelm}.

Let us now discuss an initial choice of top-level numerators for our master integrand basis.

\subsection{Explicit Numerators for the Initial Master Integrands}\label{appendix:box_triangle_numerators_detail}

As the space of master integrands defined in (\ref{initial_master_integrands_for_box_triangle}) are essentially pentaboxes, it is easy to choose reasonably-good initial numerators. In particular, we choose to take one numerator to simply be $\biketB{\b{N_1^0}}\equivR\biketB{\b{X}}$---viewed as a `contact-term' for the elliptic-double-box involving the points $(\r{\vec{a}}),(\r{b_1}\r{b_2}\b{X})$---and the other two to be the parity even/odd numerators that would appear in the basis of \cite{Bourjaily:2015jna} when the integral is viewed as a pentabox of the form shown in (\ref{box_triangle_as_pentabox}). 

Thus, our initial choice for top-level master integrand numerators will be $\x{\ell_1}{\b{N_i^0}}$ where 
\eq{\fwbox{0pt}{\hspace{-10pt}\biketB{\b{N_1^0}}\equivR\biketB{\b{X}},\;\;\biketB{\b{N_2^0}}\equivR\frac{1}{6}\epsilon^{ijk}\big|\b{Q^o_{{\color{black}ijk}}}\big),\;\;\biketB{\b{N_3^0}}\equivR\biketB{\r{b_1}}\x{\r{b_2}}{\b{X}}\x{\r{a_1}}{\r{a_3}}\x{\r{a_2}}{\b{X}}\Delta{-}\big(\r{b_1}\!\leftrightarrow\!\r{b_2}\big)}\label{initial_box_triangle_numerators}}
where $\epsilon^{ijk}$ is the totally-antisymmetric Levi-Cevita symbol, 
\eq{\begin{split}\hspace{-215pt}\big|\b{Q^o_{{\color{black}ijk}}}\big)\equivR&\fwboxL{0pt}{\Big[\big(\!(\hspace{-1.4pt}e_i\hspace{-1.4pt})\r{A_3}\big)\tcap\big(\big(\r{a_3}(\hspace{-1.4pt}e_j\hspace{-1.4pt})\!\big)\tcap\big(\!(\hspace{-1.4pt}e_k\hspace{-1.4pt})\!\big)(\hspace{-1.4pt}\r{b_1}\hspace{-1.4pt})\!\big){-}
\Big(\r{A_3}\big(\big(\r{a_3}(\hspace{-1.4pt}e_i\hspace{-1.4pt})\!\big)\tcap\big(\!(\hspace{-1.4pt}e_j\hspace{-1.4pt})\!\big)(\hspace{-1.4pt}\r{b_1}\hspace{-1.4pt})\!\big)\tcap\big(\!(\hspace{-1.4pt}e_k\hspace{-1.4pt})\!\big)\!\Big)\Big]\x{\r{b_1}}{\b{X}}}\\
\hspace{-215pt}&\fwboxL{0pt}{{-}\big((\hspace{-1.4pt}\r{b_1}\hspace{-1.4pt})\!\leftrightarrow\!(\hspace{-1.4pt}\r{b_2}\hspace{-1.4pt})\big){-}\big(\r{A_3}\!\leftrightarrow\!\r{a_3}\big)\hspace{30pt}\text{with}\qquad \{e_1,e_2,e_3\}\equivR\{(\hspace{-1.4pt}\r{a_1}\hspace{-1.4pt}),(\hspace{-1.4pt}\r{a_2}\hspace{-1.4pt}),(\hspace{-1.4pt}\b{X}\hspace{-1.4pt})\},}
\end{split}}
and
\eq{\Delta\equivR\sqrt{\left(1{-}\frac{\x{\r{a_1}}{\r{a_2}}\x{\r{a_3}}{\b{X}}}{\x{\r{a_1}}{\r{a_3}}\x{\r{a_2}}{\b{X}}}{-}\frac{\x{\b{X}}{\r{a_1}}\x{\r{a_2}}{\r{a_3}}}{\x{\r{a_1}}{\r{a_3}}\x{\r{a_2}}{\b{X}}}\right)^2{-}4\,\frac{\x{\r{a_1}}{\r{a_2}}\x{\r{a_3}}{\b{X}}}{\x{\r{a_1}}{\r{a_3}}\x{\r{a_2}}{\b{X}}}\,\frac{\x{\b{X}}{\r{a_1}}\x{\r{a_2}}{\r{a_3}}}{\x{\r{a_1}}{\r{a_3}}\x{\r{a_2}}{\b{X}}}}\,.\label{delta_normalization}}

There are several aspects of these initial numerators that deserve some remark. First, notice that while $|\b{N_{2,3}^0}\big)$ correspond to dual-conformally invariant integrands, $\biketB{\b{N_1^0}}$ is \emph{not} properly normalized: we have made this choice for simplicity of the resulting formulae below, but must emphasize that once diagonalized on (any choice) of contours, the resulting basis \emph{will} be dual-conformal. (Alternatively, a compensatory factor of $\x{\r{a_1}}{\r{a_3}}\x{\r{b_1}}{\b{X}}\x{\r{a_2}}{\r{b_2}}$ may be included in the definition of $\biketB{\b{N_1^0}}$ to render it dual-conformally invariant (while still far from pure).) 

Secondly, while the definition of $\biketB{\b{N_2^0}}$ above is unquestionably rather involved, it is simply the parity-odd combination of the chiral integrands defined in \cite{Bourjaily:2015jna} which are generated by `merging' a chiral one-loop box and a scalar triangle integrand. The $\epsilon$-symbol and anti-symmetrization is part of the definition of (the momentum-twistor version of) the parity-odd, chiral box integrand of \cite{Bourjaily:2013mma}. Interested readers can find complete details for these numerators (and their rotated forms) in the ancillary files for this work.

Finally, it is important to emphasize that the integrands defined above vanish on the parity-even contours of all double-triangle contact terms. As such, we are justified in considering the space of master integrands to be three-dimensional.

\newpage
\subsection{Explicit Form of the Period Matrix and Diagonalized Masters}\label{appendix:box_triangle_initial_period_matrix_detail}
The elliptic-containing seven-cut used to define all of the spanning set of contours can be parameterized as done in \cite{Bourjaily:2020hjv,Bourjaily:2021vyj}. The first step is to parameterize the six-cut which encircles the vanishing of each propagators \emph{except} $\x{\ell_1}{\ell_2}$ by representing the loop momenta (as lines in momentum-twistor space) by $(\ell_1)\equivR(\!\r{\hat{\,a_1\!}}\,\r{\hat{\,a_3\!}}\,)$ and $(\ell_2)\equivR(\!\r{\hat{\,b_1\!}}\,\,\b{\hat{x}})$ where 
\eq{\begin{array}{r@{}lr@{}l}\r{\hat{\,a_1\!}}\,(\g{\alpha})&\equivR z_{\r{a_1}}\!+\!\g{\alpha}\,z_{\r{A_1}}\hspace{20pt}~&\r{\hat{\,b_1\!}}\,(\g{\beta})&\equivR z_{\r{b_1}}\!+\!\g{\beta}\,z_{\r{B_1}}\\
\r{\hat{\,a_3\!}}\,(\g{\alpha})&\equivR\big(\!(\hspace{-1.4pt}\r{a_3}\hspace{-1.4pt})\!\big)\tcap\big(\!\r{\hat{\,a_1\!}}\,(\hspace{-1.4pt}\r{a_2}\hspace{-1.4pt})\!\big)&\b{\hat{x}}(\g{\beta})&\equivR\big(\!(\hspace{-1.4pt}\b{X}\hspace{-1.4pt})\!\big)\tcap\big(\!\r{\hat{\,b_1\!}}\,(\hspace{-1.4pt}\r{b_2}\hspace{-1.4pt})\!\big)\,.\end{array}}
In this parameterization, the remaining propagator of the seven-cut would be parameterized by $\x{\ell_1}{\ell_2}=\ab{\r{\hat{\,a_1\!}}\,\r{\hat{\,a_3\!}}\,\r{\hat{\,b_1\!}}\,\,\b{\hat{x}}}$---which is degree-two in both $\g{\alpha},\g{\beta}$. We can eliminate either variable---we choose to eliminate $\g{\beta}$---by taking the contour encircling $\x{\ell_1}{\ell_2}\!=\!0$ as following from 
\eq{\oint\limits_{\x{\ell_1}{\ell_2}=0}\!\!\!\!\frac{\dbar\g{\beta}}{\x{\ell_1}{\ell_2}}=\pm i\frac{1}{y(\g{\alpha})}}
where $y^2(\g{\alpha})$ is a quartic polynomial given by
\eq{\begin{split}y^2(\g{\alpha})\equivR&\phantom{{-}}\hspace{-10pt}\Big(\ab{\!\r{\hat{\,a_1\!}}\,\r{\hat{\,a_3\!}}\,\,\r{B_1}\,\big(\!(\hspace{-1.4pt}\b{X}\hspace{-1.4pt})\!\big)\tcap\big(\r{b_1}\,(\hspace{-1.4pt}\r{b_2}\hspace{-1.4pt})\!\big)}{+}\ab{\!\r{\hat{\,a_1\!}}\,\r{\hat{\,a_3\!}}\,\,\r{b_1}\,\big(\!(\hspace{-1.4pt}\b{X}\hspace{-1.4pt})\!\big)\tcap\big(\r{B_1}\,(\hspace{-1.4pt}\r{b_2}\hspace{-1.4pt})\!\big)}\Big)^2\\
&{-}4\ab{\!\r{\hat{\,a_1\!}}\,\r{\hat{\,a_3\!}}\,\,\r{B_1}\,\big(\!(\hspace{-1.4pt}\b{X}\hspace{-1.4pt})\!\big)\tcap\big(\r{B_1}\,(\hspace{-1.4pt}\r{b_2}\hspace{-1.4pt})\!\big)}\ab{\!\r{\hat{\,a_1\!}}\,\r{\hat{\,a_3\!}}\,\,\r{b_1}\,\big(\!(\hspace{-1.4pt}\b{X}\hspace{-1.4pt})\!\big)\tcap\big(\r{b_1}\,(\hspace{-1.4pt}\r{b_2}\hspace{-1.4pt})\!\big)}\end{split}\label{quartic_poly}}
and the particular solutions $\g{\beta}^*_{\pm}$ to the equation $\ab{\r{\hat{\,a_1\!}}\,\r{\hat{\,a_3\!}}\,\r{\hat{\,b_1\!}}\,\,\b{\hat{x}}}\!=\!0$ are given by 
\eq{\g{\beta}^*_{\pm}(\g{\alpha})\equivR\frac{\ab{\!\r{\hat{\,a_1\!}}\,\r{\hat{\,a_3\!}}\,\,\r{B_1}\,\big(\!(\hspace{-1.4pt}\r{b_2}\hspace{-1.4pt})\,\r{b_1}\big)\tcap\big(\!(\hspace{-1.4pt}\b{X}\hspace{-1.4pt})\!\big)}{+}\ab{\!\r{\hat{\,a_1\!}}\,\r{\hat{\,a_3\!}}\,\,\r{b_1}\,\big(\!(\hspace{-1.4pt}\r{b_2}\hspace{-1.4pt})\,\r{B_1}\big)\tcap\big(\!(\hspace{-1.4pt}\b{X}\hspace{-1.4pt})\!\big)}{\pm}y(\g{\alpha})}{2\,\ab{\!\r{\hat{\,a_1\!}}\,\r{\hat{\,a_3\!}}\,\,\r{B_1}\,\big(\!(\hspace{-1.4pt}\b{X}\hspace{-1.4pt})\!\big)\tcap\big(\r{B_1}\,(\hspace{-1.4pt}\r{b_2}\hspace{-1.4pt})\!\big)}}\,.}
For future reference, it is worth noting that $y^2(\g{\alpha})$ given in (\ref{quartic_poly}) is not monic; its leading coefficient (that of $\g{\alpha}^4$) is given by 
\eq{\begin{split}
\hspace{-214pt}y_4^2\equivR&\fwboxL{100pt}{\phantom{}\!\Big(\!\ab{\r{A_1}\!\big(\r{A_1}(\hspace{-1.4pt}\r{a_2}\hspace{-1.4pt})\!\big)\tncap\big(\!(\hspace{-1.4pt}\r{a_3}\hspace{-1.4pt})\!\big)\r{B_1}\big(\r{b_1}(\hspace{-1.4pt}\r{b_2}\hspace{-1.4pt})\!\big)\tncap\big(\!(\hspace{-1.4pt}\b{X}\hspace{-1.4pt})\!\big)}{-}\ab{\r{A_1}\!\big(\r{A_1}(\hspace{-1.4pt}\r{a_2}\hspace{-1.4pt})\!\big)\tncap\big(\!(\hspace{-1.4pt}\r{a_3}\hspace{-1.4pt})\!\big)\r{b_1}\big(\r{B_1}(\hspace{-1.4pt}\r{b_2}\hspace{-1.4pt})\!\big)\tncap\big(\!(\hspace{-1.4pt}\b{X}\hspace{-1.4pt})\!\big)}\!\Big)^2}\\
\hspace{-212pt}&\fwboxL{100pt}{{-}4\ab{\r{A_1}\!\big(\r{A_1}(\hspace{-1.4pt}\r{a_2}\hspace{-1.4pt})\!\big)\tncap\big(\!(\hspace{-1.4pt}\r{a_3}\hspace{-1.4pt})\!\big)\r{b_1}\big(\r{b_1}(\hspace{-1.4pt}\r{b_2}\hspace{-1.4pt})\!\big)\tncap\big(\!(\hspace{-1.4pt}\b{X}\hspace{-1.4pt})\!\big)}\ab{\r{A_1}\!\big(\r{A_1}(\hspace{-1.4pt}\r{a_2}\hspace{-1.4pt})\!\big)\tncap\big(\!(\hspace{-1.4pt}\r{a_3}\hspace{-1.4pt})\!\big)\r{B_1}\big(\r{B_1}(\hspace{-1.4pt}\r{b_2}\hspace{-1.4pt})\!\big)\tncap\big(\!(\hspace{-1.4pt}\b{X}\hspace{-1.4pt})\!\big)}\,.}\end{split}\label{leading_y_coeff}}

We will be interested in taking such contours on loop integrands that include other (as-yet-uncut) propagators. As these residual integrands will be rational functions of $\g{\alpha},\g{\beta}$, we can always decompose the result into the form: 
\eq{\oint\limits_{\substack{\x{\ell_1}{\r{a_i}}=0\\\x{\ell_2}{\r{b_i}}=0,\x{\ell_2}{\b{X}}=0}}\!\!\!\!\!\!\!\!\!\!\!\!\!\!\mathcal{I}(\ell_1,\ell_2)\equivL\oint\limits_{\x{\ell_1}{\ell_2}=0}\!\!\!\!\frac{\dbar\g{\beta}}{\x{\ell_1}{\ell_2}}\widehat{\mathcal{I}}(\g{\alpha},\g{\beta})\equivL\,\, i\left[R(\g{\alpha})\pm\frac{1}{y(\g{\alpha})}S(\g{\alpha})\right]\label{general_int_on_seven_cut}}
where $R(\g{\alpha})$ and $S(\g{\alpha})$ are defined by canonicalizing the square-root $y(\g{\alpha})$; thus, to expose the ellipticity, we may define $\Omega$ to be \emph{half} the difference between the contours encircling the $\g{\beta}^*_\pm$ solutions to the seventh cut equation $\x{\ell_1}{\ell_2}\!=\!0$, respectively. This has the result of isolating the term proportional to $1/y(\g{\alpha})$ in (\ref{general_int_on_seven_cut})---that part involving the geometry of the elliptic curve defined by the quartic $y^2(\g{\alpha})$. 

In the case of our box-triangle integral, the only propagator \emph{not} cut by the seven-cut is $1/\x{\ell_1}{\b{\infX}}\!=\!1/\ab{\!\r{\hat{\,a_1\!}}\,\r{\hat{\,a_3\!}}\,\,(\hspace{-1.4pt}\b{X}\hspace{-1.4pt})}$. On the six-(or seven-)cut of interest, this propagator is independent of $\g{\beta}$, and is simply a quadratic polynomial in $\g{\alpha}$. Thus, the three top-level master integrands for the box-triangle integrand given in (\ref{initial_master_integrands_for_box_triangle}) with numerators chosen according to (\ref{initial_box_triangle_numerators}) take the form\footnote{The careful reader will note that the relative${-}$sign of (\ref{initial_ints_on_seven_cut}) to that of (\ref{general_int_on_seven_cut}) follows from the fact that the six-dimensional contour involves $i^{6}\!=\!{-}1$ relative to the residue.} 
\eq{\oint\limits_{\Omega[\r{\vec{a}};(\r{b_1},\r{b_2},\b{X})]}\hspace{-17pt}\mathcal{I}_{i}^0={-}i\,\dbar\g{\alpha}\frac{\x{\ell_1(\g{\alpha})}{\b{N_i^0}}}{y(\g{\alpha})\x{\ell_1(\g{\alpha})}{\infX}}\label{initial_ints_on_seven_cut}}
on the `elliptic'-containing contour. It is worth noting that the factor $\x{\ell_1}{\b{\infX}}\!=\!\ab{\!\r{\hat{\,a_1\!}}\,\r{\hat{\,a_3\!}}\,\,(\hspace{-1.4pt}\b{X}\hspace{-1.4pt})}$ has two roots at $\g{\alpha}\!=\!q_{\pm}$ given by:
\eq{{q}_{\pm}\equivR\frac{\ab{\r{A_1}\big(\r{a_1}(\hspace{-1.4pt}\b{X}\hspace{-1.4pt})\!\big)\tcap\big(\!(\hspace{-1.4pt}\r{a_3}\hspace{-1.4pt})\!\big)\,(\hspace{-1.4pt}\r{a_2}\hspace{-1.4pt})}{-}\ab{\r{A_1}\big(\r{a_1}(\hspace{-1.4pt}\r{a_2}\hspace{-1.4pt})\!\big)\tcap\big(\!(\hspace{-1.4pt}\r{a_3}\hspace{-1.4pt})\!\big)\,(\hspace{-1.4pt}\b{X}\hspace{-1.4pt})}\pm\Delta\x{\r{a_1}}{\r{a_3}}\x{\r{a_2}}{\b{X}}}{2\,\ab{\r{A_1}(\hspace{-1.4pt}\r{a_2}\hspace{-1.4pt})\,\big(\!(\hspace{-1.4pt}\r{a_3}\hspace{-1.4pt})\!\big)\tcap\big(\r{A_1}\,(\hspace{-1.4pt}\b{X}\hspace{-1.4pt})\!\big)}}}
where $\Delta$ was defined above in (\ref{delta_normalization}). 

The precise form of these roots are not that important to us. However, the initial numerators $\biketB{\b{N_2^0}},\biketB{\b{N_3^0}}$ were chosen to have `unit residues' on contours that enclose these poles; that means that they evaluate to $\pm1$ on any contour which start from $\Omega$ which then encircles the even/odd combinations of solutions $\g{\alpha}\!=\!{q}_{\pm}$ to cutting the final propagator $\x{\ell_1}{\b{X}}\!=\!0$. Thus, we can more simply write the one-forms appearing in (\ref{initial_ints_on_seven_cut}) as
\eq{\begin{split}
\oint\limits_{\Omega[\r{\vec{a}};(\r{b_1},\r{b_2},\b{X})]}\hspace{-17pt}\mathcal{I}_1^0\equivL&\fwboxL{195pt}{{-}i\frac{\dbar\g{\alpha}}{y(\g{\alpha})}}\equivL\,\omega_1(\g{\alpha})\\
\oint\limits_{\Omega[\r{\vec{a}};(\r{b_1},\r{b_2},\b{X})]}\hspace{-17pt}\mathcal{I}_2^0\equivL&\fwboxL{195pt}{{-}i\frac{\dbar\g{\alpha}}{y(\g{\alpha})}\left[n_{2}^1\,{+}\left(\frac{y({q}_{+})}{(\g{\alpha}{-}{q}_+)}{-}\frac{y({q}_{-})}{(\g{\alpha}{-}{q}_-)}\right)\right]}\equivL\, n_2^1\,\omega_1(\g{\alpha}){+}\omega_2(\g{\alpha})\\
\oint\limits_{\Omega[\r{\vec{a}};(\r{b_1},\r{b_2},\b{X})]}\hspace{-17pt}\mathcal{I}_3^0\equivL&\fwboxL{195pt}{{-}i\frac{\dbar\g{\alpha}}{y(\g{\alpha})}\left[n_3^1\,{+}\left(\frac{y({q}_{+})}{(\g{\alpha}{-}{q}_+)}{+}\frac{y({q}_{-})}{(\g{\alpha}{-}{q}_-)}\right)\right]}\equivL\, n_3^1\,\omega_1(\g{\alpha}){+}\omega_3(\g{\alpha})\\
\end{split}\label{initial_nums_on_heptacuts}}
where we have introduced the differential forms $\omega_i(\g{\alpha})$, defined implicitly above. and the coefficients of $\omega_1$ appearing above are given by 
\eq{\begin{split}n_2^1\equivR&\frac{1}{6\,\ab{\r{A_1}\big(\!(\hspace{-1.4pt}\r{a_3}\hspace{-1.4pt})\!\big)\tcap\big(\!(\hspace{-1.4pt}\r{a_2}\hspace{-1.4pt})\r{A_1}\big)(\hspace{-1.4pt}\b{X}\hspace{-1.4pt})}}\epsilon^{ijkl}\ab{\big(\!(\hspace{-1.4pt}e_i\hspace{-1.4pt})\!\big)\tcap\big(\r{A_1}(\hspace{-1.4pt}e_j\hspace{-1.4pt})\!\big)\big(\!(\hspace{-1.4pt}e_k\hspace{-1.4pt})\!\big)\tcap\big(\r{A_1}(\hspace{-1.4pt}e_l\hspace{-1.4pt})\!\big)(\hspace{-1.4pt}\r{b_1}\hspace{-1.4pt})}\x{\r{b_2}}{\b{X}}\\
&{-}\big((\hspace{-1.4pt}\r{b_1}\hspace{-1.4pt})\leftrightarrow(\hspace{-1.4pt}\r{b_2}\hspace{-1.4pt})\big)\qquad\qquad\text{where}\quad \{e_1,e_2,e_3,e_4\}\equivR\{(\hspace{-1.4pt}\r{a_1}\hspace{-1.4pt}),(\hspace{-1.4pt}\r{a_2}\hspace{-1.4pt}),(\hspace{-1.4pt}\r{a_3}\hspace{-1.4pt}),(\hspace{-1.4pt}\b{X}\hspace{-1.4pt})\}\end{split}}
where $\epsilon$ is the totally-antisymmetric Levi-Cevita symbol, and 
\eq{n_3^1\equivR\Delta\frac{\x{\r{a_1}}{\r{a_3}}\x{\r{a_2}}{\b{X}}}{\ab{\r{A_1}(\hspace{-1.4pt}\r{a_2}\hspace{-1.4pt})\big(\!(\hspace{-1.4pt}\r{a_3}\hspace{-1.4pt})\!\big)\tcap\big(\r{A_1}(\hspace{-1.4pt}\b{X}\hspace{-1.4pt})\!\big)}}\x{\r{b_2}}{\b{X}}\ab{\r{A_1}(\hspace{-1.4pt}\r{b_2}\hspace{-1.4pt})\big(\!(\hspace{-1.4pt}\r{a_3}\hspace{-1.4pt})\!\big)\tcap\big(\!(\hspace{-1.4pt}\r{b_1}\hspace{-1.4pt})\r{A_1}\big)}{-}\big((\hspace{-1.4pt}\r{b_1}\hspace{-1.4pt})\leftrightarrow(\hspace{-1.4pt}\r{b_2}\hspace{-1.4pt})\big)
}
where $\Delta$ is the algebraic factor defined in (\ref{delta_normalization}) above. 

Already, the form of the differential forms on $\Omega[\r{\vec{a}};(\r{b_1},\r{b_2},\b{X})]$ in (\ref{initial_nums_on_heptacuts}) suggests that we may clean up our basis of master numerators by replacing them according to 
\eq{\begin{split}
\biketB{\b{N_2^0}}\mapsto\biketB{\b{N_2}}&\equivR\biketB{\b{N_2^0}}{-}n_2^1\biketB{\b{N_1}}\\
\biketB{\b{N_3^0}}\mapsto\biketB{\b{N_3}}&\equivR\biketB{\b{N_3^0}}{-}n_3^1\biketB{\b{N_1}}\end{split}\label{rotated_numerators}}
which will ensure that on the seven-cut contour $\Omega[\r{\vec{a}};(\r{b_1},\r{b_2},\b{X})]$, the new integrals will be diagonal on the three differential forms $\omega_i(\g{\alpha})$.\\

We are now ready to compute explicit period matrices for the new integrands defined by $\biketB{N_i}$ defined in (\ref{rotated_numerators}). There are four natural choices of contours available: the $a$ or $b$-cycles of the elliptic curve, and the even/odd combinations of polylogarithmic contours enclosing the simple poles $(\g{\alpha}{-}q_{\pm})$. Any choice of 3 of these cycles results in a rank-3 period matrix that can be diagonalized so that exactly one linear combination has support on each cycle; however, it is easy to confirm that all of the diagonalized integrals will have support on the integration cycle \emph{not} chosen for the diagonalization. For example, following \cite{Bourjaily:2021vyj} we may be inclined to choose the polylogarithmic contours and either the $a$ or $b$ cycle of the elliptic curve; this would result in a basis for which all three integrands had non-vanishing elliptic support---and thus non-trivial elliptic contributions. 

The most natural choice of three cycles, therefore, would be both the $a$- and $b$-cycles of the elliptic curve together with the (parity-)even polylogarithmic contour (divided by two). Let us denote these three contour integrals by $\Omega_a$, $\Omega_b$ and $\Omega_{\text{poles}}^e$, respectively. To be clear, by these we mean the 8-dimensional compact contours defined by \emph{first} encircling the vanishing the seven propagators of the double-box according to $\Omega[\r{\vec{a}};(\r{b_1},\r{b_2},\b{X})]$, and then taking one of these final contour integrals on the remaining, one-dimensional differential form. 

To be concrete about these cycles of integration, we must specify our conventions regarding the roots of the quartic $y^2(\g{\alpha})$ which defines the elliptic curve. Labeling these roots $r_i$, we may write
\eq{y^2(\g{\alpha})\equivL\,\, y_4^2\,\,(\g{\alpha}{-}r_1)(\g{\alpha}{-}r_2)(\g{\alpha}{-}r_3)(\g{\alpha}{-}r_4)\,,}
where the prefactor $y_4^2$ was given in (\ref{leading_y_coeff}). For Euclidean kinematics, the roots of the quartic come in two complex-conjugate pairs, $\{r_1,r_2\}$ and $\{r_3,r_4\}$; and we choose to order these roots so that $\mathfrak{Re}(r_1)\!>\!\mathfrak{Re}(r_3)$ and $\mathfrak{Im}(r_{1,3})\!>\!0$. With these conventions, the $a$-cycle may be defined to be the \emph{half} of the integral which encloses the branch cut between $r_{1}$ and $r_2$, and the $b$-cycle \emph{half} that which encloses the cut between $r_{1}$ and $r_3$.\footnote{These two cycles are distinguished in Euclidean kinematics, as the $a$-cycle encircles the pair of roots which are complex conjugates. In particular, this means that the $b$-cycle cannot degenerate at co-dimension one while remaining in the Euclidean domain.}
With these conventions, the period integrals of the three differential forms $\omega_i(\g{\alpha})$ appearing in (\ref{initial_nums_on_heptacuts}) may be expressed
\eq{\fwboxL{0pt}{\raisebox{-9pt}{$\hspace{44pt}\left(\rule{0pt}{29pt}\right.$}}\begin{array}{@{}l@{}|@{$\;\;$}c@{}c@{}c@{}c@{}c@{}c@{}c@{}c@{}c@{}}
\fwbox{40pt}{\mathfrak{n}(\ell_1)\!\!\,\,}&\fwbox{38pt}{{\Omega_a}}&\fwbox{38pt}{\Omega_b}&\fwbox{38pt}{\Omega_{\text{poles}}^e}
\\\hline\\[-12pt]
\x{\ell_1}{\b{N_1}}&J_1^a&J_1^b&0\\
\x{\ell_1}{\b{N_2}}&J_2^a&J_2^b&0\\
\x{\ell_1}{\b{N_3}}&J_3^a&J_3^b&1\\
\end{array}\fwboxL{0pt}{\raisebox{-9pt}{$\hspace{-6.5pt}\left.\rule{0pt}{29pt}\right)\!\!\!\equivL\mathbf{M}$}}\label{box_triangle_period_matrix}}
where 
\eq{\begin{split}
J_1^a\equivR&\oint\limits_{\text{a-cycle}}\!\!\omega_1(\g{\alpha})={-}\frac{1}{\pi\,y_4\sqrt{r_{14}\,r_{23}}}K[\phi]\\
J_2^a\equivR&\oint\limits_{\text{a-cycle}}\!\!\omega_2(\g{\alpha})=\frac{1}{\pi\,y_4\sqrt{r_{14}\,r_{23}}}\frac{y(q_+)}{q^+_4}\left(\!K[\phi]{+}\frac{r_{24}}{q^+_2}\,\Pi\!\left[\frac{q^+_4r_{12}}{q^+_2r_{14}},\phi\right]\right){-}\big(q_+\!\leftrightarrow\!q_-\big)\\
J_3^a\equivR&\oint\limits_{\text{a-cycle}}\!\!\omega_3(\g{\alpha})=\frac{1}{\pi\,y_4\sqrt{r_{14}\,r_{23}}}\frac{y(q_+)}{q^+_4}\left(\!K[\phi]{+}\frac{r_{24}}{q^+_2}\,\Pi\!\left[\frac{q^+_4r_{12}}{q^+_2r_{14}},\phi\right]\right){+}\big(q_+\!\leftrightarrow\!q_-\big)
\end{split}\label{a_cycle_periods}}
where $r_{i\,j}\equivR(r_{i}{-}r_{j})$, $q^{\pm}_i\equivR (q_{\pm}{-}r_i)$, 
\eq{\phi\equivR\frac{(r_2{-}r_1)(r_3{-}r_4)}{(r_2{-}r_3)(r_1{-}r_4)}\equivL\frac{r_{21}r_{34}}{r_{23}r_{14}}\,,}
and $K[\phi]$ and $\Pi[a,\phi]$ represent the standard, complete elliptic integrals of the first and third kinds, respectively. We have been careful to ensure that the results given here match the standard conventions of \textsc{Mathematica}'s implementation of these functions. With only one small exception, the $b$-cycle integrals can be obtained from those in (\ref{a_cycle_periods}) by simply swapping $r_2\!\leftrightarrow\!r_3$ (which has the effect of exchanging $\phi\!\leftrightarrow\!(1{-}\phi)$):
\begin{align}
J_1^b\equivR&\oint\limits_{\text{b-cycle}}\!\!\omega_1(\g{\alpha})={-}\frac{1}{\pi\,y_4\sqrt{r_{14}\,r_{32}}}K[1{-}\phi]\label{b_cycle_periods}\\
J_2^b\equivR&\oint\limits_{\text{b-cycle}}\!\!\omega_2(\g{\alpha})=\frac{1}{\pi\,y_4\sqrt{r_{14}\,r_{32}}}\frac{y(q_+)}{q^+_4}\left(\!K[1{-}\phi]{+}\frac{r_{24}}{q^+_3}\,\Pi\!\left[\frac{q^+_4r_{13}}{q^+_3r_{14}},1{-}\phi\right]\right){-}\big(q_+\!\leftrightarrow\!q_-\big)\nonumber\\
J_3^b\equivR&\oint\limits_{\text{b-cycle}}\!\!\omega_3(\g{\alpha})=\frac{1}{\pi\,y_4\sqrt{r_{14}\,r_{32}}}\frac{y(q_+)}{q^+_4}\left(\!K[\phi]{+}\frac{r_{24}}{q^+_3}\,\Pi\!\left[\frac{q^+_4r_{13}}{q^+_3r_{14}},1{-}\phi\right]\right){+}\big(q_+\!\leftrightarrow\!q_-\big){-}\frac{1}{2}\nonumber
\end{align}
where the `${-}\frac{1}{2}$' in the final expression is the result of the permuted expression including a simple pole contribution for the standard conventions defining the functions $K$ and $\Pi$ in \textsc{Mathematica}, say. 

Although this is not manifest from their representations above, it worth noting that, in the Euclidean domain, all $J_i^a$ are real and all $J_i^b$ are pure imaginary. In particular, this means that the determinant is pure imaginary:
\eq{\begin{split}\det\!\big(\mathbf{M}\big)&=J_1^a\,J_2^b{-}J_1^b\,J_2^a\\
&=\frac{i}{\pi^2y_4^2}\frac{y(q_+)r_{34}}{q_3^+q_4^+r_{14}r_{23}}K[\phi]\Pi\left[\frac{q_4^+r_{13}}{q_3^+r_{14}},1{-}\phi\right]{-}\big(q_+\!\leftrightarrow\!q_-\big){-}\big(r_2\!\leftrightarrow\!r_3\big)\,.\end{split}\label{triangle_box_period_det}}

In order to find the diagonalized master integrands, we require the inverse of the period matrix $\mathbf{M}$:
\eq{\mathbf{M}^{-1}=\frac{1}{\det\!\big(\mathbf{M}\big)}\left(\begin{array}{ccc}J_2^b&{-}J_1^b\phantom{{-}}&0\\
{-}J_2^a\phantom{{-}}&J_1^a&0\\
(J_2^a\,J_3^b{-}J_2^b\,J_3^a)&(J_1^b\,J_3^a{-}J_1^a\,J_3^b)&(J_1^a\,J_2^b{-}J_1^b\,J_2^a)\end{array}\right)}
Thus, our final, diagonalized master integrands can be written as those involving numerators $\x{\ell_1}{\b{\overline{N_i}}}$ defined by
\eq{\begin{split}
\biketB{\b{\overline{N_1}}}\equivR&\frac{1}{\det\!\big(\mathbf{M}\big)}\Big[\phantom{{-}}J_2^b\biketB{\b{N_1}}{-}J_1^b\biketB{\b{N_2}}\Big]\\
\biketB{\b{\overline{N_2}}}\equivR&\frac{1}{\det\!\big(\mathbf{M}\big)}\Big[{-}J_2^a\biketB{\b{N_1}}{+}J_1^a\biketB{\b{N_2}}\Big]\\
\biketB{\b{\overline{N_3}}}\equivR&\frac{1}{\det\!\big(\mathbf{M}\big)}\Big[(J_2^a\,J_3^b{-}J_2^b\,J_3^a)\biketB{\b{N_1}}{+}(J_1^b\,J_3^a{-}J_1^a\,J_3^b)\biketB{\b{N_2}}\Big]{+}\biketB{\b{N_3}}\,.\\
\end{split}\label{prescriptive_numerators_for_box_triangle}}

\newpage
\section{Content of the Ancillary Files}\label{appendix:ancillary_files}
Included as part of this work's submission to the \texttt{arXiv}, we have prepared ancillary files which include complete, analytic expressions for the example of the box-triangle integrands discussed in \mbox{appendix \ref{appendix:box_triangle_masters}}. In particular, we provide expressions for the initial numerators $\x{\ell_1}{\b{N_i^0}}$ in (\ref{initial_box_triangle_numerators}) (including the correctly normalized contact-terms in (\ref{normalized_alice_contact_terms})), their slightly rotated versions $\x{\ell_1}{\b{N_i}}$ in (\ref{rotated_numerators}), the expressions appearing in the period matrix (\ref{box_triangle_period_matrix}), and the diagonalized basis with numerators $\x{\ell_1}{\overline{\b{N_i}}}$ defined in (\ref{prescriptive_numerators_for_box_triangle}).

These results, together with explicit tools for numerical evaluation, may be found in the \textsc{Mathematica} package `\texttt{box\uscore triangle\uscore master\uscore integrands\uscore tools.m}', the usage of which is described and illustrated with examples in the \textsc{Mathematica} notebook `\texttt{box\uscore triangle\uscore master\uscore integrands\uscore walkthrough.nb}'. These computational tools borrow code from the packages developed in the works \cite{Bourjaily:2010wh,Bourjaily:2012gy,Bourjaily:2013mma,Bourjaily:2015jna}.

Finally, for the sake of the interested reader, we have provided Feynman-parametric representations of each of the (rotated, but not-yet diagonalized) master integrands involving numerators $\x{\ell_1}{\b{N_i}}$. These integrands were constructed using the strategy described in \cite{Bourjaily:2018aeq,Bourjaily:2019jrk,Bourjaily:2019vby} and are given as parametric integrands of the form
\eq{\int\limits_{\ell_i\in\mathbb{R}^4}\!\!\!\mathcal{I}_{\b{i}}(\ell_1,\ell_2)\equivL\int\limits_{0}^{\infty}\!d^4\!\g{\vec{\alpha}}\,d^3\!\g{\vec{\beta}}\,\,\,\mathcal{I}(\g{\vec{\alpha}},\g{\vec{\beta}})\,,}
as a five-fold Feynman-parametric integrand with parameters $\g{\alpha_i},\g{\beta_i}$ which should be integrated over all of $\mathbb{R}_{+}$.

\newpage

\providecommand{\href}[2]{#2}\begingroup\raggedright\endgroup
\end{document}